\documentclass[11pt,a4paper]{article} 
\pdfoutput=1

\usepackage{jheppub}                  

\usepackage{graphicx}
\usepackage[T1]{fontenc}
\usepackage[normalem]{ulem}	

\def\lsim{\raise0.3ex\hbox{$\;<$\kern-0.75em\raise-1.1ex
\hbox{$\sim\;$}}}
\def\gsim{\raise0.3ex\hbox{$\;>$\kern-0.75em\raise-1.1ex
\hbox{$\sim\;$}}}

\def\thetitle{ 
Standard versus Non-Standard CP Phases in Neutrino Oscillation in Matter with Non-Unitarity \\
}
\title{\thetitle}
\hypersetup{pdftitle={\thetitle}}
\author{Ivan Martinez-Soler$^{a,b,c,d}$}
\author{Hisakazu Minakata$^{a,e}$}

\affiliation{
  $^a$Instituto F\'{\i}sica Te\'{o}rica, UAM/CSIC, Calle Nicola's Cabrera 13-15, Cantoblanco E-28049 Madrid, Spain \\
  $^b$Theoretical Physics Department, Fermi National Accelerator Laboratory, P.O. Box 500, Batavia IL 60510, USA \\
  $^c$Department of Physics and Astronomy, Northwestern University, Evanston, IL 60208, USA \\
  $^d$Colegio de F\'isica Fundamental e Interdisciplinaria de las Am\'ericas (COFI), 254 Norzagaray street, San Juan, Puerto Rico 00901 \\
  $^e$Center for Neutrino Physics, Department of Physics, Virginia Tech, Blacksburg, Virginia 24061, USA \\
}

\emailAdd{ivan.martinezsoler@northwestern.edu}
\emailAdd{minakata71@vt.edu}
\date{\today}

\abstract{ 
We formulate a perturbative framework for the flavor transformation of the standard three active neutrinos but with non-unitary flavor mixing matrix, a system which may be relevant for leptonic unitarity test. We use the $\alpha$ parametrization of the non-unitary matrix and take its elements $\alpha_{\beta \gamma}$ ($\beta,\gamma = e,\mu,\tau$) and the ratio $\epsilon \simeq \Delta m^2_{21} / \Delta m^2_{31}$ as the small expansion parameters. Qualitatively new two features that hold in all the oscillation channels are uncovered in the probability formula obtained to first order in the expansion: (1) The phases of the complex $\alpha$ elements always come in into the observable in the particular combination with the $\nu$SM CP phase $\delta$ in the form $[e^{- i \delta } \bar{\alpha}_{\mu e}, ~e^{ - i \delta} \bar{\alpha}_{\tau e}, ~\bar{\alpha}_{\tau \mu}]$ under the PDG convention of unitary $\nu$SM mixing matrix. (2) The diagonal $\alpha$ parameters appear in particular combinations $\left( a/b - 1 \right) \alpha_{ee} + \alpha_{\mu \mu}$ and $\alpha_{\mu \mu} - \alpha_{\tau \tau}$, where $a$ and $b$ denote, respectively, the matter potential due to CC and NC reactions. This property holds only in the unitary evolution part of the probability, and there is no such feature in the genuine non-unitary part, while the $\delta$ - $\alpha$ parameter phase correlation exists for both. The reason for such remarkable stability of the phase correlation is discussed. 
}

\begin{document} 

\begin{flushright}
IFT-UAM/CSIC-18-061\footnote{First version of this paper was released while the both authors were at: Instituto F\'{\i}sica Te\'{o}rica, UAM/CSIC, Calle Nicola's Cabrera 13-15, Cantoblanco E-28049 Madrid, Spain.}
\end{flushright}

\maketitle

\section{Introduction} 
\label{sec:introduction}

It appears that by now the three flavor lepton mixing \cite{Maki:1962mu} is well established after the long term best endeavor by the experimentalists, which are recognized in an honorable way \cite{Kajita:2016cak,McDonald:2016ixn}. Though we do not know the value of CP phase $\delta$, the lepton Kobayashi-Maskawa (KM) phase \cite{Kobayashi:1973fv}, and the neutrino mass ordering, there appeared some hints toward identifying these unknowns. Continuing search for CP violation in T2K \cite{Abe:2018wpn} entailed a strong indication of it at nearly $3\sigma$ confidence level (CL) \cite{Abe:2019vii} with the best fit around $\delta \sim \frac{3\pi}{2}$. It is awaiting confirmation by NO$\nu$A \cite{Acero:2019ksn}. The value of $\delta$ is the best possible one for determination of the mass ordering, as can be seen clearly by the bi-probability plot introduced in ref.~\cite{Minakata:2001qm}. 

In fact, quite consistently, preference of the normal mass ordering over the inverted one has already been seen in the atmospheric neutrino observation by Super-Kamiokande~\cite{Abe:2017aap}. Currently, it is strengthened by the ongoing long-baseline (LBL) experiments~\cite{Abe:2018wpn,Acero:2019ksn} only modestly, but there exists a tantalizing possibility that coming high-statistics results could change the scene.
A recent global analysis~\cite{Capozzi:2018ubv} showed for the first time that preference of the normal ordering can be claimed at 3$\sigma$ CL. See also refs.~\cite{Esteban:2018azc,deSalas:2018bym}. 
The apparent convergence of various results from dozens of experiments suggests that we may reach a stage of knowing the remaining unknowns in a definitive way, possibly at a time earlier than we thought, for example, by T2HK \cite{Abe:2015zbg}, DUNE \cite{Abi:2020evt}, ESS$\nu$SB \cite{Baussan:2013zcy}, JUNO \cite{An:2015jdp}, T2KK\footnote{
A possible acronym for the setting, ``Tokai-to-Kamioka observatory-Korea neutrino observatory'', an updated version of the one used in ref.~\cite{Kajita:2006bt}.} \cite{Abe:2016ero}, 
INO \cite{Kumar:2017sdq}, IceCube-Gen2/PINGU \cite{TheIceCube-Gen2:2016cap}, or KM3NeT/ORCA \cite{Adrian-Martinez:2016zzs}. 

It then prompts us to think about how to conclude the era of discovery of neutrino mass and the lepton flavor mixing. One of the most important key elements is the paradigm test, that is, to verify the standard three flavor mixing scheme of neutrinos to a high precision. As in the quark sector, unitarity test is the most popular, practical way of carrying this out.

A favourable way of performing a leptonic unitarity test is to formulate a model independent generic framework in which unitarity is violated, and confront it to the experimental data. It was attempted in a pioneering work by Antusch et al. \cite{Antusch:2006vwa}, which indeed provided such a framework in the context of high-scale unitarity violation (UV).\footnote{
We are aware that in the physics literature UV usually means ``ultraviolet''. But, in this paper UV is used as an abbreviation for ``unitarity violation'' or ``unitarity violating''.  }
In low-scale UV, on the other hand, the currently available model is essentially unique, the 3 active plus $N_{s}$ sterile model, see e.g., refs.~\cite{Schechter:1980gr,Barger:1980tfa} for a partial list of the early references. In the present context, low and high scales imply, typically, energy scales of new physics much lower and higher than the electroweak scale, respectively. Recently, within the $(3+N_{s})$ model, a model-independent framework is created to describe neutrino propagation in vacuum \cite{Fong:2016yyh} and in matter \cite{Fong:2017gke} in such a way that the observable quantities are insensitive to details of the sterile sector, e.g., its mass spectrum and active-sterile mixing.

In this paper, we construct a perturbative framework by which we can derive a simple expression of the neutrino oscillation probability in matter in the presence of UV. The framework has the two kind of expansion parameters, the ratio $\epsilon \approx \Delta m^2_{21} / \Delta m^2_{31}$ (precise definition is in eq.~\eqref{renem-param}), and the UV parameters, hence dubbed as the ``helio-UV perturbation theory'' in this paper. It can be regarded as an extension of the ``renormalized helio perturbation theory''~\cite{Minakata:2015gra} to include non-unitarity, which allows us to discuss UV flavor transition of neutrinos with sizable matter effect. Our experience tells us that the relatively simple and at the same time revealing formulas for the oscillation probability, e.g., as in refs.~\cite{Arafune:1997hd,Cervera:2000kp}, help us to understand physics of the three-flavor mixing. 

Then, the natural question would be: What is the qualitatively revealing aspects of the approximate formulas for the probability we derive? To answer the question, we quote the following two features:  

\begin{itemize}

\item 
Universal channel-independent correlation between the neutrino-mass embedded Standard Model ($\nu$SM) CP phase $\delta$ and the complex UV parameter phases. 

\item 
Universal correlation between the diagonal $\alpha$ parameters in the unitary part of the oscillation probability, and absence of such correlation in the non-unitary part. 

\end{itemize}
\noindent
For the first point, if we use the $\alpha$ matrix to parametrize the UV effect \cite{Escrihuela:2015wra} (see section~\ref{sec:convention-dependence} for definition) it takes the form $[e^{- i \delta } \bar{\alpha}_{\mu e}, ~e^{ - i \delta} \bar{\alpha}_{\tau e}, ~\bar{\alpha}_{\tau \mu}]$ under the Particle Data Group (PDG) convention \cite{Tanabashi:2018oca} of the flavor mixing MNS matrix $U_{\text{\tiny MNS}}$.\footnote{
The phase correlation is $U_{\text{\tiny MNS}}$ convention dependent, see sections~\ref{sec:convention-dependence} and \ref{sec:convention-dep}. For example, in the ATM ($e^{ \pm i \delta}$ attached to $s_{23}$) convention of $U_{\text{\tiny MNS}}$, it takes the form $[e^{- i \delta } \alpha_{\mu e}, ~~ \alpha_{\tau e}, ~~ e^{i \delta} \alpha_{\tau \mu}]$. 
}
It generalizes the earlier observation of $e^{- i \delta } \bar{\alpha}_{\mu e}$ correlation done in analyzing the data \cite{Miranda:2016wdr,Abe:2017jit} (see also \cite{Escrihuela:2015wra}) to all the complex $\alpha$ matrix elements in an analytic way. 
The second statement necessitates as a prerequisite a clear separation of unitary and non-unitary parts of the $S$ matrix as well as the oscillation probability. 

One can argue that neutrino evolution must be unitary even in the presence of non-unitary mixing matrix, because only the three active neutrinos span the complete state space of neutral leptons at low energies. There is no way to go outside the complete state space during propagation, assuming absence of inelastic scattering, absorption, etc. Then, the question is: Why and how does the oscillation probability not respect unitarity? We will answer this question in section~\ref{sec:basis-relations}. 

One of the appealing features of our formalism is that it applies both to the high-scale UV and the $3 \times 3$ active three-flavor subspace in low-scale UV, the leading order in $W$ expansion where $W$ is the mixing matrix elements connecting the active and the sterile sectors, see refs.~\cite{Fong:2016yyh} and \cite{Fong:2017gke}. 
High-scale unitarity violation is a well studied subject with many references, only part of which is quoted here \cite{Blennow:2016jkn,Antusch:2006vwa,FernandezMartinez:2007ms,Goswami:2008mi,Antusch:2009pm,Antusch:2009gn,Antusch:2014woa,Escrihuela:2015wra,Fernandez-Martinez:2016lgt,Miranda:2016wdr,Ge:2016xya,Dutta:2016vcc,Dutta:2016czj,Pas:2016qbg,Escrihuela:2016ube,Rout:2017udo,Abe:2017jit,Li:2018jgd}. 
Low-scale and high-scale UV have a number of characteristic features that distinguishes with each other. They include presence (high-scale) or absence (low-scale) of flavor non-universality and zero-distance flavor transition. Recent additions to this list are the presence of the probability leaking term in the oscillation probability \cite{Fong:2016yyh} and possible detection of UV perturbative corrections \cite{Fong:2017gke}, both of which would testify for the low-scale UV. See refs.~\cite{Parke:2015goa,Blennow:2016jkn} for the current constraints on unitarity violation in low-scale UV scenario. 

In section~\ref{sec:phase-correlation}, we start with a pedagogical discussion to argue that the phase correlation between $\nu$SM and UV variables is quite natural in non-unitary subsystems inside the larger unitary theory. 
In section~\ref{sec:formulating-helio-UV}, we construct our perturbative framework with UV in matter in a step-by-step manner. 
In section~\ref{sec:P-e-mu-sector}, we compute the neutrino oscillation probabilities in the $\nu_{e} - \nu_{\mu}$ sector to first order in the helio-UV expansion, and discuss $\delta$ - $\alpha$ parameter phase correlation and diagonal $\alpha_{\beta \beta}$ correlations. In section~\ref{sec:stability-convention-dep}, the stability and the $U_{\text{\tiny MNS}}$ convention dependence of the phase correlation are discussed. It is followed by a brief note on the nature of the correlation.
Accuracy of the helio-UV expansion and relative importance of unitary and non-unitary parts of the probability are examined in section~\ref{sec:accuracy}. 
The oscillation probabilities in the $\nu_{\mu} - \nu_{\tau}$ sector are calculated in appendix~\ref{sec:mu-tau-sector}. Table~\ref{tab:P-eq-numbers} summarizes the equation numbers of all the oscillation probabilities.

\section{Correlation between the CP phase of $\nu$SM and the UV parameter phases} 
\label{sec:phase-correlation}

Focus of our physics discussion in this paper is on the correlation between the CP phase $\delta$ of $\nu$SM and the phases associated with the $\alpha$ parameters which will be introduced to describe the effect of UV, see eq.~\eqref{alpha-matrix-PDG}. This section~\ref{sec:phase-correlation} is meant to be an introduction to the topics, aiming at motivating the readers to this relatively unfamiliar subject. We try to illuminate, though only intuitively, why the phase correlation is relevant and can be {\em characteristic} in the theories with UV. 

To make our discussion transparent, but only for an explanatory purpose in this section, we rely on the particular model of leptonic UV with 3 active and $N_{s}$ sterile neutrinos, the $(3+N_{s})$ space unitary model. See \cite{Fong:2016yyh,Fong:2017gke} for an exposition of the properties of the model in the context of leptonic unitarity test. The model has the $ U = (3+N_{s}) \times (3+N_{s})$ unitary mixing matrix
\begin{equation}
  U = \left(\begin{array}{cc} N &W\\ Z &V\\\end{array}\right)
\end{equation}
which satisfies $UU^{\dagger}=U^{\dagger}U = {\bf 1}_{(3+N_{s})\times(3+N_{s})}$. For our present discussion, the relevant parts of it are the $3 \times 3$ $N$ matrix in the active neutrino subspace, and $3 \times N_{s}$ $W$ matrix which bridges between the active and the sterile neutrino subspaces.
The active $3 \times 3$ part of the unitarity relation in the whole state space can be written as\footnote{
The discussion here is not on a particular parametrization of non-unitarity such as the one we introduce in section~\ref{sec:formulating-helio-UV}. The similar argument may also be done with another type of $3 \times 3$ unitarity $\sum_{\beta= e,\mu,\tau} N^*_{\beta i} N_{\beta j} + \sum_{J=4}^{N_s + 3} Z^*_{J i} Z_{J j} = \delta_{ij}$ \cite{Fong:2017gke}.
}
\begin{eqnarray} 
\delta_{\alpha \beta} = 
\sum_{j=1}^{3} N_{\alpha j} N^{*}_{\beta j} + \sum_{J=4}^{N_{s}+3} W_{\alpha J} W^{*}_{\beta J}, 
\label{unitarity}
\end{eqnarray}
where $j=1,2,3$ and the sterile space index $J$ runs over $J=4,5, \cdot \cdot, N_{s}+3$. The off-diagonal elements of \eqref{unitarity} are complex numbers and it can be written as polygons with $3+N_{s}$ sides, the unitarity polygons, an extension of the well-known unitarity triangle. Therefore, the CP violating phases associated with $N$ and $W$ are all inter-related with each other. It is the basic reason why correlation between phases could be a relevant question in such models of the non-unitarity caused in the subsystem even though the whole system is unitary. 

In this paper, we deal with the active $3 \times 3$ subspace by restricting to the leading order in the $W$ perturbation theory \cite{Fong:2017gke}. Even in this case, the correlation among the phases in the $N$ and $W$ matrices could leave a part of its memory to the restricted $3 \times 3$ active neutrino space. Assuming $N_{s}$ is large and the magnitudes and phases of $W_{\alpha J}$ are random, it is conceivable that cancellation takes place among the terms in the $W W^{\dagger}$ term in eq.~\eqref{unitarity}, leaving a small non-unitarity effects. In such cases the major part of the phase correlation occurs inside the $3 \times 3$ active subspace. It could lead to correlations between $\nu$SM phase and the phases which originates from the new physics that cause UV. 

\section{Formulating the helio-unitarity violation (UV) perturbation theory} 
\label{sec:formulating-helio-UV}

In this paper, we investigate the dynamics of active three-flavor neutrino system with non-unitary mixing matrix in matter. In the context of the $(3+N_{s})$ space unitary model mentioned in the previous section, it is nothing but the leading (zeroth) order term in the $W$ expansion. Following the observation in ref.~\cite{Fong:2017gke}, we work with the neutrino evolution in $3 \times 3$ active neutrino space in the vacuum mass eigenstate basis, see eq.~\eqref{evolution-check-basis} below. It is conceivable that the system also describes high-mass limit of sterile states, with effective decoupling $W \approx 0$ of the sterile sector, or more generically the system of high-scale UV. Therefore, it should not come as a surprise that the equivalent evolution equation is used by Blennow {\it et al} \cite{Blennow:2016jkn}, for example, to describe high-scale UV. 

\subsection{Unitary evolution of neutrinos in the mass eigenstate basis }
\label{sec:mass-basis}

We discuss neutrino evolution in high-scale UV and (leading-order in $W$) low-scale UV in a same footing. The three active neutrino evolution in matter in the presence of non-unitary flavor mixing can be described by the Schr\"odinger equation in the vacuum mass eigenstate basis \cite{Fong:2017gke,Blennow:2016jkn}\footnote{
We must note that it is a highly nontrivial statement. A naive extension of the unitary system to the non-unitary one with the Hamiltonian in the flavor basis 
\begin{eqnarray}
H = 
\frac{1}{2E} 
\left\{  
N \left[
\begin{array}{ccc}
0 & 0 & 0 \\
0 & \Delta m^2_{21} & 0 \\
0 & 0 & \Delta m^2_{31} \\
\end{array}
\right] N^{\dagger} + 
\left[
\begin{array}{ccc}
a - b & 0 & 0 \\
0 & -b & 0 \\
0 & 0 & -b \\
\end{array}
\right] 
\right\} 
\end{eqnarray}
is {\em not} equivalent with the one in eq.~\eqref{evolution-check-basis} due to non-unitarity of the $N$ matrix \cite{Antusch:2006vwa}.
}
\begin{eqnarray}
i \frac{d}{dx} \check{\nu} = 
\frac{1}{2E} 
\left\{  
\left[
\begin{array}{ccc}
0 & 0 & 0 \\
0 & \Delta m^2_{21} & 0 \\
0 & 0 & \Delta m^2_{31} \\
\end{array}
\right] + 
N^{\dagger} \left[
\begin{array}{ccc}
a - b & 0 & 0 \\
0 & -b & 0 \\
0 & 0 & -b \\
\end{array}
\right] N 
\right\} 
\check{\nu}. 
\label{evolution-check-basis}
\end{eqnarray}
In this paper, we denote the vacuum mass eigenstate basis as the ``check basis''. In eq.~(\ref{evolution-check-basis}), $N$ denotes the $3 \times 3$ non-unitary flavor mixing matrix which relates the flavor neutrino states to the vacuum mass eigenstates as 
\begin{eqnarray}
\nu_{\alpha} = N_{\alpha i} \check{\nu}_{i}. 
\label{N-def}
\end{eqnarray}
Hereafter, the subscript Greek indices $\alpha$, $\beta$, or $\gamma$ run over $e, \mu, \tau$, and the Latin indices $i$, $j$ run over the mass eigenstate indices $1,2,$ and $3$. $E$ is neutrino energy and $\Delta m^2_{ji} \equiv m^2_{j} - m^2_{i}$. The usual phase redefinition of neutrino wave function is done to leave only the mass squared differences.

The functions $a(x)$ and $b(x)$ in eq.~(\ref{evolution-check-basis}) denote the
Wolfenstein matter potential \cite{Wolfenstein:1977ue} due to charged current (CC) and neutral current (NC) reactions, respectively.
\begin{eqnarray} 
a &=&  
2 \sqrt{2} G_F N_e E \approx 1.52 \times 10^{-4} \left( \frac{Y_e \rho}{\rm g\,cm^{-3}} \right) \left( \frac{E}{\rm GeV} \right) {\rm eV}^2, 
\nonumber \\
b &=& \sqrt{2} G_F N_n E = \frac{1}{2} \left( \frac{N_n}{N_e} \right) a. 
\label{matt-potential}
\end{eqnarray}
Here, $G_F$ is the Fermi constant, $N_e$ and $N_n$ are the electron
and neutron number densities in matter. $\rho$ and $Y_e$ denote,
respectively, the matter density and number of electrons per nucleon
in matter. For simplicity and clarity we will work with the uniform
matter density approximation in this paper. But, it is not difficult
to extend our treatment to varying matter density case if adiabaticity
holds.

By writing the evolution equation as in eq.~(\ref{evolution-check-basis}) with the hermitian Hamiltonian, the neutrino evolution is obviously unitary, which is in agreement with our discussion given in section~\ref{sec:introduction}. Then, the answer to the remaining question, ``how the effect of non-unitarity comes in into the observables as a consequence of non-unitary mixing matrix'' is given in section~\ref{sec:basis-relations}.

\subsection{$\alpha$ parametrization of the non-unitary mixing matrix and its  convention dependence}
\label{sec:convention-dependence}

To parametrize the non-unitary $N$ matrix we use the so-called $\alpha$ parametrization \cite{Escrihuela:2015wra}, $N = \left( \bf{1} - \alpha \right) U$, where $U \equiv U_{\text{\tiny MNS}}$ denotes the $\nu$SM $3 \times 3$ unitary flavor mixing matrix.\footnote{
For early references for parametrizing the UV effect, see e.g., \cite{Valle:1987gv,Broncano:2002rw,FernandezMartinez:2007ms,Xing:2007zj,Xing:2011ur}. It must be remarked that the authors of ref.~\cite{Blennow:2016jkn} made an important point in explaining why the $\alpha$ parametrization is more superior than their traditional way of using the hermitian $\eta$ matrix. 
}
To define the $\alpha$ matrix, however, we must specify the phase convention by which $U$ matrix is defined. 

We start from the most commonly used form, the PDG convention \cite{Tanabashi:2018oca} of the MNS matrix, 
\begin{eqnarray} 
U_{\text{\tiny PDG}} = 
\left[
\begin{array}{ccc}
1 & 0 &  0  \\
0 & c_{23} & s_{23} \\
0 & - s_{23} & c_{23} \\
\end{array}
\right] 
\left[
\begin{array}{ccc}
c_{13}  & 0 & s_{13} e^{- i \delta} \\
0 & 1 & 0 \\
- s_{13} e^{ i \delta} & 0 & c_{13}  \\
\end{array}
\right] 
\left[
\begin{array}{ccc}
c_{12} & s_{12}  &  0  \\
- s_{12} & c_{12} & 0 \\
0 & 0 & 1 \\
\end{array}
\right], 
\label{MNS-PDG}
\end{eqnarray}
with the obvious notations $s_{ij} \equiv \sin \theta_{ij}$ etc. and $\delta$ being the CP violating phase. Then, we define the non-unitary mixing matrix $N_{\text{\tiny PDG}}$ as 
\begin{eqnarray} 
N_{\text{\tiny PDG}} &=& 
\left( \bf{1} - \bar{\alpha} \right) U_{\text{\tiny PDG}} = 
\left\{ 
\bf{1} - 
\left[ 
\begin{array}{ccc}
\bar{\alpha}_{ee} & 0 & 0 \\
\bar{\alpha}_{\mu e} & \bar{\alpha}_{\mu \mu}  & 0 \\
\bar{\alpha}_{\tau e}  & \bar{\alpha}_{\tau \mu} & \bar{\alpha}_{\tau \tau} \\
\end{array}
\right] 
\right\}
U_{\text{\tiny PDG}}.
\label{alpha-matrix-PDG}
\end{eqnarray}
By inserting $N = N_{\text{\tiny PDG}}$ in eqs.~(\ref{alpha-matrix-PDG}) to (\ref{evolution-check-basis}), we define the neutrino evolution equation in the vacuum mass eigenstate basis. 

\subsubsection{Neutrino evolution with general convention of the MNS matrix}
\label{sec:convention-dep}

After reducing the standard three-flavor mixing matrix to $U_{\text{\tiny PDG}}$, which has four degrees of freedom, we still have freedom of phase redefinition 
\begin{eqnarray} 
\check{\nu} 
\rightarrow 
\left[
\begin{array}{ccc}
1 & 0 &  0  \\
0 & e^{ i \beta} & 0 \\
0 & 0 & e^{ i \gamma} \\
\end{array}
\right] \check{\nu} 
\equiv 
\Gamma \left( \beta, \gamma \right) \check{\nu} 
\end{eqnarray}
without affecting physics of the system. Then, the evolution equation in the $\Gamma \left( \beta, \gamma \right)$ transformed basis, 
\begin{eqnarray}
&& i \frac{d}{dx} \check{\nu} 
= 
\frac{1}{2E} 
\left[
\begin{array}{ccc}
0 & 0 & 0 \\
0 & \Delta m^2_{21} & 0 \\
0 & 0 & \Delta m^2_{31} \\
\end{array}
\right] \check{\nu}
\nonumber \\
&+&
\frac{1}{2E} 
U (\beta, \gamma)^{\dagger}
\left\{ \bf{1} - \alpha (\beta, \gamma) \right\}^{\dagger}  
\left[
\begin{array}{ccc}
a - b & 0 & 0 \\
0 & -b & 0 \\
0 & 0 & -b \\
\end{array}
\right] 
\left\{ \bf{1} - \alpha (\beta, \gamma) \right\} 
U (\beta, \gamma) \check{\nu}, 
\label{evolution-Gamma-transf}
\end{eqnarray}
describes the same physics. In (\ref{evolution-Gamma-transf}), $U (\beta, \gamma)$ and $\alpha (\beta, \gamma)$ denote, respectively, the $\Gamma \left( \beta, \gamma \right)$ transformed MNS matrix and $\bar{\alpha}$ matrix: 
\begin{eqnarray}
U (\beta, \gamma) &\equiv& 
\left[
\begin{array}{ccc}
1 & 0 &  0  \\
0 & e^{ - i \beta} & 0 \\
0 & 0 & e^{ - i \gamma} \\
\end{array}
\right] 
U_{\text{\tiny PDG}} 
\left[
\begin{array}{ccc}
1 & 0 &  0  \\
0 & e^{ i \beta} & 0 \\
0 & 0 & e^{ i \gamma} \\
\end{array}
\right] 
\nonumber \\
\alpha (\beta, \gamma) &\equiv& 
\left[
\begin{array}{ccc}
1 & 0 &  0  \\
0 & e^{ - i \beta} & 0 \\
0 & 0 & e^{ - i \gamma} \\
\end{array}
\right] 
\bar{\alpha}
\left[
\begin{array}{ccc}
1 & 0 &  0  \\
0 & e^{ i \beta} & 0 \\
0 & 0 & e^{ i \gamma} \\
\end{array}
\right] 
\label{G-transformed-U-A}
\end{eqnarray}
That is, we can use different convention of the MNS matrix $U (\beta, \gamma)$, but then our $\alpha$ matrix has to be changed accordingly, as in (\ref{G-transformed-U-A}).

\subsubsection{The three useful conventions of the MNS matrix}
\label{sec:3-conventions}

Among general conventions defined in (\ref{G-transformed-U-A}), practically, there exist the three useful conventions of the MNS matrix. In addition to $U_{\text{\tiny PDG}}$ in (\ref{MNS-PDG}), they are $U (0, \delta)$ and $U (\delta, \delta)$:
\begin{eqnarray} 
&&U_{\text{\tiny ATM}} \equiv U (0, \delta) 
= 
\left[
\begin{array}{ccc}
1 & 0 &  0  \\
0 & c_{23} & s_{23} e^{ i \delta} \\
0 & - s_{23} e^{- i \delta} & c_{23} \\
\end{array}
\right] 
\left[
\begin{array}{ccc}
c_{13}  & 0 & s_{13}  \\
0 & 1 & 0 \\
- s_{13} & 0 & c_{13} \\
\end{array}
\right] 
\left[
\begin{array}{ccc}
c_{12} & s_{12}  &  0  \\
- s_{12} & c_{12} & 0 \\
0 & 0 & 1 \\
\end{array}
\right],  
\nonumber \\
&&U_{\text{\tiny SOL}} \equiv U (\delta, \delta) 
= \left[
\begin{array}{ccc}
1 & 0 &  0  \\
0 & c_{23} & s_{23} \\
0 & - s_{23} & c_{23} \\
\end{array}
\right] 
\left[
\begin{array}{ccc}
c_{13}  & 0 & s_{13} \\
0 & 1 & 0 \\
- s_{13} & 0 & c_{13} \\
\end{array}
\right] 
\left[
\begin{array}{ccc}
c_{12} & s_{12} e^{ i \delta}  &  0  \\
- s_{12} e^{- i \delta} & c_{12} & 0 \\
0 & 0 & 1 \\
\end{array}
\right]. 
\label{MNS-ATM-SOL}
\end{eqnarray}
The reason for our terminology of $U_{\text{\tiny ATM}}$ and $U_{\text{\tiny SOL}}$ in (\ref{MNS-ATM-SOL}) is because CP phase $\delta$ is attached to the ``atmospheric angle'' $s_{23}$ in $U_{\text{\tiny ATM}}$, and to the ``solar angle'' $s_{12}$ in $U_{\text{\tiny SOL}}$, respectively. Whereas in $U_{\text{\tiny PDG}}$, $\delta$ is attached to $s_{13}$. 

Accordingly, we have the three different definition of the $\alpha$ matrix. In addition to 
$N_{\text{\tiny PDG}} = \left( \bf{1} - \bar{\alpha} \right) U_{\text{\tiny PDG}}$ as in (\ref{alpha-matrix-PDG}), we have 
$N_{\text{\tiny ATM}} = \left( \bf{1} - \alpha_{\text{\tiny ATM}} \right) U_{\text{\tiny ATM}}$, and 
$N_{\text{\tiny SOL}} = \left( \bf{1} - \alpha_{\text{\tiny SOL}} \right) U_{\text{\tiny SOL}}$. The latter two and their relations to $\bar{\alpha}$ are given by 
\begin{eqnarray}
&& 
\alpha_{\text{\tiny ATM}} = \alpha (0, \delta) 
\equiv 
\left[ 
\begin{array}{ccc}
\alpha_{ee} & 0 & 0 \\
\alpha_{\mu e} & \alpha_{\mu \mu}  & 0 \\
\alpha_{\tau e}  & \alpha_{\tau \mu} & \alpha_{\tau \tau} \\
\end{array}
\right] 
= 
\left[ 
\begin{array}{ccc}
\bar{\alpha}_{ee} & 0 & 0 \\
\bar{\alpha}_{\mu e} & 
\bar{\alpha}_{\mu \mu} & 
0 \\
\bar{\alpha}_{\tau e} e^{ - i \delta} & 
\bar{\alpha}_{\tau \mu} e^{ - i \delta} & 
\bar{\alpha}_{\tau \tau}  \\
\end{array}
\right], 
\nonumber \\
&&
\alpha_{\text{\tiny SOL}} = \alpha (\delta, \delta)
\equiv 
\left[ 
\begin{array}{ccc}
\tilde{\alpha}_{ee} & 0 & 0 \\
\tilde{\alpha}_{\mu e} & \tilde{\alpha}_{\mu \mu}  & 0 \\
\tilde{\alpha}_{\tau e}  & \tilde{\alpha}_{\tau \mu} & \tilde{\alpha}_{\tau \tau} \\
\end{array}
\right] 
=
\left[ 
\begin{array}{ccc}
\bar{\alpha}_{ee} & 0 & 0 \\
\bar{\alpha}_{\mu e} e^{ - i \delta} & 
\bar{\alpha}_{\mu \mu} & 
0 \\
\bar{\alpha}_{\tau e} e^{ - i \delta} & 
\bar{\alpha}_{\tau \mu} & 
\bar{\alpha}_{\tau \tau}  \\
\end{array}
\right]. 
\label{alpha-alpha-bar-matrix}
\end{eqnarray}
In this paper, for convenience of the calculations, we take the ATM convention with $U_{\text{\tiny ATM}}$ and $\alpha_{\text{\tiny ATM}}$. But, the translation of our results to the PDG or the SOL conventions can be done easily by using eq.~(\ref{alpha-alpha-bar-matrix}). 

Notice that, because of the structure $N_{\text{\tiny ATM}} = \left( \bf{1} - \alpha_{\text{\tiny ATM}} \right) U_{23} U_{13} U_{12}$, the $\alpha$ matrix is always attached to $U_{23}$. Then, the correlation between the lepton KM phase $\delta$ and the UV parameter phases becomes more transparent if $e^{ \pm i \delta }$ is attached to $U_{23}$. This is the reason why we take the MNS matrix convention $U_{\text{\tiny ATM}}$ in (\ref{MNS-ATM-SOL}) in our following calculation.

\subsection{Preliminary step toward perturbation theory: Tilde-basis}
\label{sec:tilde-basis} 

Taking the $U_{\text{\tiny ATM}}$ convention with $\alpha_{\text{\tiny ATM}}$ matrix, we formulate our helio-UV perturbation theory. We assume that deviation from unitarity is small, so that $\alpha_{ \beta \gamma } \ll 1$ hold for all flavor indices $\beta$ and $\gamma$ including the diagonal ones. Therefore, we are able to use the two kind of expansion parameters, $\epsilon \approx \Delta m^2_{21} / \Delta m^2_{31}$ (see eq.~(\ref{renem-param}) below) and the $\alpha$ parameters in our helio-UV perturbation theory. 
Though our presentation partly repeats the same procedure as used in ref.~\cite{Minakata:2015gra}, we go through the steps to make this paper self-contained. 

We define the following notations for simplicity to be used in the discussions hereafter in this paper:
\begin{eqnarray}
\Delta_{ji} \equiv \frac{\Delta m^2_{ji}}{2E},
\hspace{8mm}
\Delta_{a} \equiv \frac{ a }{2E}, 
\hspace{8mm}
\Delta_{b} \equiv \frac{ b }{2E}. 
\label{Delta-def}
\end{eqnarray}

For convenience in formulating the helio-UV perturbation theory, we move from the check basis to an intermediate basis, which we call the ``tilde basis'', $\tilde{\nu} = ( U_{13} U_{12} ) \check{\nu}$, with Hamiltonian
\begin{eqnarray} 
&& \tilde{H} = 
( U_{13} U_{12} ) \check{H} ( U_{13} U_{12} )^{\dagger} 
= 
( U_{13} U_{12} ) 
\left[
\begin{array}{ccc}
0 & 0 & 0 \\
0 & \Delta_{21} & 0 \\
0 & 0 & \Delta_{31} \\
\end{array}
\right] 
( U_{13} U_{12} )^{\dagger} 
\nonumber \\
&+& 
U_{23}^{\dagger} 
\left\{
{\bf 1} - 
\left[
\begin{array}{ccc}
\alpha_{ee} & \alpha_{\mu e}^* & \alpha_{\tau e}^*  \\
0 & \alpha_{\mu \mu} & \alpha_{\tau \mu}^* \\
0 & 0 & \alpha_{\tau \tau} \\
\end{array}
\right] 
\right\}
\left[
\begin{array}{ccc}
\Delta_{a} - \Delta_{b} & 0 & 0 \\
0 & - \Delta_{b} & 0 \\
0 & 0 & - \Delta_{b} \\
\end{array}
\right] 
\left\{ 
\bf{1} - 
\left[ 
\begin{array}{ccc}
\alpha_{ee} & 0 & 0 \\
\alpha_{\mu e} & \alpha_{\mu \mu}  & 0 \\
\alpha_{\tau e}  & \alpha_{\tau \mu} & \alpha_{\tau \tau} \\
\end{array}
\right] 
\right\}
U_{23} 
\nonumber \\ 
&\equiv& 
\tilde{H}_{ \text{vac} } + \tilde{H}_\text{ UV }.
\label{tilde-H}
\end{eqnarray}
In the last line, we have denoted the first and the second terms in
eq.~\eqref{tilde-H} as $\tilde{H}_{ \text{vac} }$ and
$\tilde{H}_\text{ UV }$, respectively. The explicit form of the
$\tilde{H}_{ \text{vac} }$ in a form decomposed into the unperturbed
and perturbed parts is given by
\begin{eqnarray} 
\tilde{H}_{ \text{vac} }^{(0)} (x) &=& 
\Delta_{ \text{ren} } 
\left\{
\left[
\begin{array}{ccc}
s^2_{13} & 0 & c_{13} s_{13} \\
0 & 0 & 0 \\
c_{13} s_{13}  & 0 & c^2_{13} 
\end{array}
\right] 
+
\epsilon 
\left[
\begin{array}{ccc}
s^2_{12} & 0 & 0 \\
0 & c^2_{12} & 0 \\
0 & 0 & s^2_{12} 
\end{array}
\right]  \right\}, 
\label{Hvac-zeroth}
\end{eqnarray} 
\begin{eqnarray} 
\tilde{H}_{ \text{vac} }^{(1)} (x) &=& 
\epsilon c_{12} s_{12} 
\Delta_{ \text{ren} } 
\left[
\begin{array}{ccc}
0 & c_{13} & 0 \\
c_{13} & 0 & - s_{13}  \\
0 & - s_{13} & 0 
\end{array}
\right],
\label{Hvac-first}
\end{eqnarray} 
where 
\begin{eqnarray}
&& \Delta_{ \text{ren} } \equiv \frac{ \Delta m^2_{ \text{ren} } }{2E}, 
\hspace{10mm} 
\Delta m^2_{ \text{ren} } \equiv \Delta m^2_{31} - s^2_{12} \Delta m^2_{21}, 
\nonumber \\ 
&& \epsilon \equiv \frac{ \Delta m^2_{21} }{ \Delta m^2_{ \text{ren} } }.
\label{renem-param}
\end{eqnarray} 
The superscripts $(0)$ and $(1)$ in eqs.~\eqref{Hvac-zeroth}~and~\eqref{Hvac-first}, respectively, show that they are zeroth and first order in $\epsilon$. 
An order $\epsilon$ term is intentionally absorbed into the zeroth-order term in $\tilde{H}_{ \text{vac} }^{(0)}$ as in eq.~\eqref{Hvac-zeroth} to make the formulas of the oscillation probabilities simple and compact.

We note that the matter term $\tilde{H}_\text{ UV }$ in eq.~\eqref{tilde-H} can be decomposed into the zeroth, first and the second order terms in $\alpha$ (or $\tilde{\alpha}$) matrix elements as $\tilde{H}_\text{ UV } = \tilde{H}_{ \text{matt} }^{(0)} + \tilde{H}_\text{ UV }^{(1)} + \tilde{H}_\text{ UV }^{(2)} $: 
\begin{eqnarray}
&& \tilde{H}_{ \text{matt} }^{(0)} = 
\left[
\begin{array}{ccc}
\Delta_{a} - \Delta_{b} & 0 & 0 \\
0 & - \Delta_{b} & 0 \\
0 & 0 & - \Delta_{b} \\
\end{array}
\right], 
\nonumber \\ 
\tilde{H}_\text{ UV }^{(1)} &=& 
U_{23}^{\dagger} 
\left\{ 
\Delta_{b} 
\left[ 
\begin{array}{ccc}
2 \alpha_{ee} \left( 1 - \frac{ \Delta_{a} }{ \Delta_{b} } \right) & \alpha_{\mu e}^* & \alpha_{\tau e}^* \\
\alpha_{\mu e} & 2 \alpha_{\mu \mu}  & \alpha_{\tau \mu}^* \\
\alpha_{\tau e}  & \alpha_{\tau \mu} & 2 \alpha_{\tau \tau} \\
\end{array}
\right] 
\right\} 
U_{23}, 
\nonumber \\ 
\tilde{H}_\text{ UV }^{(2)} &=& 
- U_{23}^{\dagger} 
\left\{ 
\Delta_{b} 
\left[
\begin{array}{ccc}
\alpha_{ee}^2 \left( 1 - \frac{ \Delta_{a} }{ \Delta_{b} } \right) + |\alpha_{\mu e}|^2 + |\alpha_{\tau e}|^2 & 
\alpha_{\mu e}^* \alpha_{\mu \mu} + \alpha_{\tau e}^* \alpha_{\tau \mu} & 
\alpha_{\tau e}^* \alpha_{\tau \tau} \\
\alpha_{\mu e} \alpha_{\mu \mu} + \alpha_{\tau e} \alpha_{\tau \mu}^* & 
\alpha_{\mu \mu}^2 + |\alpha_{\tau \mu}|^2 & 
\alpha_{\tau \mu}^* \alpha_{\tau \tau} \\
\alpha_{\tau e} \alpha_{\tau \tau} & 
\alpha_{\tau \mu} \alpha_{\tau \tau} & 
\alpha_{\tau \tau}^2 \\
\end{array}
\right] 
\right\} 
U_{23}. 
\nonumber \\ 
\label{tilde-H1-matt} 
\end{eqnarray}
As it stands, the subscript ``UV'' indicates that the quantity contains the unitarity-violating $\alpha$ parameters. The total Hamiltonian in the tilde basis is, therefore, given by
$\tilde{H}=\tilde{H}_{ \text{vac} } + \tilde{H}_\text{ UV }$, where
$\tilde{H}_{ \text{vac} } = \tilde{H}_{ \text{vac} }^{(0)} +
\tilde{H}_{ \text{vac} }^{(1)}$.

\subsection{Unperturbed and perturbed Hamiltonian in the tilde basis}
\label{sec:unper-perturbed-H}

To formulate the helio-UV perturbation theory, we decompose the tilde
basis Hamiltonian in the following way:
\begin{eqnarray} 
\tilde{H} = \tilde{H}^{(0)} + \tilde{H}^{(1)}. 
\end{eqnarray}
The unperturbed (zeroth-order) Hamiltonian is given by
$\tilde{H}^{(0)}=\tilde{H}_{ \text{vac} }^{(0)} + \tilde{H}_\text{
  matt }^{(0)}$.

We make a phase redefinition 
\begin{eqnarray}
\tilde{\nu} = \exp{ [ i \int^{x} dx^{\prime} \Delta_{b} (x^{\prime}) ] } \tilde{\nu}^{\prime}
\end{eqnarray}
which is valid even for non-uniform matter density. Then, the
Schr\"odinger equation for $\tilde{\nu}^{\prime}$ becomes the form in
eq.~\eqref{N-def} with unperturbed part of the Hamiltonian $(
\tilde{H}^{(0)} )^{\prime}$ as given in
\begin{eqnarray} 
( \tilde{H}^{(0)} )^{\prime} &=& 
\Delta_{ \text{ren} } 
\left\{
\left[
\begin{array}{ccc}
\frac{ a(x) }{ \Delta m^2_{ \text{ren} }}  + s^2_{13} & 0 & c_{13} s_{13} \\
0 & 0 & 0 \\
c_{13} s_{13}  & 0 & c^2_{13} 
\end{array}
\right] 
+
\epsilon 
\left[
\begin{array}{ccc}
s^2_{12} & 0 & 0 \\
0 & c^2_{12} & 0 \\
0 & 0 & s^2_{12} 
\end{array}
\right]  \right\}. 
\label{tilde-H0}
\end{eqnarray}
namely, without NC matter potential terms. It is evident that the
phase redefinition does not affect the physics of flavor
change. Hereafter, we omit the prime symbol and use the zeroth-order
Hamiltonian eq.~\eqref{tilde-H0} without NC term.
This is nothing but the zeroth order Hamiltonian used in
\cite{Minakata:2015gra}, which led to the ``simple and compact''
formulas of the oscillation probabilities in the standard three-flavor
mixing.\footnote{
Having the same zeroth order Hamiltonian as the one in \cite{Minakata:2015gra} was not expected a priori because the NC reaction is involved, but it came out quite naturally as described here. 
}

The perturbed Hamiltonian is then given by
\begin{eqnarray} 
\tilde{H}^{(1)} = \tilde{H}_{ \text{vac} }^{(1)} + \tilde{H}_\text{ UV }^{(1)} + \tilde{H}_\text{ UV }^{(2)}
\label{tilde-H1}
\end{eqnarray}
where each term in eq.~\eqref{tilde-H1} is defined in eqs.~\eqref{Hvac-first}~and~\eqref{tilde-H1-matt}. In the actual computation, we drop the second-order term (the last term) in eq.~\eqref{tilde-H1} because we confine ourselves into the zeroth and first order terms in the UV parameters in this paper.

\subsection{Diagonalization of zeroth-order Hamiltonian and the hat basis}
\label{sec:zeroth-order-H}

To carry out perturbative calculation, it is convenient to transform
to a basis which diagonalizes $\tilde{H}^{(0)}$, which we call the
``hat basis''. $\tilde{H}^{(0)}$ is diagonalized by the unitary
transformation as follows:
\begin{eqnarray} 
\hat{H}_{0} &=& U^{\dagger}_{\phi} \tilde{H}_{0} U_{\phi} 
= 
\left[
\begin{array}{ccc}
h_{1} & 0 & 0 \\
0 & h_{2} & 0 \\
0 & 0 & h_{3} 
\end{array}
\right], 
\label{hat-hamiltonian1}
\end{eqnarray}
where the eigenvalues $h_{i}$ are given by
\begin{eqnarray} 
h_{1} &=& 
\frac{ 1 }{ 2 } \left[
\left( \Delta_{ \text{ren} } + \Delta_{ a } \right) - {\rm sign}(\Delta m^2_{ \text{ren} }) \sqrt{ \left( \Delta_{ \text{ren} } - \Delta_{ a } \right)^2 + 4 s^2_{13} \Delta_{ \text{ren} } \Delta_{ a }  }
\right] 
+ \epsilon \Delta_{ \text{ren} } s^2_{12},
\nonumber\\[3mm]
h_{2} &=&  c^2_{12} ~\epsilon ~\Delta_{ \text{ren} },
\label{eq:lambda-pm0}  
\\[3mm]
h_{3} &=& 
\frac{ 1 }{ 2 } \left[
\left( \Delta_{ \text{ren} } + \Delta_{ a } \right) + {\rm sign}(\Delta m^2_{ \text{ren} }) \sqrt{ \left( \Delta_{ \text{ren} } - \Delta_{ a } \right)^2 + 4 s^2_{13} \Delta_{ \text{ren} } \Delta_{ a }  }
\right] 
+ \epsilon \Delta_{ \text{ren} } s^2_{12}.
\nonumber
\end{eqnarray}
See eqs.~\eqref{Delta-def} and \eqref{renem-param} for the
definitions of $\Delta_{ \text{ren} }$, $\Delta_{ a }$ etc. By the
convention with ${\rm sign}(\Delta m^2_{ \text{ren} })$, we can treat
the normal and the inverted mass orderings in a unified way.
The foregoing and the following treatment of the system without the UV
$\alpha$ parameters in this section, which recapitulates the one in
ref.~\cite{Minakata:2015gra}, is to make description in this paper
self-contained.

$U_{\phi}$ is parametrized as 
\begin{eqnarray} 
U_{\phi} =
\left[
\begin{array}{ccc}
\cos \phi & 0 & \sin \phi \\
0 & 1 & 0 \\
- \sin \phi & 0 & \cos \phi 
\end{array}
\right].  
\label{eq:U-phi}
\end{eqnarray}
where $\phi$ is nothing but the mixing angle $\theta_{13}$ in
matter. With the definitions of the eigenvalues
eq.~\eqref{eq:lambda-pm0}, the following mass-ordering independent
expressions for cosine and sine $2 \phi$ are obtained:
\begin{eqnarray} 
\cos 2 \phi &=& 
\frac{ \Delta_{ \text{ren} } \cos 2\theta_{13} - \Delta_{ a } }{ h_{3} - h_{1} },
\nonumber \\
\sin 2 \phi &=& \frac{ \Delta_{ \text{ren} } \sin 2\theta_{13} }{ h_{3} - h_{1} }.
\label{eq:cos-sin-2phi}
\end{eqnarray}

The perturbing Hamiltonian in vacuum in the tilde basis,
$\tilde{H}^{(1)}_{ \text{vac} }$, has a simple form such that the
positions of ``zeros'' are kept after transformed into the hat basis:
\begin{eqnarray} 
&& \hat{H}^{(1)}_{ \text{vac} } 
= U^{\dagger}_{\phi} \tilde{H}^{(1)}_{ \text{vac} } U_{\phi} 
\nonumber\\
&=&
\epsilon c_{12} s_{12} 
\Delta_{\rm ren} 
\left[
\begin{array}{ccc}
0 & \cos \left( \phi - \theta_{13} \right) & 0   \\
\cos \left( \phi - \theta_{13} \right)  & 0 & \sin \left( \phi - \theta_{13} \right) \\
0  & \sin \left( \phi - \theta_{13} \right) & 0 
\end{array}
\right]. 
\label{hatH-vac-1}
\end{eqnarray}
In fact, $ \hat{H}_{1}$ is identical to $\tilde{H}_1$ with
$\theta_{13}$ replaced by $(\theta_{13}-\phi)$.
However, the form of $\hat{H}^{(1)}_\text{ UV }$ is somewhat
complicated,
\begin{eqnarray} 
\hat{H}^{(1)}_\text{ UV } 
&=& U^{\dagger}_{\phi} \tilde{H}^{(1)}_\text{ UV } U_{\phi} 
\equiv 
\Delta_{b} 
U^{\dagger}_{\phi} H U_{\phi} 
\label{hatH-matt-1} 
\end{eqnarray}
where we have defined $H$ matrix 
\begin{eqnarray} 
&& 
H \equiv 
\left[
\begin{array}{ccc}
H_{11} & H_{12} & H_{13} \\
H_{21} & H_{22} & H_{23} \\
H_{31} & H_{32} & H_{33} \\
\end{array}
\right] 
=
U_{23}^{\dagger} 
\left[ 
\begin{array}{ccc}
2 \alpha_{ee} \left( 1 - \frac{ \Delta_{a} }{ \Delta_{b} } \right) & \alpha_{\mu e}^* & \alpha_{\tau e}^* \\
\alpha_{\mu e} & 2 \alpha_{\mu \mu}  & \alpha_{\tau \mu}^* \\
\alpha_{\tau e}  & \alpha_{\tau \mu} & 2 \alpha_{\tau \tau} \\
\end{array}
\right] 
U_{23}. 
\label{Hij-def}
\end{eqnarray}
The explicit expressions of the elements $H_{ij}$ are given in
appendix~\ref{sec:Hij-Phi-ij}.

\subsection{Flavor basis, the tilde and hat bases, the $S$ and $\hat{S}$ matrices, and their relations}
\label{sec:basis-relations}

We summarize the relationship between the flavor basis, the check
(vacuum mass eigenstate) basis, the tilde, and the hat basis (with the zeroth order diagonalized hamiltonian). The convention dependence of the flavor mixing matrix and the $\alpha$ matrix are understood though not displayed explicitly. 

Only the unitary transformations are involved in changing from the hat
basis to the tilde basis, and from the tilde basis to the check basis:
\begin{eqnarray} 
&& \hat{H} = U^{\dagger}_{\phi} \tilde{H} U_{\phi}, 
\hspace{8mm}
\text{or}
\hspace{8mm}
\tilde{H} = U_{\phi} \hat{H} U^{\dagger}_{\phi}, 
\nonumber\\
&&
\tilde{H} = 
( U_{13} U_{12} ) \check{H} 
( U_{13} U_{12} )^{\dagger}, 
\hspace{8mm}
\text{or}
\hspace{8mm}
\check{H} = ( U_{13} U_{12} )^{\dagger} \tilde{H} ( U_{13} U_{12} ).
\label{hat-tilde-check}
\end{eqnarray}
The non-unitary transformation is involved from the check basis to the
flavor basis:
\begin{eqnarray} 
\nu_{\alpha} = N_{\alpha i} \check{\nu}_{i} 
= \left\{ ( 1 - \alpha ) U \right\}_{\alpha i} \check{\nu}_{i}. 
\label{flavor-check}
\end{eqnarray}
The relationship between the flavor basis Hamiltonian $H_{
  \text{flavor} }$ and the hat basis one $\hat{H}$ is
\begin{eqnarray}
  H_{ \text{flavor} } &=& \left\{ ( 1 - \alpha ) U \right\} \check{H} \left\{ ( 1 - \alpha ) U \right\}^{\dagger} 
\nonumber \\
&=& 
( 1 - \alpha ) U_{23} U_{\phi} \hat{H} U^{\dagger}_{\phi} U_{23}^{\dagger} ( 1 - \alpha )^{\dagger}. 
\label{flavor-hat}
\end{eqnarray}
Then, the flavor basis $S$ matrix is related to $\hat{S}$ and
$\tilde{S}$ matrices as
\begin{eqnarray} 
S &=& 
( 1 - \alpha ) U_{23} U_{\phi} \hat{S} U^{\dagger}_{\phi} U_{23}^{\dagger} ( 1 - \alpha )^{\dagger} 
= 
( 1 - \alpha ) U_{23} \tilde{S} U_{23}^{\dagger} ( 1 - \alpha )^{\dagger}. 
\label{S-flavor-hat}
\end{eqnarray}
Notice that both $\hat{S}$ and $\tilde{S}$ are unitary, but $S$ is {\em not} because of non-unitarity of the $(1-\alpha)$ matrix. 

This is the answer to the question we posed in section~\ref{sec:introduction}. Namely, the non-unitarity of $S$ matrix in the flavor basis, whose square is the observable, comes only from the initial projection from the flavor- to mass-basis and the final projection back from the mass- to flavor-eigenstate. There is no other way, because neutrino evolution has to be unitary, as discussed in sections~\ref{sec:introduction} and \ref{sec:mass-basis}. Though it is nothing more than a clarifying discussion, we believe it worthwhile to note this point. 

Now, we want to make our notations unambiguous in such a way that which part of the $S$ matrix conserves unitarity and which part does not. We first define the ``evolution-$S$ matrix'' which is unitary, 
\begin{eqnarray} 
S_{ \text{EV} } \equiv U_{23} \tilde{S} U_{23}^{\dagger}
\label{evolution-S}
\end{eqnarray}
and decompose the flavor basis $S$ matrix in \eqref{S-flavor-hat} as 
\begin{eqnarray} 
S = S_{ \text{EV} } + S_{ \text{UV} }, 
\hspace{10mm}
S_{ \text{UV} } \equiv 
- \alpha S_{ \text{EV} } - S_{ \text{EV} } \alpha^{\dagger} 
+ \alpha S_{ \text{EV} } \alpha^{\dagger}, 
\label{S-EV-UV}
\end{eqnarray}
where $S_{ \text{UV} }$ denotes the non-unitary part of the $S$ matrix. The evolution-$S$ matrix $S_{ \text{EV} }$ can be written in terms of the $\tilde{S}$ elements \cite{Minakata:2015gra}:
\begin{eqnarray} 
\left( S_{ \text{EV} } \right)_{ee} &=& \tilde{ S }_{ee}, 
\nonumber \\
\left( S_{ \text{EV} } \right)_{e \mu} &=& c_{23} \tilde{ S }_{e \mu} + s_{23} e^{ - i \delta} \tilde{ S }_{e \tau}, 
\nonumber \\ 
\left( S_{ \text{EV} } \right)_{e \tau} &=& c_{23} \tilde{ S }_{e \tau} - s_{23} e^{ i \delta} \tilde{ S }_{e \mu},
\nonumber \\
\left( S_{ \text{EV} } \right)_{\mu e} &=& c_{23} \tilde{ S }_{\mu e} + s_{23} e^{ i \delta} \tilde{ S }_{\tau e} 
\nonumber \\
\left( S_{ \text{EV} } \right)_{\mu \mu} &=& c^2_{23} \tilde{ S }_{\mu \mu} + s^2_{23} \tilde{ S }_{\tau \tau} + c_{23} s_{23} ( e^{ - i \delta} \tilde{ S }_{\mu \tau} + e^{ i \delta}  \tilde{ S }_{\tau \mu} ), 
\nonumber \\
\left( S_{ \text{EV} } \right)_{\mu \tau} &=& c^2_{23} \tilde{ S }_{\mu \tau} - s^2_{23} e^{ 2 i \delta}  \tilde{ S }_{\tau \mu} + c_{23} s_{23} e^{ i \delta}  ( \tilde{ S }_{\tau \tau} - \tilde{ S }_{\mu \mu} ), 
\nonumber \\ 
\left( S_{ \text{EV} } \right)_{\tau e} &=& c_{23} \tilde{ S }_{\tau e} - s_{23} e^{ - i \delta} \tilde{ S }_{\mu e} 
\nonumber \\ 
\left( S_{ \text{EV} } \right)_{\tau \mu} &=& c^2_{23} \tilde{ S }_{\tau \mu } - s^2_{23} e^{ - 2 i \delta} \tilde{ S }_{ \mu \tau} + c_{23} s_{23} e^{ - i \delta} ( \tilde{ S }_{\tau \tau} - \tilde{ S }_{\mu \mu} ) 
\nonumber \\
\left( S_{ \text{EV} } \right)_{\tau \tau} &=& s^2_{23} \tilde{ S }_{\mu \mu} + c^2_{23} \tilde{ S }_{\tau \tau} - c_{23} s_{23} ( e^{ - i \delta} \tilde{ S }_{\mu \tau} + e^{ i \delta} \tilde{ S }_{\tau \mu} ).
\label{SU-elements}
\end{eqnarray}
Notice that the leading zeroth-order contribution to $S_{ \text{EV} }$ is nothing but the one $S_{ \text{EV} }^{(0)} = U_{23} \tilde{S}^{(0)} U_{23}^{\dagger}$ to be given in section~\ref{sec:hatS-tildeS-helio}. 
Notice also that the first order corrections to the evolution-$S$ matrix $S_{ \text{EV} }$ contain contributions both from the helio- and UV $\alpha$ parameter terms in the Hamiltonian, $\hat{H}^{(1)}_{ \text{vac} }$ and $\hat{H}^{(1)}_\text{ UV}$, respectively. We hereafter denote them as $S_{ \text{helio} }^{(1)}$ and $S_{ \text{EV} }^{(1)}$, only whose latter contains the $\alpha$ parameters.

Finally, the oscillation probabilities are simply given by 
\begin{eqnarray} 
P(\nu_{\beta} \rightarrow \nu_{\alpha}; x) = \vert S_{\alpha \beta} (x)
\vert^2. 
\label{probability-S-squared}
\end{eqnarray}

\subsection{Calculation of $\hat{S}$ matrix}

To calculate $\hat {S} (x)$ we define $\Omega(x)$ as
\begin{eqnarray} 
\Omega(x) = e^{i \hat{H}_{0} x} \hat{S} (x).
\label{def-omega}
\end{eqnarray}
Then, $\Omega(x)$ obeys the evolution equation
\begin{eqnarray} 
i \frac{d}{dx} \Omega(x) = H_{1} \Omega(x) 
\label{eq:omega-evolution}
\end{eqnarray}
where
\begin{eqnarray} 
H_{1} \equiv e^{i \hat{H}_{0} x} \hat{H}_{1} e^{-i \hat{H}_{0} x} .
\label{eq:def-H1}
\end{eqnarray}
where $\hat{H}_{1} = \hat{H}^{(1)}_{ \text{vac} } + \hat{H}^{(1)}_\text{ UV }$. See eqs.~\eqref{hatH-vac-1}~and~\eqref{hatH-matt-1}. 
Then, $\Omega(x)$ can be computed perturbatively as
\begin{eqnarray} 
\Omega(x) &=& 1 + 
(-i) \int^{x}_{0} dx' H_{1} (x') + 
(-i)^2 \int^{x}_{0} dx' H_{1} (x') \int^{x'}_{0} dx'' H_{1} (x'') 
+ \cdot \cdot \cdot,
\label{Omega-expansion}
\end{eqnarray}
and the $\hat{S}$ matrix is given by
\begin{eqnarray} 
\hat{S} (x) =  
e^{-i \hat{H}_{0} x} \Omega(x). 
\label{hat-Smatrix}
\end{eqnarray}

\subsection{Recapitulating the leading order $\tilde{S}$ matrix and the first order helio corrections }
\label{sec:hatS-tildeS-helio} 

Since all the relevant quantities are computed for the leading order
and the helio corrections in ref.~\cite{Minakata:2015gra}, we just
recapitulate them in below. The zeroth order result of $\tilde{S}$
matrix is given by
\begin{eqnarray} 
\tilde{S}^{(0)} 
&=& 
U_{\phi} 
e^{-i \hat{H}_{0} x} U_{\phi}^{\dagger} 
= \left[
\begin{array}{ccc}
c_{\phi}^2 e^{ - i h_{1} x } + s_{\phi}^2 e^{ - i h_{3} x } & 
0 & 
c_{\phi} s_{\phi} \left( e^{ - i h_{3} x } - e^{ - i h_{1} x } \right) \\
0 & 
e^{ - i h_{2} x } & 
0 \\
c_{\phi} s_{\phi} \left( e^{ - i h_{3} x } - e^{ - i h_{1} x } \right) & 
0 & 
c_{\phi}^2 e^{ - i h_{3} x } + s_{\phi}^2 e^{ - i h_{1} x } \\
\end{array}
\right] 
\label{tilde-Smatrix-0th}
\end{eqnarray}
where $c_{\phi} \equiv \cos \phi$ and $s_{\phi} \equiv \sin \phi$.
The non-vanishing first order helio corrections (order $\sim
\epsilon$) to $\tilde{S}$ matrix are given by
\begin{eqnarray} 
\left( \tilde{S}^{(1)}_{ \text{helio} } \right)_{e \mu} &=& 
\left( \tilde{S}^{(1)}_{ \text{helio} } \right)_{\mu e} = 
\epsilon \Delta_{ren} c_{12} s_{12} 
\left[
c_{\phi} c_{\left( \phi - \theta_{13} \right)} \frac{e^{ - i h_{2} x } - e^{ - i h_{1} x }}{ h_{2} - h_{1} } + s_{\phi} s_{ \left( \phi - \theta_{13} \right)} \frac{e^{ - i h_{3} x } - e^{ - i h_{2} x }}{ h_{3} - h_{2} } 
\right], 
\nonumber \\
\left( \tilde{S}^{(1)}_{ \text{helio} } \right)_{\mu \tau} &=& 
\left( \tilde{S}^{(1)}_{ \text{helio} } \right)_{\tau \mu} = 
\epsilon \Delta_{ren} c_{12} s_{12} 
\left[
- s_{\phi} c_{\left( \phi - \theta_{13} \right)} \frac{e^{ - i h_{2} x } - e^{ - i h_{1} x }}{ h_{2} - h_{1} } + c_{\phi} s_{\left( \phi - \theta_{13} \right)} \frac{e^{ - i h_{3} x } - e^{ - i h_{2} x }}{ h_{3} - h_{2} } 
\right], 
\nonumber \\
\label{tildeS-elements-1st} 
\end{eqnarray}
and all the other elements vanish. The elements of the evolution-$S$ matrix $S_{ \text{EV} } = U_{23} \tilde{S} U_{23}^{\dagger}$ can be obtained from $\tilde{S}$ matrix elements by using eq.~\eqref{SU-elements}.

\subsection{Calculation of $\hat{S}^{(1)}$ and $\tilde{S}^{(1)}$ for the evolution part of the $S$ matrix}
\label{sec:hatS-tildeS-matt}

In this paper, we restrict ourselves to the perturbative calculation
to first order in $\epsilon \equiv \Delta m^2_{21}/ \Delta m^2_{
  \text{ren} } \approx \Delta m^2_{21}/ \Delta m^2_{31} $ and to first
order in the UV parameters $\alpha_{\beta \gamma}$. Then, the form of
$S$ matrix and the oscillation probability in zeroth and the
first-order helio corrections are identical with those computed in
ref.~\cite{Minakata:2015gra}. Therefore, we only calculate, in the
rest of this section, the matter part which produces the unitary evolution part (and eventually the genuine UV part in higher orders) of the $S$ matrix. 

By inserting $U_{\phi}^{\dagger} U_{\phi}$, $H_{1}$ (hereafter the
matter part only) can be written as
\begin{eqnarray} 
&& H_{1} \equiv 
\Delta_{b} 
U_{\phi}^{\dagger} U_{\phi}
e^{i \hat{H}_{0} x} 
U^{\dagger}_{\phi} H 
U_{\phi}
e^{-i \hat{H}_{0} x} 
U_{\phi}^{\dagger} U_{\phi}
\nonumber \\
&=&
\Delta_{b} 
U_{\phi}^{\dagger} 
\left[
\begin{array}{ccc}
c_{\phi}^2 e^{ i h_{1} x } + s_{\phi}^2 e^{ i h_{3} x } & 
0 & 
c_{\phi} s_{\phi} \left( e^{ i h_{3} x } - e^{ i h_{1} x } \right) \\
0 & 
e^{ i h_{2} x } & 
0 \\
c_{\phi} s_{\phi} \left( e^{ i h_{3} x } - e^{ i h_{1} x } \right) & 
0 & 
c_{\phi}^2 e^{ i h_{3} x } + s_{\phi}^2 e^{ i h_{1} x } \\
\end{array}
\right] 
\left[
\begin{array}{ccc}
H_{11} & H_{12} & H_{13} \\
H_{21} & H_{22} & H_{23} \\
H_{31} & H_{32} & H_{33} \\
\end{array}
\right] 
\nonumber \\
&\times&
\left[
\begin{array}{ccc}
c_{\phi}^2 e^{- i h_{1} x } + s_{\phi}^2 e^{- i h_{3} x } & 
0 & 
c_{\phi} s_{\phi} \left( e^{- i h_{3} x } - e^{- i h_{1} x } \right) \\
0 & 
e^{- i h_{2} x } & 
0 \\
c_{\phi} s_{\phi} \left( e^{- i h_{3} x } - e^{- i h_{1} x } \right) & 
0 & 
c_{\phi}^2 e^{- i h_{3} x } + s_{\phi}^2 e^{- i h_{1} x } \\
\end{array}
\right] 
U_{\phi} 
\nonumber \\
&\equiv& 
\Delta_{b} 
U_{\phi}^{\dagger} 
\left[
\begin{array}{ccc}
\Phi_{11} & \Phi_{12} & \Phi_{13} \\
\Phi_{21} & \Phi_{22} & \Phi_{23} \\
\Phi_{31} & \Phi_{32} & \Phi_{33} \\
\end{array}
\right] 
U_{\phi}
\equiv
\Delta_{b} 
U_{\phi}^{\dagger} 
\Phi U_{\phi}, 
\label{H1-matter}
\end{eqnarray}
where we have introduced another simplifying matrix notation $\Phi$ and its elements $\Phi_{ij}$. The explicit expressions of $\Phi_{ij}$ are given in appendix~\ref{sec:Hij-Phi-ij}.

Assuming the uniform matter density, we obtain the $\alpha$ parameter related part of the first order evolution $\hat{S}$ matrix 
\begin{eqnarray} 
&& \hat{S} (x)_{ \text{EV} }^{(1)} = 
e^{-i \hat{H}_{0} x} \Omega(x)^{(1)}_{ \text{matt} } 
= 
\Delta_{b} U_{\phi}^{\dagger} U_{\phi} 
e^{-i \hat{H}_{0} x} 
U_{\phi}^{\dagger} 
\left[ (-i) \int^{x}_{0} dx' \Phi (x') \right] 
U_{\phi} 
\nonumber \\
&=& 
\Delta_{b} U_{\phi}^{\dagger}
\left[
\begin{array}{ccc}
c_{\phi}^2 e^{- i h_{1} x } + s_{\phi}^2 e^{- i h_{3} x } & 
0 & 
c_{\phi} s_{\phi} \left( e^{- i h_{3} x } - e^{- i h_{1} x } \right) \\
0 & 
e^{- i h_{2} x } & 
0 \\
c_{\phi} s_{\phi} \left( e^{- i h_{3} x } - e^{- i h_{1} x } \right) & 
0 & 
c_{\phi}^2 e^{- i h_{3} x } + s_{\phi}^2 e^{- i h_{1} x } \\
\end{array}
\right] 
\nonumber \\
&\times& 
(-i) \int^{x}_{0} dx' 
\left[
\begin{array}{ccc}
\Phi_{11} (x') & \Phi_{12} (x') & \Phi_{13} (x') \\
\Phi_{21} (x') & \Phi_{22} (x') & \Phi_{23} (x') \\
\Phi_{31} (x') & \Phi_{32} (x') & \Phi_{33} (x') \\
\end{array}
\right] U_{\phi}. 
\label{hat-Smatrix-1st}
\end{eqnarray}
At first order in the evolution S matrix, we are using the subscript ``EV'' just to specify the first order in the $\alpha$ matrix terms. Since $\tilde{S}_{ \text{EV} }^{(1)} = U_{\phi} \hat{S}_{
  \text{EV} }^{(1)} U_{\phi}^{\dagger}$, one can obtain $\tilde{S}_{
  \text{EV} }^{(1)}$ by removing $U_{\phi}^{\dagger}$ and $U_{\phi}$
from eq.~\eqref{hat-Smatrix-1st}. Their explicit forms are given in
appendix~\ref{sec:tilde-S-UV-1st}.
Then, the $\alpha$ parameter related part of the first order contribution to the $S_{ \text{EV} }^{(1)}$ matrix can readily be calculated as $S_{ \text{EV} }^{(1)} = U_{23} \tilde{S}_{ \text{EV} }^{(1)} U_{23}^{\dagger}$. 

\section{Neutrino oscillation probability to first order: $\nu_{e} - \nu_{\mu}$ sector} 
\label{sec:P-e-mu-sector}

In this section, we present the expressions of the oscillation probability in the $\nu_{\mu} \rightarrow \nu_{e}$ and $\nu_{\mu} \rightarrow \nu_{\mu}$ channels. They may be the most relevant ones in the accelerator and atmospheric neutrino experiments. The oscillation probabilities in the $\nu_{\mu} - \nu_{\tau}$ sector are given in appendix~\ref{sec:mu-tau-sector}. For convenience of the readers, the locations of the oscillation probability in these and the other channels are tabulated in Table~\ref{tab:P-eq-numbers} at the end of section~\ref{sec:P-mue-mumu}. 
We start from the general expression to first order. 

\subsection{General expression of the oscillation probability to first order} 
\label{sec:P-general}

To first order in $\epsilon$ or UV $\alpha$ parameters, it is convenient to categorize $P(\nu_{\beta} \rightarrow \nu_{\alpha})$ into the three
types of terms:
\begin{eqnarray} 
P(\nu_{\beta} \rightarrow \nu_{\alpha}) &=& 
P(\nu_{\beta} \rightarrow \nu_{\alpha})_{ \text{ helio} }^{(0+1)} 
+ P(\nu_{\beta} \rightarrow \nu_{\alpha})_{ \text{EV} }^{(1)} 
+ P(\nu_{\beta} \rightarrow \nu_{\alpha})_{ \text{ UV } }^{(1)}, 
\label{S-flavor-all}
\end{eqnarray}
where 
\begin{eqnarray} 
P(\nu_{\beta} \rightarrow \nu_{\alpha})_{ \text{ helio} }^{(0+1)} 
&=& 
\vert S^{(0)}_{\alpha \beta} \vert^2 
+ 2 \mbox{Re} \left[
\left( S^{(0)}_{\alpha \beta} \right)^* 
\left( S^{(1)}_{ \text{ helio} } \right)_{\alpha \beta} 
\right], 
\nonumber \\
P(\nu_{\beta} \rightarrow \nu_{\alpha})_{ \text{EV} }^{(1)}
&=& 
2 \mbox{Re} \left[
\left( S^{(0)}_{\alpha \beta} \right)^* 
\left( S^{(1)}_{ \text{EV} } \right)_{\alpha \beta} 
\right], 
\nonumber \\
P(\nu_{\beta} \rightarrow \nu_{\alpha})_{ \text{ UV } }^{(1)}
&=& 
2 \mbox{Re} \left[
\left( S^{(0)}_{\alpha \beta} \right)^* 
\left( S_{ \text{UV} }^{(1)} \right)_{\alpha \beta} 
\right]
=
- 2 \mbox{Re} \left[
\left( S^{(0)}_{\alpha \beta} \right)^* 
\left( \alpha S^{(0)} + S^{(0)} \alpha^{\dagger} \right)_{\alpha \beta} 
\right]. 
\label{P-three-types}
\end{eqnarray}

The first term in eq.~\eqref{S-flavor-all}, $P(\nu_{\beta} \rightarrow \nu_{\alpha})_{ \text{ helio} }^{(0+1)}$, is the $\nu$SM part \cite{Minakata:2015gra}, the second and third terms denote the $\alpha$-parameter related unitary evolution part, and the non-unitary part due to initial and final multiplication of the $\alpha$ matrix, respectively. Notice that $S^{(0)}$ is obtained as $S^{(0)} \equiv S_{ \text{EV} }^{(0)} = U_{23} \tilde{S}^{(0)} U_{23}^{\dagger}$ by using \eqref{tilde-Smatrix-0th}. In the last equality in \eqref{S-EV-UV}, we have restricted to the first order terms, omitting the last term as $S_{ \text{UV} } = - \alpha S_{ \text{EV} } - S_{ \text{EV} } \alpha^{\dagger}$. 

\subsection{The oscillation probabilities $P(\nu_{\mu} \rightarrow \nu_{e})$ and $P(\nu_{\mu} \rightarrow \nu_{\mu})$ in first order}
\label{sec:P-mue-mumu} 

We now present the oscillation probability $P(\nu_{\mu} \rightarrow \nu_{\beta})$ ($\beta=e, \mu$) using the notations in eq.~\eqref{P-three-types}. Here, we restrict ourselves to the pieces which contain the UV $\alpha$ parameters, $P(\nu_{\mu} \rightarrow \nu_{\beta})_{ \text{EV} }^{(1)}$ and $P(\nu_{\mu} \rightarrow \nu_{\beta})_{ \text{ UV } }^{(1)}$, because they contain information on UV. The $\nu$SM part, $P(\nu_{\mu} \rightarrow \nu_{\beta})_{ \text{ helio} }^{(0+1)}$, which was derived in  ref.~\cite{Minakata:2015gra} is recapitulated in appendix~\ref{sec:e-mu-sector} to make this paper self-contained. 

The $\alpha$ parameter related first-order unitary and non-unitary contributions to $P(\nu_{\mu} \rightarrow \nu_{\beta})$ ($\beta=e, \mu$) read: 
\begin{eqnarray} 
&& 
P(\nu_{\mu} \rightarrow \nu_{e})_{ \text{EV} }^{(1)} 
= 2 \mbox{Re} \left[
\left( S^{(0)}_{e \mu} \right)^* 
\left( S^{(1)}_{ \text{EV} } \right)_{e \mu} 
\right] 
\nonumber \\
&=& 
s^2_{23} 
\sin^2 2\phi \cos 2 \phi 
%
\left[ 
\left\{ \left( \frac{ \Delta_{a} }{ \Delta_{b} } - 1 \right) \alpha_{ee} + \alpha_{\mu \mu} \right\} 
- c_{23}^2 ( \alpha_{\mu \mu} - \alpha_{\tau \tau} ) 
+ c_{23} s_{23} \mbox{Re} \left( e^{i \delta } \alpha_{\tau \mu} \right)
\right] 
\nonumber \\
&\times&
( \Delta_{b} x) \sin ( h_{3} - h_{1} ) x 
\nonumber \\
&+& 
s^2_{23} \sin^3 2\phi 
\left[
s_{23} \mbox{Re} \left( e^{- i \delta } \alpha_{\mu e} \right) 
+ c_{23} \mbox{Re} \left( \alpha_{\tau e} \right) 
\right]
( \Delta_{b} x) \sin ( h_{3} - h_{1} ) x 
\nonumber \\
&+& 
\sin 2\theta_{23} \sin 2\phi 
\nonumber \\ 
&\times& 
\biggl\{
c_{\phi}^2 \left[
c_{23} \mbox{Re} \left( e^{- i \delta } \alpha_{\mu e} \right) - s_{23} \mbox{Re} \left( \alpha_{\tau e} \right) 
\right] 
- \cos 2\theta_{23} c_{\phi} s_{\phi} \mbox{Re} \left( e^{ i \delta } \alpha_{\tau \mu} \right) 
- \sin 2\theta_{23} c_{\phi} s_{\phi} ( \alpha_{\mu \mu} - \alpha_{\tau \tau} ) 
\biggr\}
\nonumber \\ 
&\times& 
\frac{ \Delta_{b} }{ ( h_{2} - h_{1} ) }
\left\{
- \sin^2 \frac{ ( h_{3} - h_{2} ) x }{2} 
+ \sin^2 \frac{ ( h_{3} - h_{1} ) x }{2} 
+ \sin^2 \frac{ ( h_{2} - h_{1} ) x }{2} 
\right\} 
\nonumber \\
&+& 
\sin 2\theta_{23} \sin 2\phi 
\nonumber \\
&\times& 
\biggl\{
s_{\phi}^2 \left[
c_{23} \mbox{Re} \left( e^{- i \delta } \alpha_{\mu e} \right) - s_{23} \mbox{Re} \left( \alpha_{\tau e} \right) 
\right]
+ \cos 2\theta_{23} c_{\phi} s_{\phi} \mbox{Re} \left( e^{ i \delta } \alpha_{\tau \mu} \right) 
+ \sin 2\theta_{23} c_{\phi} s_{\phi} ( \alpha_{\mu \mu} - \alpha_{\tau \tau} ) 
\biggr\}
\nonumber \\
&\times&
\frac{ \Delta_{b} }{ ( h_{3} - h_{2} ) }
\left\{
- \sin^2 \frac{ ( h_{2} - h_{1} ) x }{2}
+ \sin^2 \frac{ ( h_{3} - h_{2} ) x }{2} 
+ \sin^2 \frac{ ( h_{3} - h_{1} ) x }{2} 
\right\} 
\nonumber \\
&-& 
4 s^2_{23} 
\sin 2\phi \cos 2 \phi 
\biggl\{ 
\sin 2\phi
\left[ 
\left\{ \left( \frac{ \Delta_{a} }{ \Delta_{b} } - 1 \right) \alpha_{ee} + \alpha_{\mu \mu} \right\} 
- c_{23}^2 ( \alpha_{\mu \mu} - \alpha_{\tau \tau} ) 
+ c_{23} s_{23} \mbox{Re} \left( e^{i \delta } \alpha_{\tau \mu} \right)
\right] 
\nonumber \\
&-& 
\cos 2 \phi 
\left[
s_{23} \mbox{Re} \left( e^{- i \delta } \alpha_{\mu e} \right) + c_{23} \mbox{Re} \left( \alpha_{\tau e} \right) 
\right]
\biggr\}
\frac{ \Delta_{b} }{ h_{3} - h_{1} } 
\sin^2 \frac{ ( h_{3} - h_{1} ) x }{2} 
\nonumber \\
&+& 
2 \sin 2\theta_{23} \sin 2\phi 
\biggl\{
c_{\phi}^2 \left[ 
c_{23} \mbox{Im} \left( e^{- i \delta } \alpha_{\mu e} \right) - s_{23} \mbox{Im} \left( \alpha_{\tau e} \right) 
\right] 
+ c_{\phi} s_{\phi} \mbox{Im} \left( e^{ i \delta } \alpha_{\tau \mu} \right) 
\biggr\}
\nonumber \\
&\times& 
\frac{ \Delta_{b} }{ ( h_{2} - h_{1} ) } 
\sin \frac{( h_{3} - h_{1} ) x}{2} 
\sin \frac{( h_{1} - h_{2} ) x}{2} 
\sin \frac{( h_{2} - h_{3} ) x}{2} 
\nonumber \\
&-& 
2 \sin 2\theta_{23} \sin 2\phi 
\biggl\{
s_{\phi}^2 \left[
c_{23} \mbox{Im} \left( e^{- i \delta } \alpha_{\mu e} \right) - s_{23} \mbox{Im} \left( \alpha_{\tau e} \right) 
\right] 
- c_{\phi} s_{\phi} \mbox{Im} \left( e^{ i \delta } \alpha_{\tau \mu} \right) 
\biggr\}
\nonumber \\
&\times&
\frac{ \Delta_{b} }{ ( h_{3} - h_{2} ) }
\sin \frac{( h_{3} - h_{1} ) x}{2} 
\sin \frac{( h_{1} - h_{2} ) x}{2} 
\sin \frac{( h_{2} - h_{3} ) x}{2}, 
\label{P-mue-intrinsic-UV}
\end{eqnarray}
\begin{eqnarray} 
&& 
P(\nu_{\mu} \rightarrow \nu_{e})_{ \text{ UV } }^{(1)} 
= 
2 \mbox{Re} \left[
\left( S^{(0)}_{e \mu} \right)^* 
\left( S_{ \text{UV} }^{(1)} \right)_{e \mu} 
\right]
\nonumber \\
&=& 
2 s_{23} \sin 2\phi 
\left[ 
\cos 2 \phi 
\mbox{Re} \left( e^{- i \delta } \alpha_{\mu e} \right)  
- s_{23} \sin 2\phi 
\left( \alpha_{e e} + \alpha_{\mu \mu} \right) 
\right]
\sin^2 \frac{ ( h_{3} - h_{1} ) x }{2} 
\nonumber \\
&-& 
s_{23} \sin 2\phi  
\mbox{Im} \left( e^{- i \delta } \alpha_{\mu e} \right) 
\sin ( h_{3} - h_{1} ) x.  
\label{P-mue-extrinsic-UV}
\end{eqnarray}
\begin{eqnarray} 
&& 
P(\nu_{\mu} \rightarrow \nu_{\mu})_{ \text{EV} }^{(1)} 
= 2 \mbox{Re} \left[
\left( S^{(0)}_{\mu \mu} \right)^* 
\left( S^{(1)}_{ \text{EV} } \right)_{\mu \mu} 
\right] 
\nonumber \\
&=& 
- \sin^2 2\theta_{23} 
\left[
\cos 2 \theta_{23} \left( \alpha_{\tau \tau} - \alpha_{\mu \mu} \right) 
+ \sin 2 \theta_{23} \mbox{Re} \left( e^{i \delta } \alpha_{\tau \mu} \right) 
\right]
( \Delta_{b} x ) 
\biggl\{
c_{\phi}^2 \sin ( h_{3} - h_{2} ) x 
- s_{\phi}^2 \sin ( h_{2} - h_{1} ) x 
\biggr\}
\nonumber \\
&+& 
s^2_{23} \sin^2 2\phi 
\left[ 
\left\{ \left( \frac{ \Delta_{a} }{ \Delta_{b} } - 1 \right) \alpha_{ee} + \alpha_{\mu \mu} \right\} 
- c_{23}^2 ( \alpha_{\mu \mu} - \alpha_{\tau \tau} ) 
+ c_{23} s_{23} \mbox{Re} \left( e^{i \delta } \alpha_{\tau \mu} \right)
\right] 
\nonumber \\
&\times&
( \Delta_{b} x ) 
\biggl\{
c^2_{23} \sin ( h_{3} - h_{2} ) x 
- c^2_{23} \sin ( h_{2} - h_{1} ) x 
- s^2_{23} \cos 2\phi \sin ( h_{3} - h_{1} ) x
\biggr\}
\nonumber \\
&-& 
2 s^2_{23} \sin 2\phi   
\left[
s_{23} \mbox{Re} \left( e^{- i \delta } \alpha_{\mu e} \right) 
+ c_{23} \mbox{Re} \left( \alpha_{\tau e} \right) 
\right]
\nonumber \\
&\times&
( \Delta_{b} x ) 
\biggl\{ 
c^2_{23} c_{\phi}^2 \sin ( h_{3} - h_{2} ) x 
+ c^2_{23} s_{\phi}^2 \sin ( h_{2} - h_{1} ) x 
+ 2 s^2_{23} c_{\phi}^2 s_{\phi}^2 \sin ( h_{3} - h_{1} ) x
\biggr\} 
\nonumber \\
&+& 
4 s^2_{23} \sin 2\phi
\biggl\{ 
\sin 2\phi
\left[ 
\left\{ \left( \frac{ \Delta_{a} }{ \Delta_{b} } - 1 \right) \alpha_{ee} + \alpha_{\mu \mu} \right\} 
- c_{23}^2 ( \alpha_{\mu \mu} - \alpha_{\tau \tau} ) 
+ c_{23} s_{23} \mbox{Re} \left( e^{i \delta } \alpha_{\tau \mu} \right)
\right] 
\nonumber \\
&-& 
\cos 2 \phi 
\left[
s_{23} \mbox{Re} \left( e^{- i \delta } \alpha_{\mu e} \right) 
+ c_{23} \mbox{Re} \left( \alpha_{\tau e} \right) 
\right]
\biggr\} 
\nonumber \\
&\times&
\frac{ \Delta_{b} }{ h_{3} - h_{1} }
\biggl\{ 
s^2_{23} \cos 2 \phi \sin^2 \frac{ ( h_{3} - h_{1} ) x }{2}  
- c^2_{23} \sin^2 \frac{ ( h_{3} - h_{2} ) x }{2} 
+ c^2_{23} \sin^2 \frac{ ( h_{2} - h_{1} ) x }{2} 
\biggr\}
\nonumber \\
&-& 
4 \sin 2\theta_{23}
\biggl\{ 
c_{\phi} s_{\phi} 
\left[ 
c_{23} \mbox{Re} \left( e^{- i \delta } \alpha_{\mu e} \right) - s_{23} \mbox{Re} \left( \alpha_{\tau e} \right) \right] 
- \cos 2\theta_{23} s_{\phi}^2 \mbox{Re} \left( e^{ i \delta } \alpha_{\tau \mu} \right) 
- \sin 2\theta_{23} s_{\phi}^2 ( \alpha_{\mu \mu} - \alpha_{\tau \tau} ) 
\biggr\} 
\nonumber \\
&\times&
\frac{ \Delta_{b} }{ ( h_{2} -  h_{1} ) } 
\biggl\{
s^2_{23} c_{\phi}^2 \sin^2 \frac{ ( h_{3} - h_{1} ) x }{2} 
- s^2_{23} c_{\phi}^2 \sin^2 \frac{ ( h_{3} - h_{2} ) x }{2}
+ \left( c^2_{23} - s^2_{23} s_{\phi}^2 \right) 
\sin^2 \frac{ ( h_{2} - h_{1} ) x }{2} 
\biggr\}
\nonumber \\
&-& 
4 \sin 2\theta_{23}
\biggl\{ 
c_{\phi} s_{\phi} 
\left[ c_{23} \mbox{Re} \left( e^{- i \delta } \alpha_{\mu e} \right) - s_{23} \mbox{Re} \left( \alpha_{\tau e} \right) \right] 
+ \cos 2\theta_{23} c_{\phi}^2 \mbox{Re} \left( e^{ i \delta } \alpha_{\tau \mu} \right) 
+ \sin 2\theta_{23} c_{\phi}^2 ( \alpha_{\mu \mu} - \alpha_{\tau \tau} )
\biggr\} 
\nonumber \\
&\times&
\frac{ \Delta_{b} }{ ( h_{3} - h_{2} ) } 
\biggl\{ 
s^2_{23} s_{\phi}^2 \sin^2 \frac{ ( h_{3} - h_{1} ) x }{2} 
+ \left( c^2_{23} - s^2_{23} c_{\phi}^2 \right) \sin^2 \frac{ ( h_{3} - h_{2} ) x }{2} 
- s^2_{23} s_{\phi}^2 \sin^2 \frac{ ( h_{2} - h_{1} ) x }{2} 
\biggr\}. 
\label{P-mumu-intrinsic-UV}
\end{eqnarray}
\begin{eqnarray} 
&& P(\nu_{\mu} \rightarrow \nu_{\mu})^{(1)}_{ \text{UV} } 
= 2 \mbox{Re} \left[
\left( S^{(0)}_{\mu \mu} \right)^* 
\left( S^{(1)}_{ \text{UV} } \right)_{\mu \mu} 
\right] 
\nonumber \\
&=& 
4 s_{23} \sin 2\phi 
\mbox{Re} \left( e^{- i \delta } \alpha_{\mu e} \right) 
\left[
- s^2_{23} \cos 2 \phi 
\sin^2 \frac{ ( h_{3} - h_{1} ) x }{2} 
+ c^2_{23} 
\left\{  
\sin^2 \frac{ ( h_{3} - h_{2} ) x }{2} 
- \sin^2 \frac{ ( h_{2} - h_{1} ) x }{2} 
\right\}  
\right]
\nonumber \\
&+& 
4 \alpha_{\mu \mu}
\biggl[ 
-1 
+ s^4_{23} \sin^2 2\phi 
\sin^2 \frac{ ( h_{3} - h_{1} ) x }{2} 
+ \sin^2 2\theta_{23}
\left\{ 
c^2_{\phi} \sin^2 \frac{ ( h_{3} - h_{2} ) x }{2} 
+ s^2_{\phi} \sin^2 \frac{ ( h_{2} - h_{1} ) x }{2} 
\right\}   
\biggr], 
\label{P-mumu-extrinsic-UV}
\end{eqnarray}

In looking into the expressions of $P(\nu_{\mu} \rightarrow \nu_{\beta})_{ \text{EV} }^{(1)}$ and $P(\nu_{\mu} \rightarrow \nu_{\beta})_{ \text{ UV } }^{(1)}$ ($\beta=e, \mu$) above, we observe the following two key features which are qualitatively new:\footnote{
A perturbative treatment using the similar expansion parameters is presented in ref.~\cite{Li:2015oal} within the framework of $3+3$ model, in which the calculation of the oscillation probabilities of the first order are carried out. However, due to different implementation of UV, it is essentially impossible to compare our formulas to theirs. As a consequence, none of the points of our emphasis, the canonical phase combination, diagonal $\alpha$ parameter correlation, and unitarity of neutrino propagation in matter, is not reached in their paper.
}

\begin{itemize}

\item
The complex $\alpha$ parameters are accompanied by the $\nu$SM CP phase $\delta$ in the particular way, $[e^{- i \delta } \alpha_{\mu e}$, $\alpha_{\tau e}$, $e^{i \delta } \alpha_{\tau \mu}]$, which we call the ``{\em canonical phase combination''} in this paper.\footnote{
Non-association of $e^{ \pm i \delta }$ to $\alpha_{\tau e}$ must be understood as a particular ``correlation'', and naturally there is no association of $\delta$ in the diagonal $\alpha$ parameters, $\alpha_{\beta \beta}$ ($\beta=e,\mu,\tau$).
}

\item 
The diagonal $\alpha$ parameters have a universal correlation 
\begin{eqnarray} 
\left( \frac{ \Delta_{a} }{ \Delta_{b} } - 1 \right) \alpha_{ee} + \alpha_{\mu \mu}, 
\hspace{8mm}
\text{and}
\hspace{8mm}
\alpha_{\mu \mu} - \alpha_{\tau \tau}, 
\end{eqnarray} 
in the unitary evolution part of the probability $P(\nu_{\beta} \rightarrow \nu_{\gamma})_{ \text{EV} }^{(1)}$, but {\em not} in the UV part  $P(\nu_{\beta} \rightarrow \nu_{\gamma})_{ \text{ UV } }^{(1)}$. 

\end{itemize}

\noindent 
We remark that the above two properties hold in all the oscillation channels, as we will confirm by seeing the rest of the expressions of the oscillation probability in appendices~\ref{sec:e-mu-sector} and \ref{sec:mu-tau-sector}. That is why the flavor indices are elevated to the generic ones in the above, $\beta, \gamma = e, \mu, \tau$ in advance. 
The crucial difference, is however, that the canonical phase combination prevails even in the non-unitary part of the probability, while the diagonal $\alpha$ parameter correlation does not. 

The readers may ask ``Why is such canonical phase combination realized so universally?''. We plan to give an organized discussion of the mechanism for generating the canonical phase combination and its stability in section~\ref{sec:stability-convention-dep}.

As noticed in the above description the properties of $P(\nu_{\beta} \rightarrow \nu_{\gamma})_{ \text{ UV } }^{(1)}$ are vastly different from those of $P(\nu_{\beta} \rightarrow \nu_{\gamma})_{ \text{ EV } }^{(1)}$. Another notable feature of the UV part of the disappearance probability is that there is a constant term which is proportional to the $\alpha$ parameter, $\alpha_{ee}$ and $\alpha_{\mu \mu}$ in $P(\nu_{e} \rightarrow \nu_{e})_{ \text{ UV } }^{(1)}$ and $P(\nu_{\mu} \rightarrow \nu_{\mu})_{ \text{ UV } }^{(1)}$, respectively. 

A few comments are ready on both of the two characteristic features itemized above. It is good to know that a part of our canonical phase combination, $e^{ - i \delta } \alpha_{\mu e}$, has been observed in the foregoing studies~\cite{Escrihuela:2015wra,Miranda:2016wdr,Abe:2017jit}.\footnote{
Our result is consistent with theirs if the authors of refs.~\cite{Escrihuela:2015wra,Miranda:2016wdr,Abe:2017jit} have used the $U_{\text{\tiny PDG}}$ or $U_{\text{\tiny ATM}}$ conventions because the correlation $e^{ - i \delta } \alpha_{\mu e}$ holds in the both conventions. See section~\ref{sec:convention-dep}.
} 
Our result places these observation into more generic setting which includes all the complex $\alpha$ parameters, and clarify its nature using the first order analytic formulas. See section~\ref{sec:stability-convention-dep} for more about it.

The diagonal $\alpha$ parameter correlation that holds for the unitary part of the probability $P(\nu_{\mu} \rightarrow \nu_{\beta})_{ \text{EV} }^{(1)}$ stems from an invariance under phase redefinition, which would manifest if one goes to the ``would-be flavor'' basis (the one which would be the flavor basis in the absence of $\alpha$) 
\begin{eqnarray} 
H_{ \text{wb-flavor} } \equiv U \check{H} U^{\dagger} 
= U_{23} \tilde{H} U_{23}^{\dagger}. 
\label{wb-flavor-H}
\end{eqnarray}
The UV part of the $H_{ \text{wb-flavor} }$ includes the form of an inversely $U_{23}$ rotated $H$ matrix $\propto  (\alpha + \alpha^{\dagger}) $ as seen in eq.~\eqref{Hij-def}. 
One should note that the simple form of the diagonal $\alpha$ parameter correlation above ceases to hold if the terms in second order in $\alpha$ is included. Of course, there is no reason to expect the same diagonal $\alpha$ parameter correlation hold in the non-unitary part $P(\nu_{\mu} \rightarrow \nu_{\beta})_{ \text{UV} }^{(1)}$.\footnote{
An exceptional situation occurs in the $\nu_{\mu} \rightarrow \nu_{e}$ channel under the charge neutral iso-singlet medium in which the $\alpha_{ee}$ diagonal correlation takes the same form $\alpha_{ee} + \alpha_{\mu \mu}$ in both the unitary and the non-unitary parts $P(\nu_{\mu} \rightarrow \nu_{\beta})_{ \text{EV} }^{(1)}$ and $P(\nu_{\mu} \rightarrow \nu_{\beta})_{ \text{UV} }^{(1)}$ in the first order probability formulas. It is just accidental. }

The diagonal element correlation due to re-phasing invariance is a familiar feature in theories with so called the propagation NSI, where ``NSI'' stands for Non-Standard Interactions.
See ref.~\cite{Kikuchi:2008vq} for an explicit demonstration to third order in the NSI parameters. In fact, the authors of ref.~\cite{Blennow:2016jkn} derived a one-to-one correspondence between the propagation NSI and the $\alpha$ parameters under the assumption of charge neutral iso-singlet medium, in which $N_{p} = N_{n}$ and $\Delta_{a} / \Delta_{b} = a/b =2$ hold. Then, one can translate the diagonal NSI parameter correlation into 
the diagonal $\alpha$ parameter correlation. In their treatment~\cite{Blennow:2016jkn}, the correlation involving $\alpha_{ee}$ takes a simpler form $\alpha_{ee} + \alpha_{\mu \mu}$. Our second itemized result above can be regarded as a generalization of theirs.  

In this paper, the expressions of the oscillation probabilities are scattered into various places in this paper. Therefore, for the readers' convenience, we tabulate in table~\ref{tab:P-eq-numbers} the equation numbers for $P(\nu_{\beta} \rightarrow \nu_{\alpha})_{ \text{ EV } }^{(1)}$ and $P(\nu_{\beta} \rightarrow \nu_{\alpha})_{ \text{ UV } }^{(1)}$ in various channels. 

\begin{table}[h!]
\vglue -0.2cm
\begin{center}
\caption{ 
The equation numbers for 
$P(\nu_{\beta} \rightarrow \nu_{\alpha})_{ \text{ EV } }^{(1)}$ and 
$P(\nu_{\beta} \rightarrow \nu_{\alpha})_{ \text{ UV } }^{(1)}$ 
(see eq.~\eqref{P-three-types} for definitions), if available, are summarized. For the ones that the explicit expressions are not available the methods for calculation are briefly mentioned.
}
\label{tab:P-eq-numbers} 
\vglue 0.2cm
\begin{tabular}{c|c|c}
\hline 
~~channel~~ & 
~~~~$P(\nu_{\beta} \rightarrow \nu_{\alpha})_{ \text{ EV } }^{(1)}$~~~~ & 
~~~~$P(\nu_{\beta} \rightarrow \nu_{\alpha})_{ \text{ UV } }^{(1)}$~~~~ \\
\hline 
$\nu_{\mu} \rightarrow \nu_{e}$ &
eq.~\eqref{P-mue-intrinsic-UV} in section~\ref{sec:P-int-ext-UV-1st-ee-mue} &
eq.~\eqref{P-mue-extrinsic-UV} in section~\ref{sec:P-int-ext-UV-1st-ee-mue} \\
\hline 
$\nu_{e} \rightarrow \nu_{e}$ & 
eq.~\eqref{Pee-intrinsic-UV-1st} in section~\ref{sec:P-int-ext-UV-1st-ee-mue} & 
eq.~\eqref{Pee-extrinsic-UV-1st} in section~\ref{sec:P-int-ext-UV-1st-ee-mue} \\
\hline
$\nu_{e} \rightarrow \nu_{\mu}$ & 
T transformation of eq.~\eqref{P-mue-intrinsic-UV} & 
T transformation of eq.~\eqref{P-mue-extrinsic-UV} \\
\hline
$\nu_{e} \rightarrow \nu_{\tau}$ & 
$\theta_{23}$ transformation \eqref{23-rotation} of $\nu_{e} \rightarrow \nu_{\mu}$ & 
the same $\theta_{23}$ transformation \\
\hline 
$\nu_{\mu} \rightarrow \nu_{\mu}$ & 
eq.~\eqref{P-mumu-intrinsic-UV} in section~\ref {sec:P-e-mu-sector} & 
eq.~\eqref{P-mumu-extrinsic-UV} in section~\ref {sec:P-e-mu-sector} \\
\hline 
$\nu_{\mu} \rightarrow \nu_{\tau}$ & 
eq.~\eqref{P-mutau-intrinsic-UV} in appendix~\ref{sec:mu-tau-sector} & 
eq.~\eqref{P-mutau-extrinsic-UV} in appendix~\ref{sec:mu-tau-sector} \\
\hline 
\end{tabular}
\end{center}
\vglue -0.4cm
\end{table}

\subsection{Symmetry of the oscillation probability} 
\label{sec:symmetry}

We have pointed out in ref.~\cite{Martinez-Soler:2019nhb} that the unitary part of our present system possesses the symmetry under the transformations involving angle $\phi$.\footnote{
While we believe that $\phi$ symmetry is first pointed out in ref.~\cite{Martinez-Soler:2019nhb}, the $\varphi$ symmetry, its brother for the matter-dressed $\theta_{12}$, had an ancestor in ref.~\cite{Denton:2016wmg} to which a proper attention was not paid regrettably in \cite{Martinez-Soler:2019nhb}. 
}
Here, we give a more complete discussion including the UV effect. 

One can observe that the oscillation probability given in this section, $P(\nu_{e} \rightarrow \nu_{e})^{(1)}$ and $P(\nu_{\mu} \rightarrow \nu_{e})^{(1)}$, including both the EV and UV parts are invariant under the transformation 
\begin{eqnarray} 
&& 
\phi \rightarrow \phi + \frac{\pi}{2}. 
\label{phi-transformation}
\end{eqnarray}
It implies the simultaneous transformations 
\begin{eqnarray} 
&&
c_{\phi} \rightarrow - s_{\phi}, 
\hspace{8mm}
s_{\phi} \rightarrow c_{\phi}, 
\hspace{8mm}
\cos 2\phi \rightarrow - \cos 2\phi, 
\hspace{8mm}
\sin 2\phi \rightarrow - \sin 2\phi, 
\nonumber \\
&& 
h_{1} \rightarrow h_{3},  
\hspace{10mm}
h_{3} \rightarrow h_{1}, 
\label{all-transformation}
\end{eqnarray}
whose last two, exchanging eigenvalues $h_{1} \leftrightarrow h_{3}$, is enforced by \eqref{phi-transformation}. It can be understood by noticing that 
$\cos (\phi - \theta_{12}) \rightarrow - \sin ( \phi - \theta_{12} )$ and 
$\sin (\phi - \theta_{12}) \rightarrow \cos ( \phi - \theta_{12} )$ under \eqref{phi-transformation}, together with an alternative expressions of the eigenvalues 
\begin{eqnarray} 
h_{1} &=& 
\sin^2 \left( \phi - \theta_{13} \right) \Delta_{ \text{ren} } 
+ c^2_{\phi} \Delta_{a} 
+ \epsilon \Delta_{ \text{ren} } s^2_{12},
\nonumber \\
h_{3} &=&
\cos^2 \left( \phi - \theta_{13} \right) \Delta_{ \text{ren} } 
+ s^2_{\phi} \Delta_{a} 
+ \epsilon \Delta_{ \text{ren} } s^2_{12}.
\label{eigenvalues}
\end{eqnarray}
It is also easy to verify that the invariance under \eqref{phi-transformation} holds in all the oscillation probabilities listed in table~\ref{tab:P-eq-numbers}. The $\phi \rightarrow \phi + \frac{\pi}{2}$ symmetry can be characterized as a ``dynamical symmetry'' as discussed in ref.~\cite{Martinez-Soler:2019nhb}. 
It is interesting to note that in both the ``solar resonance'' perturbation theory and the helio-UV perturbation theory in this paper, the dynamical symmetry involves the matter enhanced mixing angles, $\theta_{12}$ in the former and $\theta_{13}$ in the latter. 

\subsection{Unitarity of neutrino evolution with first order UV corrections: $\nu_{e}$ row}
\label{sec:unitarity-nue}

We briefly discuss unitarity of neutrino oscillation probability in $\nu_{e}$ row. The $\nu$SM part is discussed in ref.~\cite{Minakata:2015gra} so that we do not repeat this part. As we discussed in section~\ref{sec:basis-relations} the $S$ matrix is decomposed into the unitary evolution part and the initial- and final-$\alpha$ matrix multiplied non-unitary part. Unitary holds only in the former, so that 
\begin{eqnarray} 
&& 
\sum_{\alpha = e, \mu, \tau} P(\nu_{\beta}  \rightarrow \nu_{\alpha})_{ \text{ EV } }^{(1)} = \mathcal{O} (\epsilon^2) 
\label{unitarity-nue-row}
\end{eqnarray}
must hold. For simplicity of notation, we have assumed that $\alpha_{\beta \gamma} \sim \epsilon$ in \eqref{unitarity-nue-row}. 

To prove unitarity in $\nu_{e}$ row, we need to prepare the following three oscillation probabilities at first order, 
$P(\nu_{e} \rightarrow \nu_{e})_{ \text{ EV } }^{(1)}$, 
$P(\nu_{e} \rightarrow \nu_{\mu})_{ \text{ EV } }^{(1)}$, and 
$P(\nu_{e} \rightarrow \nu_{\tau})_{ \text{ EV } }^{(1)}$. Now, $P(\nu_{e} \rightarrow \nu_{e})_{ \text{ EV } }^{(1)}$ is given in eq.~(\ref{Pee-intrinsic-UV-1st}). 
$P(\nu_{e} \rightarrow \nu_{\mu})_{ \text{ EV } }^{(1)}$, can be obtained by generalized T transformation\footnote{
The T transformation of the probability from the channel $\nu_{\beta} \rightarrow \nu_{\alpha}$ to $\nu_{\alpha} \rightarrow \nu_{\beta}$ can be done by taking complex conjugate of all the complex parameters, $e^{ \pm i \delta }$ and the complex $\alpha$ parameters. 
} \cite{Fong:2017gke}
of eq.~\eqref{P-mue-intrinsic-UV}. Finally, the oscillation probability of the $\nu_e \rightarrow \nu_\tau$ channel can be obtained by using the $\theta_{23}$ transformation \cite{Akhmedov:2004ny} 
\begin{eqnarray} 
&& 
P(\nu_e \rightarrow \nu_\tau) = P(\nu_e \rightarrow \nu_\mu: c_{23} \rightarrow - s_{23}, s_{23} \rightarrow c_{23}).
\label{23-rotation}
\end{eqnarray}
Thus, one can easily prepare all the necessary ingredients. Then, it is a straightforward exercise to show that 
$P(\nu_{e}  \rightarrow \nu_{e})_{ \text{ EV } }^{(1)} 
+ P(\nu_{e}  \rightarrow \nu_{\mu})_{ \text{ EV } }^{(1)} 
+ P(\nu_{e}  \rightarrow \nu_{\tau})_{ \text{ EV } }^{(1)} = 0$, 
where ``0'' in the right-hand side implies absence of first order terms in the UV parameters. It is also easy to show that the unitarity relation does not hold for the non-unitary counterpart, $\sum_{\beta = e, \mu, \tau} P(\nu_{e}  \rightarrow \nu_{\beta})_{ \text{ UV } }^{(1)} \neq 0$. The similar results corresponding to \eqref{unitarity-nue-row} and its UV counterpart will be shown to hold in $\nu_{\mu}$ row in appendix~\ref{sec:unitarity-nu-mu-row}.

With the similar proof of unitarity in the $\nu_{\mu}$ row in appendix~\ref{sec:unitarity-nu-mu-row}, to our knowledge, they constitute the first explicit proof of the unitarity in evolution part of the probability with the UV parameters. It is nice to see that the general argument on unitarity of neutrino evolution in section~\ref{sec:mass-basis} holds even in the presence of non-unitary mixing matrix.

\section{Canonical phase combination: Stability and phase convention dependence} 
\label{sec:stability-convention-dep}

Given the universal feature of the correlation between the $\nu$SM CP phase and the ones of the UV parameters, a natural question would be: How it comes about and why it is so stable. In this section, we answer these  questions. We also clarify its dependence on the convention of the MNS matrix, and address the related question of physical reality of the phase correlation. 

\subsection{Secret of the stability and universality of the canonical phase combination}
\label{sec:stability}

Let us introduce the terminology {\em ``lozenge position $e^{ \pm i \delta }$ matrix''}, or simply ``lozenge matrix'', or ``lozenge structure'' for short. We define $Y$ as lozenge position $e^{ \pm i \delta }$ matrix if $Y$ has the following form  
\begin{eqnarray}
Y \equiv 
\left[
\begin{array}{ccc}
Y_{e e} & Y_{e \mu} e^{- i \delta} & Y_{e \tau} \\
Y_{\mu e} e^{ i \delta} & Y_{\mu \mu} & Y_{\mu \tau} e^{ i \delta} \\
Y_{\tau e} & Y_{\tau \mu} e^{- i \delta} & Y_{\tau \tau} \\
\end{array}
\right], 
\label{canonical-form-Y}
\end{eqnarray}
where the elements $Y_{\beta \gamma}$ contain $\delta$ and the UV phases only in the form of canonical phase combination. We will show below the stability of the ``lozenge structure'' under multiplications and $U_{23}$ rotation, and it provide the key to the stability and universality of the canonical phase combination. 

One can easily prove the following properties of the lozenge position $e^{ \pm i \delta }$ matrix. 

\begin{itemize}

\item 
If $Y_{1}$ and $Y_{2}$ are both lozenge matrices, $Y_{1} Y_{2}$ is also a lozenge position $e^{ \pm i \delta }$ matrix. 

\item 
$U_{23} Y$ and $Y U_{23}^{\dagger}$ are both lozenge matrices, and hence $U_{23} Y U_{23}^{\dagger}$ is also a lozenge position matrix.

\item 
$\alpha Y$ and $Y \alpha^{\dagger}$ are both lozenge position matrices. See appendix~\ref{sec:alpha-multiplication} for the explicit forms. 

\end{itemize}
\noindent

Now, the most important statement here is that if the oscillation probability is written with use of the two lozenge position matrices as 
$P(\nu_\gamma \rightarrow \nu_\beta) = ( Y_{1} )_{\beta \gamma}^* ( Y_{2} )_{\beta \gamma}$ (i.e., its real or imaginary parts), then the probability contains $\delta$ and the UV phases only in the form of the canonical phase combination. 


We note first that the zeroth-order $S$ matrix has the lozenge position $e^{ \pm i \delta }$ structure:
\begin{eqnarray} 
&& S^{(0)} = U_{23} \tilde{S}^{(0)} U_{23}^{\dagger} 
\nonumber \\
&&
\hspace{-24mm} 
=
\left[
\begin{array}{ccc}
c_{\phi}^2 e^{- i h_{1} x } + s_{\phi}^2 e^{- i h_{3} x } & 
s_{23} e^{- i \delta} c_{\phi} s_{\phi} \left( e^{- i h_{3} x } - e^{- i h_{1} x } \right) & 
c_{23} c_{\phi} s_{\phi} \left( e^{- i h_{3} x } - e^{- i h_{1} x } \right) \\
s_{23} e^{ i \delta} c_{\phi} s_{\phi} \left( e^{- i h_{3} x } - e^{- i h_{1} x } \right) & 
c^2_{23} e^{- i h_{2} x } + s^2_{23} \left( c_{\phi}^2 e^{- i h_{3} x } + s_{\phi}^2 e^{- i h_{1} x } \right) & 
c_{23} s_{23} e^{ i \delta} 
\left( c_{\phi}^2 e^{- i h_{3} x } + s_{\phi}^2 e^{- i h_{1} x } - e^{- i h_{2} x } \right)  \\
c_{23} c_{\phi} s_{\phi} \left( e^{- i h_{3} x } - e^{- i h_{1} x } \right) & 
c_{23} s_{23} e^{ - i \delta} 
\left( c_{\phi}^2 e^{- i h_{3} x } + s_{\phi}^2 e^{- i h_{1} x } - e^{- i h_{2} x } \right) & 
s^2_{23} e^{- i h_{2} x } + c^2_{23} \left( c_{\phi}^2 e^{- i h_{3} x } + s_{\phi}^2 e^{- i h_{1} x } \right) \\
\end{array}
\right] 
\nonumber \\
\label{flavorS-0th}
\end{eqnarray}
but in a special manner such that there is no phase dependence inside the $Y_{\beta \gamma}$ elements. 
It is also not difficult to observe that the first-order unitary evolution amplitude $S^{(1)}_{ \text{ EV } } = U_{23} \tilde{S}_{ \text{EV} }^{(1)} U_{23}^{\dagger}$ is the lozenge position matrix because $\tilde{S}_{ \text{EV} }^{(1)}$ has a lozenge structure. See appendix~\ref{sec:tilde-S-UV-1st}. Then, the first order UV related unitary part of the probability $P(\nu_{\beta} \rightarrow \nu_{\gamma})_{ \text{EV} }^{(1)} = 2 \mbox{Re} \left[ \left( S^{(0)}_{\gamma \beta} \right)^* \left( S^{(1)}_{ \text{EV} } \right)_{\gamma \beta} \right]$ respects the canonical phase combination. 

What about the UV contribution to the probability $P(\nu_{\beta} \rightarrow \nu_{\gamma})_{ \text{UV} }^{(1)}$? Due to the third itemized property above it also respects the canonical phase combination. 

This is to prove that the UV $\alpha$ parameter related first order probability $P(\nu_{\beta} \rightarrow \nu_{\gamma})^{(1)}$, both the unitary and non-unitary parts, respects the canonical phase combination in all the flavor oscillation channels. 
With first itemized property of the lozenge matrix, it is not difficult to extend this argument to show that arbitrary order purely EV or purely UV as well as any mixed EV-UV contributions to the oscillation probability respect the canonical phase combination to all orders in the helio-UV parturbation theory.

The remaining terms in higher order corrections in the $S$ matrix are the ones in which arbitrary number of blobs composed of the $\nu$SM ``helio'' part  are inserted into the above pure EV, UV, or mixed EV-UV contributions, which should be done in computing the $\Omega$ matrix. Though it might look to disturb the canonical phase combination but actually it does {\em not}. It is because the helio blobs depend only on $\delta$. Therefore, the correlation between $\nu$SM CP phase and the UV $\alpha$ parameter phases prevails in higher orders.

\subsection{Phase convention dependence of the canonical phase combination}
\label{sec:convention-dep}

It must be obvious from our discussion in section~\ref{sec:convention-dependence} that the $\alpha$ matrix, and hence the $\alpha_{\beta \gamma}$ parameters, depend on the phase convention of the MNS matrix. Then, the form of the canonical phase combination also depends on the $U_{\text{\tiny MNS}}$ phase convention. But, since the relationship between the $\alpha$ parameters belonging to the three different phase conventions is explicitly given in eq.~\eqref{alpha-alpha-bar-matrix}, it is straightforward to translate the form of canonical phase combination from one convention to another. 

We have obtained the canonical phase combination with the $U_{\text{\tiny ATM}}$ phase convention as 
\begin{eqnarray} 
e^{- i \delta } \alpha_{\mu e}, ~~ \alpha_{\tau e}, ~~ e^{i \delta} \alpha_{\tau \mu}.
\label{C-combination-ATM}
\end{eqnarray}
It can be translated into the one with the $U_{\text{\tiny PDG}}$ phase convention 
\begin{eqnarray} 
e^{- i \delta } \bar{\alpha}_{\mu e}, ~~
e^{ - i \delta} \bar{\alpha}_{\tau e}, ~~
\bar{\alpha}_{\tau \mu}, 
\label{C-combination-PDG}
\end{eqnarray}
for the $\bar{\alpha}$ parameters and 
the one with the $U_{\text{\tiny SOL}}$ phase convention 
\begin{eqnarray} 
\tilde{\alpha}_{\mu e}, ~~
\tilde{\alpha}_{\tau e}, ~~
\tilde{\alpha}_{\tau \mu}, 
\label{C-combination-SOL}
\end{eqnarray}
for the $\tilde{\alpha}$ parameters. See eqs.~\eqref{alpha-matrix-PDG} and \eqref{alpha-alpha-bar-matrix} for the definitions of $\bar{\alpha}$ and $\tilde{\alpha}$ parameters, and their relationship with the $U_{\text{\tiny ATM}}$ convention $\alpha$ parameters.
That is, under the $U_{\text{\tiny SOL}}$ phase convention, no correlation between $\nu$SM phase $\delta$ and the UV $\tilde{\alpha}$ parameter phases exists. Conversely, one can easily show that the $U_{\text{\tiny SOL}}$ phase convention is the unique case without phase correlations.\footnote{
What the canonical phase combination matters in the actual analysis? 
Suppose we do analysis by taking the $U_{\text{\tiny ATM}}$ convention by defining magnitudes and phases of the complex $\alpha$ parameters as $\alpha_{\mu e} = \vert \alpha_{\mu e} \vert e^{ i \phi_{\mu e} }$, and obtain the best fit values of $\alpha_{\mu e}$ and $\phi_{\mu e}$. If one does the same analysis by using $U_{\text{\tiny PDG}}$ or $U_{\text{\tiny SOL}}$ conventions with $\bar{\alpha}$ or $\tilde{\alpha}$ parameters, respectively, with the similar parametrizations, the best-fit (and the allowed region) of the phases in each convention obeys the relation $\phi_{\mu e} = \bar{\phi}_{\mu e} = \tilde{\phi}_{\mu e} + \delta$. The similar discussion follows for the phases of the other $\alpha_{\beta \gamma}$ as well. 
Therefore, one can use any conventions of $U_{\text{\tiny MNS}}$ in doing analysis, but has to keep in mind that the best fit phase is convention dependent. 
}

\subsection{$\nu$SM CP and UV parameter phase correlation: Real or superficial?} 
\label{sec:interpretation}

If there is a $U_{\text{\tiny MNS}}$ phase convention in which the phase correlation vanishes in entire kinematical phase space, one may be able to conclude that the correlation is not physical but superficial. Then, the question is whether the absence of $\delta$ - complex $\alpha$ parameter phase correlation in the $U_{\text{\tiny SOL}}$ phase convention fits to this interpretation. 

To answer the question we have performed a separate study of the same system with non-unitarity in a different kinematical space, the region of solar-scale enhanced oscillations. We have formulated a valid perturbative framework in this region \cite{Martinez-Soler:2019noy}, an extension of the solar resonance perturbation theory \cite{Martinez-Soler:2019nhb} (mentioned in section~\ref{sec:symmetry}) with UV effect included. The result of this calculation shows that there exists a $\delta - \alpha$ parameter phase correlation in the $U_{\text{\tiny SOL}}$ convention. In fact, notion of phase correlation is somewhat enlarged, and it looks more like the correlation between $\delta$ and the certain blobs of $\alpha$ parameters. See ref.~\cite{Martinez-Soler:2019noy} for more details. Therefore, it appears that the correlation between the $\nu$SM CP phase $\delta$ and phases of the complex $\alpha$ parameters we see in the present paper is physical.\footnote{
Therefore, technically, it might be easier to go through all the calculation with the $U_{\text{\tiny SOL}}$ convention, and translate the results into the other $U_{\text{\tiny MNS}}$ convention. We did {\em not} choose this way because the phase correlation is physical and its presence is universal apart from the unique exception, the $U_{\text{\tiny SOL}}$ convention.
}

\section{Accuracy of the probability formula and $P_{ \text{ EV } } - P_{ \text{ UV } }$ comparison } 
\label{sec:accuracy}

The principal objective of constructing our helio-UV perturbation theory is to understand the qualitative features of the oscillation probability with UV.\footnote{
If one seeks numerical accuracy under the uniform matter density approximation, we recommend, as we will do in this section, to use an exact analytic formula of the probability derived in ref.~\cite{Fong:2017gke}, a UV extended version of the three-flavor KTY formula \cite{Kimura:2002wd}. Or, if more accurate perturbative formula could be obtained by improving the way of 1-2 level crossing, it may be worthwhile to follow the line taken in refs.~\cite{Denton:2016wmg,Agarwalla:2013tza}. 
}
The numerical accuracy of the first order formula is in a sense trivial, as far as the normalization is concerned, because the correction terms just scale with $|\alpha_{\beta \gamma}|$: The smaller the expansion parameters, the better the approximation. Yet, it may be useful to have an idea of how good is the approximation it can offer, in particular, to see whether the characteristic structure such as the MSW enhancement~\cite{Mikheev:1986gs,Wolfenstein:1977ue} is correctly reproduced. 

We are interested in the UV parameter related part, 
$P(\nu_{\beta} \rightarrow \nu_{\gamma})^{(1)} \equiv P(\nu_{\beta} \rightarrow \nu_{\gamma})_{ \text{ EV } }^{(1)} +
P(\nu_{\beta} \rightarrow \nu_{\gamma})_{ \text{ UV } }^{(1)}$,
because the accuracies of 
$P(\nu_{\beta} \rightarrow \nu_{\gamma})_{ \text{ helio} }^{(0+1)}$ have been examined in \cite{Minakata:2015gra}. Notice that it corresponds to the quantity
\begin{eqnarray} 
&& 
\Delta P (\nu_{\beta} \rightarrow \nu_{\gamma}) \equiv 
P(\nu_{\beta} \rightarrow \nu_{\gamma})_{ \text{non-unitary} }  
- P(\nu_{\beta} \rightarrow \nu_{\gamma})_{ \text{standard} }, 
\label{Delta-P-mualpha}
\end{eqnarray}
which is numerically computed with high precision and is plotted in the lower panels of figures 1-3 in \cite{Fong:2017gke} for $(\beta, \gamma) = \mu e$, $\mu \tau$, and $\mu \mu$. For convenience of the plot given below, we flipped the sign of $\Delta P (\nu_{\beta} \rightarrow \nu_{\gamma})$ as in \eqref{Delta-P-mualpha}. Then, we can confront our first order formulas of $P(\nu_{\beta} \rightarrow \nu_{\gamma})^{(1)}$ to $\Delta P (\nu_{\beta} \rightarrow \nu_{\gamma})$ given in \cite{Fong:2017gke}.

\begin{figure}[h!]
\begin{center}
\vspace{4mm}
\includegraphics[width=0.6\textwidth]{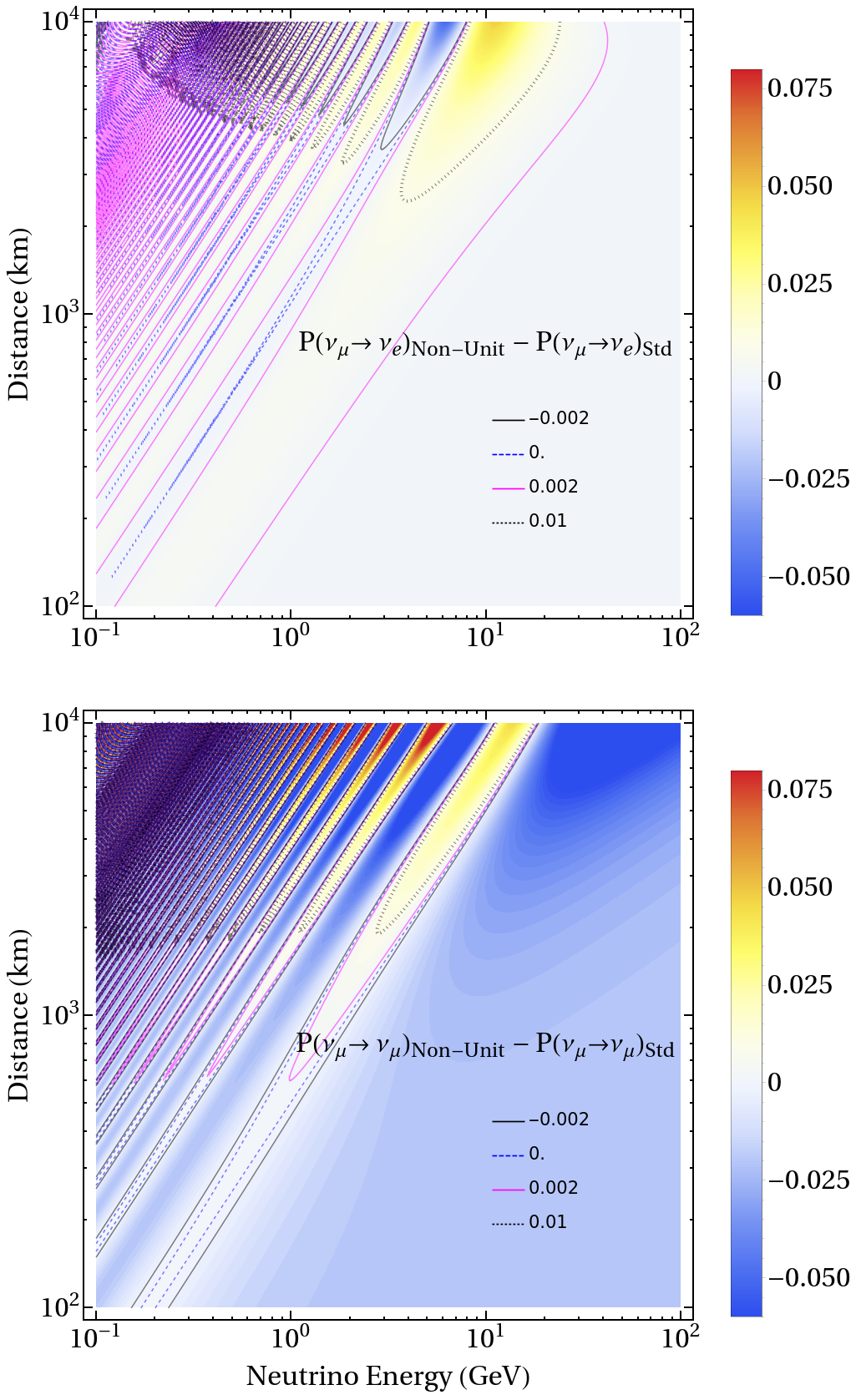}
\end{center}
\vspace{-3mm}
\caption{ The iso-contour of $P(\nu_{\mu} \rightarrow \nu_{\gamma})^{(1)} \equiv P(\nu_{\mu} \rightarrow \nu_{\gamma})_{ \text{ EV } }^{(1)} + P(\nu_{\mu} \rightarrow \nu_{\gamma})_{ \text{ UV } }^{(1)}$ is presented by using the color grading in space of neutrino energy $E$ and baseline $L$ for $\gamma = e$ (upper panel) and $\gamma = \mu$ (lower panel). 
To compare $P(\nu_{\mu} \rightarrow \nu_{\gamma})^{(1)}$ to the corresponding quantity $\Delta P (\nu_{\mu} \rightarrow \nu_{\gamma}) \equiv P(\nu_{\mu} \rightarrow \nu_{\gamma})_{ \text{non-unitary} } - P(\nu_{\mu} \rightarrow \nu_{\gamma})_{ \text{standard} }$ calculated in ref.~\cite{Fong:2017gke}, the latter results are presented as contours of the four colored lines, -0.002, 0, 0.002, 0.01 (taken from the lower panel of figures 1 ($\gamma = e$) and 3 ($\gamma = \mu$) of ref.~\cite{Fong:2017gke}).
  In this calculation, the same values for the standard mixing
  parameters as well as the UV $\alpha$ parameters as in
  \cite{Fong:2017gke} are used: $\alpha_{ee} = 0.01$, $\alpha_{\mu e}
  = 0.0141$, $\alpha_{\mu \mu} = 0.005$, $\alpha_{\tau e} = 0.0445$,
  $\alpha_{\tau \mu} = 0.0316$, $\alpha_{\tau \tau} = 0.051$. The
  matter density is taken to be $\rho = 3.2~{\rm g\,cm}^{-3}$ over the
  entire baseline.  }
\label{fig:Pmue-mumu} 
\end{figure}

In figure~\ref{fig:Pmue-mumu}, plotted are the iso-contours of $P(\nu_{\mu} \rightarrow \nu_{\gamma})^{(1)}$ as a function of energy $E$ and baseline $L$ for $\gamma = e$ (upper panel) and $\gamma = \mu$ (lower panel). We have used the same values for the $\nu$SM mixing parameters as well as the UV $\alpha$ parameters as in ref.~\cite{Fong:2017gke}. Notice that in figure~\ref{fig:Pmue-mumu}, the four colored lines  labelled as -0.002, 0, 0.002, 0.01 show the contours obtained by the exact numerical computation in \cite{Fong:2017gke}, whereas the values of $P(\nu_{\mu} \rightarrow \nu_{\gamma})^{(1)}$ are presented by the color grading. 
We see overall agreement, not only qualitatively but also quantitatively to a certain level, between the color-graded $P(\nu_{\mu} \rightarrow \nu_{\gamma})^{(1)}$ and the iso-contours of  $\Delta P (\nu_{\mu} \rightarrow \nu_{\gamma})$ taken from ref.~\cite{Fong:2017gke}. There are some disagreement in region of the solar scale enhancement partly due to difference in mesh in the numerical computations, but we do not enter this issue because it is outside of validity of our perturbation theory, as discussed in \cite{Minakata:2015gra}.

\begin{figure}[h!]
\begin{center}
\vspace{4mm}
\includegraphics[width=1.0\textwidth]{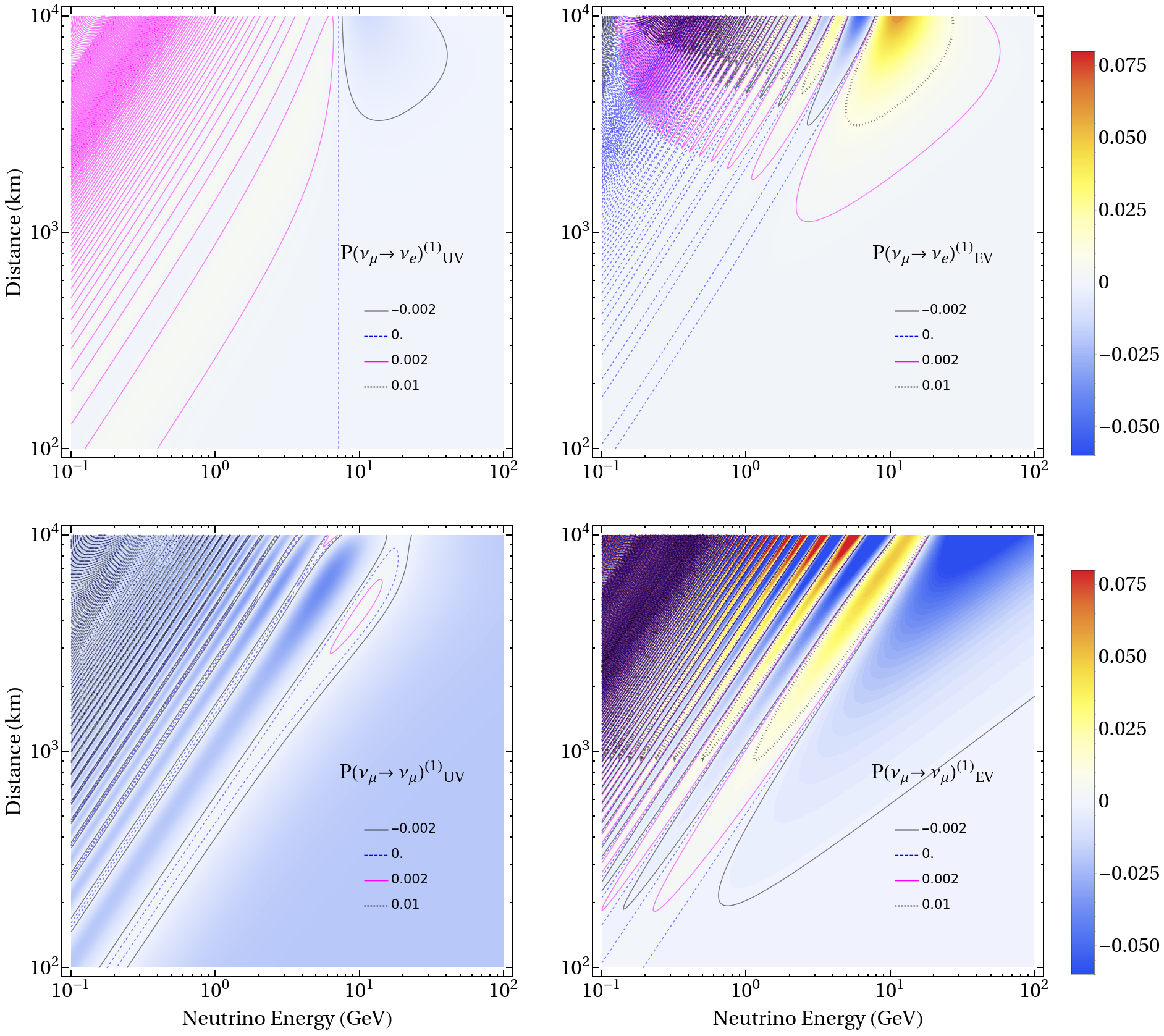}
\end{center}
\vspace{-3mm}
\caption{ 
Plotted is the iso-contours of 
$P(\nu_{\mu} \rightarrow \nu_{e})_{ \text{ UV } }^{(1)}$ (upper left panel), $P(\nu_{\mu} \rightarrow \nu_{e})_{ \text{ EV } }^{(1)}$ (upper right panel), 
$P(\nu_{\mu} \rightarrow \nu_{\mu})_{ \text{ UV } }^{(1)}$ (lower left panel), and $P(\nu_{\mu} \rightarrow \nu_{\mu})_{ \text{ EV } }^{(1)}$ (lower right panel). Unlike in figure~\ref{fig:Pmue-mumu}, the four colored lines labelled as -0.002, 0, 0.002, 0.01 in this figure show the probability contours of the first order term calculated in this paper. They are placed to guide the eye. The values of $\nu$SM and the $\alpha$ parameters taken are the same as in figure~\ref{fig:Pmue-mumu}.
 }
\label{fig:ProbEVUV} 
\end{figure}

In figure \ref{fig:ProbEVUV}, plotted in the upper two panels are the iso-contours of $P(\nu_{\mu} \rightarrow \nu_{e})_{ \text{ UV } }^{(1)}$ (left panel) and $P(\nu_{\mu} \rightarrow \nu_{e})_{ \text{ EV } }^{(1)}$ (right panel), while the lower two panels show 
$P(\nu_{\mu} \rightarrow \nu_{\mu})_{ \text{ UV } }^{(1)}$ (left panel) and $P(\nu_{\mu} \rightarrow \nu_{\mu})_{ \text{ EV } }^{(1)}$ (right panel). They serve for understanding relative contributions from the unitary and non-unitary parts of the first order probability, $P(\nu_{\mu} \rightarrow \nu_{e})^{(1)}$ and $P(\nu_{\mu} \rightarrow \nu_{\mu})^{(1)}$. 

By comparing figure \ref{fig:ProbEVUV} to figure~\ref{fig:Pmue-mumu}, one can see that overall the contribution from unitary evolution labeled as ``EV'' is dominating over the ``UV'' contribution in the both channels, $\nu_{\mu} \rightarrow \nu_{e}$ and $\nu_{\mu} \rightarrow \nu_{\mu}$. Looking into a closer detail, we observe that 
$P(\nu_{\mu} \rightarrow \nu_{\gamma})_{ \text{ EV } }^{(1)}$ has a brighter contrast both in positive and negative region than the sum $P(\nu_{\mu} \rightarrow \nu_{\gamma})^{(1)}$ in figure~\ref{fig:Pmue-mumu}, so that $P(\nu_{\mu} \rightarrow \nu_{\gamma})_{ \text{ UV } }^{(1)}$ partially cancels $P(\nu_{\mu} \rightarrow \nu_{\gamma})_{ \text{ EV } }^{(1)}$, for both $\gamma = \nu_{e}$ and $\nu_{\mu}$. An exception to this statement is the behavior in high energy region beyond the atmospheric-scale enhanced strip in the $\nu_{\mu} \rightarrow \nu_{\mu}$ channel, where a constant term $- 4 \alpha_{\mu \mu}$ in the UV probability $P(\nu_{\mu} \rightarrow \nu_{\mu})_{ \text{ UV } }^{(1)}$ is very visible.

\section{Some additional remarks} 
\label{sec:add-remarks}

\subsection{Vacuum limit} 
\label{sec:vacuum-limit}

The vacuum limit in our helio-UV perturbation theory can be taken in a straightforward manner. With vanishing matter potentials, the Hamiltonian in eq.~(\ref{evolution-check-basis}) in the vacuum mass eigenstate basis reduces to the free Hamiltonian. Then, all the UV-related unitary contributions $P(\nu_{\beta} \rightarrow \nu_{\alpha})^{(1)}_{ \text{EV} }$ vanish, and the neutrino oscillation probability coincides with the vacuum limit of $P(\nu_{\beta} \rightarrow \nu_{\alpha})^{(0+1)}_{ \text{helio} } + P(\nu_{\beta} \rightarrow \nu_{\alpha})^{(1)}_{ \text{UV} }$ to first order in the $\alpha$ parameters. Notice that the vacuum limit of the probabilities implies to take the following limits:
\begin{eqnarray} 
&& \phi \rightarrow \theta_{13}, 
\hspace{10mm}
h_{1} \rightarrow 0, 
\hspace{10mm}
h_{2} \rightarrow \frac{\Delta m^2_{21}}{2E}, 
\hspace{10mm}
h_{3} \rightarrow \frac{\Delta m^2_{31}}{2E}.
\label{vacuum limit}
\end{eqnarray}
See eq.~\eqref{eq:cos-sin-2phi} to understand the first one. Since it is straightforward to take the vacuum limit in the expressions of the helio and the extrinsic UV contributions, $P(\nu_{\beta} \rightarrow \nu_{\alpha})^{(0+1)}_{ \text{helio} } + P(\nu_{\beta} \rightarrow \nu_{\alpha})^{(1)}_{ \text{UV} }$, we do not write the explicit forms of the oscillation probabilities in vacuum.

\subsection{Non-unitarity and Non-standard interactions (NSI)} 
\label{sec:NSI}

A question is often raised: What is the relationship between non-unitarity and NSI. See refs.~\cite{Ohlsson:2012kf,Miranda:2015dra,Farzan:2017xzy} for reviews of NSI. A short answer is that starting from a generic situation which include not only NSI in propagation but also the ones in production and detection (total 27 parameters), our framework could be reproduced by placing appropriate relations between the three kind of NSI parameters. Notice that the latter two introduce non-unitarity \cite{GonzalezGarcia:2001mp}. 
A caution is, therefore, that the bound on the NSI $\epsilon_{\beta \gamma}$ parameters obtained by analysis only with the propagation NSI cannot be directly translated to the $\alpha_{\beta \gamma}$ parameter bound by the mapping rule. It is because the system with non-unitarity always carry the non-unitary UV part of the probability.

As far as the propagation NSI is concerned, the one-to-one mapping from the NSI to the $\alpha$ parameters, or vice versa, is set up, which leads to $\varepsilon_{ee} \leftrightarrow - 2 \alpha_{ee}$ and the similar ones for the other $\alpha$ parameters~\cite{Blennow:2016jkn}. It can be naturally generalized to the case $N_{e} = N_{p} = r N_{n}$ ($r$ is a constant), with a modified mapping $\varepsilon_{ee} \leftrightarrow - 2 \alpha_{ee} (2r - 1)$, where $\Delta_{a} / \Delta_{b} = a/b = 2r$, as we saw in section~\ref{sec:P-mue-mumu}. These mappings are possible only because neutrino evolution is unitary even with non-unitary mixing matrix. Yet, one should also keep in mind that the condition of constant ratio $N_{e} / N_{n} = r$ (= constant) does not hold in the sun and in the entire interior of the earth including the core.

\section{Concluding remarks} 
\label{sec:conclusion}

In this paper, we have formulated a perturbative framework, dubbed as the ``helio-UV perturbation theory'', to derive the analytic formula for the three-flavor neutrino oscillation probability in matter with non-unitary mixing matrix. It utilizes the two expansion parameters $\epsilon \approx \Delta m^2_{21} / \Delta m^2_{31}$ and the $\alpha$ parameters introduced to describe UV effects. To our knowledge, it is currently the unique perturbative analytic formula for the oscillation probability in matter with an explicit $\alpha$ parametrization of non-unitarity. This feature is in contrast to the exact analytic formula of the probability with non-unitary mixing matrix elements $N_{\alpha \beta}$~\cite{Fong:2017gke} derived by using the KTY technique \cite{Kimura:2002wd}. 
 
As an outcome of first-order computation of the oscillation probability, we were able to obtain the following qualitatively new results: 

\begin{itemize}

\item
The phases of the complex UV parameters always come in into the observable in the particular combination with the $\nu$SM CP phase $\delta$. With the PDG ($e^{ \pm i \delta}$ attached to $s_{13}$) convention of $U_{\text{\tiny MNS}}$, it takes the form $[e^{- i \delta } \bar{\alpha}_{\mu e}, ~e^{ - i \delta} \bar{\alpha}_{\tau e}, ~\bar{\alpha}_{\tau \mu}]$.\footnote{
Though we called it as the ``canonical phase combination'', it is $U_{\text{\tiny MNS}}$ convention dependent. In the ATM ($e^{ \pm i \delta}$ attached to $s_{23}$) convention it takes the form $[ e^{- i \delta } \alpha_{\mu e}$, $\alpha_{\tau e}$, and $e^{i \delta } \alpha_{\tau \mu} ]$. In the SOL ($e^{ \pm i \delta}$ attached to $s_{12}$) convention no $e^{ \pm i \delta}$ is attached to the $\alpha$ parameters. See section~\ref{sec:stability-convention-dep}.
}

\item
The diagonal $\alpha$ parameters appear in particular combinations $\left( a/b- 1 \right) \alpha_{ee} + \alpha_{\mu \mu}$ and $\alpha_{\mu \mu} - \alpha_{\tau \tau}$, where $a$ and $b$ denote the matter potential due to CC and NC reactions respectively, in the unitary evolution part of the probability. In contrast, there is no such feature in the unitarity violating part of the probability.

\end{itemize}
\noindent
Both of the above properties hold in all the flavor oscillation channels. But, only the former is true in both unitary and non-unitary pieces of the UV-parameter related part of the probability.
We have given a clarifying discussion on the unitary nature of neutrino evolution in the presence of non-unitary mixing matrix, which is utilized to define unitary and non-unitary parts of the probability. See, in particular, section~\ref{sec:basis-relations}.

The channel independence of the canonical phase combination and its universal character, validity in both the unitary and non-unitary parts of the probability, requires some explanation. We have identified the cause of stability and universal nature of the phase correlation between the $\nu$SM CP and the UV phases as due to the special structure of the $S$ matrices that we call the ``lozenge position $e^{ \pm i \delta }$ matrix''. We have also discussed the $U_{\text{\tiny MNS}}$ convention dependence of the $\alpha$ matrix and the phase correlation in section~\ref{sec:stability-convention-dep}. 
We gave, in section~\ref{sec:phase-correlation}, a heuristic discussion on why the phase correlation is motivated naturally in theories with non-unitarity which is embedded into a larger unitary theory. The phase correlation we discussed above may be understood to lie along the line of thought. 

We also note a practical utility of our formalism gained by knowing the correlations between $\delta$ and the $\alpha$ parameter phases, and among the diagonal $\alpha$ parameters. Simply, analytic demonstration of parameter correlation makes interpretation of the results of data analysis transparent \cite{Abe:2017jit,Miranda:2016wdr}, and makes the analysts prepared for the other types of correlations. It is similar to what happened in the observation of $\sin \theta_{13}$-NSI parameter confusion \cite{Huber:2002bi}, and its analytic understanding in terms of the correlated variable $\sin \theta_{13} - \varepsilon_{e \mu} -\varepsilon_{e \tau}$ in the NSI-extended version~\cite{Kikuchi:2008vq} of the Cervera {\it et al.} formula \cite{Cervera:2000kp}. 
Even though low order calculations may not be so accurate numerically, given the current status of the bound on the $\alpha$ parameters, utility of the first order formula must increase in the precision measurement era in which constraints on UV would reach to $\vert \alpha \vert \lsim 10^{-2} - 10^{-3}$. 

Finally, we emphasize that, to the best of our knowledge, figure \ref{fig:ProbEVUV} in section~\ref{sec:accuracy} is the unique figure available in the literature in that the ``EV'' (unitary evolution) part and the ``UV'' (unitarity violation) part of the oscillation probability are plotted separately. It was only possible by a clear separation of the $S$ matrix into the unitary and non-unitary parts, as done in section~\ref{sec:basis-relations}.

\acknowledgments

We thank Enrique Fernandez-Martinez for illuminating discussions about theories with non-unitarity, and Hiroshi Nunokawa for raising the issue discussed in appendix~\ref{sec:relevant-variables}.

H.M. expresses a deep gratitude to Instituto F\'{\i}sica Te\'{o}rica, UAM/CSIC in Madrid, for its support via ``Theoretical challenges of new high energy, astro and cosmo experimental data'' project, Ref: 201650E082. 
He is grateful to Center for Neutrino Physics, Virginia Tech, for support where the final version of this paper was completed.

I.M.S. acknowledges support from the Spanish grant FPA2015-65929-P
(MINECO/FEDER, UE), the Spanish Research Agency (``Agencia Estatal de
Investigacion'') grants IFT ``Centro de Excelencia Severo Ochoa''
SEV2012-0249 and SEV-2016-0597 and from the Colegio de F\'isica
Fundamental e Interdisciplinaria de las Am\'ericas (COFI).  Fermilab is
operated by the Fermi Research Alliance, LLC under contract
No. DE-AC02-07CH11359 with the United States Department of Energy.
This work has received funding/support from the European Union's
Horizon 2020 research and innovation programme under the Marie
Sklodowska-Curie grant agreement No 674896 and No 690575.

\appendix 

\section{Expression of $H$ and $\Phi$ matrix elements } 
\label{sec:Hij-Phi-ij}

The expressions of the elements $H_{ij}$ defined in eq.~\eqref{Hij-def} are given in the form in which the canonical phase combination is respected, and the additional $e^{\pm i \delta }$ dependence is isolated.
\begin{eqnarray} 
&& H_{11} =
2 \alpha_{ee} \left( 1 - \frac{ \Delta_{a} }{ \Delta_{b} } \right), 
\nonumber \\
&& H_{12} =
e^{- i \delta } \left\{ c_{23} \left( e^{- i \delta } \alpha_{\mu e} \right)^* - s_{23}  \alpha_{\tau e}^* \right\}, 
\nonumber \\
&& H_{13} = 
s_{23} \left( e^{- i \delta } \alpha_{\mu e} \right)^* + c_{23} \alpha_{\tau e}^*, 
\nonumber \\
&& H_{21} =
e^{i \delta } \left( c_{23} e^{- i \delta } \alpha_{\mu e} - s_{23}  \alpha_{\tau e} \right), 
\nonumber \\
&& H_{22} =
2 \left[ 
c_{23}^2 \alpha_{\mu \mu} + s_{23}^2 \alpha_{\tau \tau} - c_{23} s_{23} \mbox{Re} \left( e^{i \delta } \alpha_{\tau \mu} \right) 
\right], 
\nonumber \\
&& H_{23} =
e^{i \delta } \left[ 2 c_{23} s_{23} ( \alpha_{\mu \mu} - \alpha_{\tau \tau} ) + c_{23}^2 \left( e^{ i \delta } \alpha_{\tau \mu} \right)^* - s_{23}^2 e^{ i \delta } \alpha_{\tau \mu} \right], 
\nonumber \\
&& H_{31} =
s_{23} e^{- i \delta } \alpha_{\mu e} + c_{23} \alpha_{\tau e}, 
\nonumber \\
&& H_{32} =
e^{- i \delta } \left[ 2 c_{23} s_{23} ( \alpha_{\mu \mu} - \alpha_{\tau \tau} ) + c_{23}^2 e^{ i \delta } \alpha_{\tau \mu} - s_{23}^2 \left( e^{ i \delta } \alpha_{\tau \mu} \right)^* \right], 
\nonumber \\
&& H_{33} =
2 \left[
s_{23}^2 \alpha_{\mu \mu} + c_{23}^2 \alpha_{\tau \tau} + c_{23} s_{23} \mbox{Re} \left( e^{i \delta } \alpha_{\tau \mu} \right) 
\right]. 
\label{Hij-elements}
\end{eqnarray}

The expressions of $\Phi_{ij}$ defined by eq.~\eqref{H1-matter} in section~\ref{sec:hatS-tildeS-matt} is given by 
\begin{eqnarray} 
\Phi_{11} &=& 
\left\{ 
H_{11} 
-2 c_{\phi}^2 s_{\phi}^2 
\left( H_{11} - H_{33} \right) 
- c_{\phi} s_{\phi} \cos 2 \phi 
\left( H_{13} + H_{31} \right) 
\right\}  
\nonumber \\
&+& 
e^{i ( h_{3} - h_{1} ) x } 
\left\{    
c_{\phi}^2 s_{\phi}^2 \left( H_{11} - H_{33} \right) + 
c_{\phi} s_{\phi} \left( - s_{\phi}^2 H_{13} + c_{\phi}^2 H_{31} \right) 
\right\} 
\nonumber \\
&+&
e^{ - i ( h_{3} - h_{1} ) x } 
\left\{  
c_{\phi}^2 s_{\phi}^2 \left( H_{11} - H_{33} \right) + 
c_{\phi} s_{\phi} \left( c_{\phi}^2 H_{13} - s_{\phi}^2 H_{31} \right) 
\right\},  
\nonumber \\ 
\Phi_{12} &=& 
e^{ - i ( h_{2} - h_{1} ) x } 
\left( c_{\phi}^2 H_{12} - c_{\phi} s_{\phi} H_{32} \right) 
+ e^{ i ( h_{3} - h_{2} ) x } 
\left( s_{\phi}^2 H_{12} + c_{\phi} s_{\phi} H_{32} \right),  
\nonumber \\ 
\Phi_{13} &=& 
\left\{ - c_{\phi} s_{\phi} \cos 2 \phi 
\left( H_{11} - H_{33} \right) +
2 c_{\phi}^2 s_{\phi}^2 
\left( H_{31} + H_{13} \right) 
\right\} 
\nonumber \\
&+& 
e^{ i ( h_{3} - h_{1} ) x } 
\left\{ - c_{\phi} s_{\phi}^3 \left( H_{11} - H_{33} \right) 
+ s_{\phi}^4 H_{13} 
- c_{\phi}^2 s_{\phi}^2 H_{31} \right\} 
\nonumber \\
&+& 
e^{ - i ( h_{3} - h_{1} ) x } 
\left\{ 
c_{\phi}^3 s_{\phi} \left( H_{11} - H_{33} \right) 
+ c_{\phi}^4 H_{13} 
- c_{\phi}^2 s_{\phi}^2 H_{31} 
\right\},  
\nonumber \\
\Phi_{21} &=& 
e^{ i ( h_{2} - h_{1} ) x } 
\left( c_{\phi}^2 H_{21} - c_{\phi} s_{\phi} H_{23} \right) 
+ e^{ - i ( h_{3} - h_{2} ) x } \left( s_{\phi}^2 H_{21} + c_{\phi} s_{\phi} H_{23} \right), 
\nonumber \\
\Phi_{22} &=& H_{22}, 
\nonumber \\
\Phi_{23} &=& 
e^{ i ( h_{2} - h_{1} ) x } 
\left\{ - c_{\phi} s_{\phi} H_{21} + s_{\phi}^2 H_{23} \right\} 
+ e^{ - i ( h_{3} - h_{2} ) x } 
\left\{ c_{\phi} s_{\phi} H_{21} + c_{\phi}^2 H_{23} \right\},  
\nonumber \\
\Phi_{31} &=& 
\left\{ 
- c_{\phi} s_{\phi} \cos 2 \phi 
\left( H_{11} - H_{33} \right) + 
2 c_{\phi}^2 s_{\phi}^2 
\left( H_{13} + H_{31} \right) 
\right\} 
\nonumber \\
&+& 
e^{ i ( h_{3} - h_{1} ) x } 
\left\{ c_{\phi}^3 s_{\phi} \left( H_{11} - H_{33} \right) - c_{\phi}^2 s_{\phi}^2 H_{13} + c_{\phi}^4 H_{31} \right\} 
\nonumber \\
&+& 
e^{ - i ( h_{3} - h_{1} ) x } 
\left\{
- c_{\phi} s_{\phi}^3 \left( H_{11} - H_{33} \right) 
- c_{\phi}^2 s_{\phi}^2 H_{13} + s_{\phi}^4 H_{31} 
\right\}, 
\nonumber \\ 
\Phi_{32} &=& 
e^{ - i ( h_{2} - h_{1} ) x } 
\left\{ - c_{\phi} s_{\phi} H_{12} + s_{\phi}^2 H_{32} \right\} 
+ 
e^{ i ( h_{3} - h_{2} ) x } 
\left\{ c_{\phi} s_{\phi} H_{12} + c_{\phi}^2 H_{32} \right\},  
\nonumber \\
\Phi_{33} 
&=& 
\left\{ H_{33} + 2 c_{\phi}^2 s_{\phi}^2 
\left( H_{11} - H_{33} \right) 
+ c_{\phi} s_{\phi} \cos 2 \phi 
\left( H_{13} + H_{31} \right) 
\right\}
\nonumber \\
&+& 
e^{i ( h_{3} - h_{1} ) x } 
\left\{ 
- c_{\phi}^2 s_{\phi}^2 \left( H_{11} - H_{33} \right) 
+ c_{\phi} s_{\phi}^3 H_{13} 
- c_{\phi}^3 s_{\phi} H_{31} 
\right\} 
\nonumber \\
&+& 
e^{ - i ( h_{3} - h_{1} ) x } 
\left\{ 
- c_{\phi}^2 s_{\phi}^2 \left( H_{11} - H_{33} \right) 
- c_{\phi}^3 s_{\phi} H_{13} 
+ c_{\phi} s_{\phi}^3 H_{31} 
\right\}. 
\label{Phi-element2} 
\end{eqnarray}

\section{Expressions of $\tilde{S}_{ \text{EV} }^{(1)}$ matrix elements } 
\label{sec:tilde-S-UV-1st}

The expressions of $\tilde{S}_{ \text{EV} }^{(1)}$ matrix elements computed with $\tilde{S}_{ \text{EV} }^{(1)} = U_{\phi} \hat{S}_{ \text{EV} }^{(1)} U_{\phi}^{\dagger}$ where $\hat{S}_{ \text{EV} }^{(1)}$ is defined by eq.~\eqref{hat-Smatrix-1st} in section~\ref{sec:hatS-tildeS-matt} is given by
\begin{eqnarray} 
&& \left( \tilde{S}_{ \text{EV} }^{(1)} \right)_{11} =  
(-i) ( \Delta_{b} x ) 
\biggl[ 
\left( s_{\phi}^2 e^{ - i h_{3} x } + c_{\phi}^2 e^{ - i h_{1} x } \right) H_{11} 
- c_{\phi}^2 s_{\phi}^2 \left( e^{ - i h_{3} x } + e^{ - i h_{1} x } \right) \left( H_{11} - H_{33} \right) 
\nonumber \\
&+& 
c_{\phi} s_{\phi} 
\left( s_{\phi}^2 e^{ - i h_{3} x } - c_{\phi}^2 e^{ - i h_{1} x } \right) \left( H_{13} + H_{31} \right) 
\biggr] 
\nonumber \\
&+& \frac{ \Delta_{b} }{ h_{3} - h_{1} }
\left( e^{ - i h_{3} x } - e^{ - i h_{1} x } \right) 
\biggl[
2 c_{\phi}^2 s_{\phi}^2 \left( H_{11} - H_{33} \right) + c_{\phi} s_{\phi} \cos 2 \phi \left( H_{13} + H_{31} \right) 
\biggr]. 
\label{Xi-11-result}
\end{eqnarray}
\begin{eqnarray} 
&& \left( \tilde{S}_{ \text{EV} }^{(1)} \right)_{21} = 
\Delta_{b} \left[
\frac{ e^{ - i h_{2} x } - e^{ -i h_{1} x } }{ ( h_{2} - h_{1} ) } 
\left( c_{\phi}^2 H_{21} - c_{\phi} s_{\phi} H_{23} \right) 
+ 
\frac{ e^{ -i h_{3} x } - e^{ - i h_{2} x }  }{ ( h_{3} - h_{2} ) } 
\left( s_{\phi}^2 H_{21} + c_{\phi} s_{\phi} H_{23} \right) 
\right].
\nonumber \\
\label{Xi-21-result}
\end{eqnarray}
\begin{eqnarray} 
&& \left( \tilde{S}_{ \text{EV} }^{(1)} \right)_{31} = 
(-i) ( \Delta_{b} x )
\biggl[ 
c_{\phi} s_{\phi} 
\left( e^{ - i h_{3} x } - e^{ - i h_{1} x } \right) H_{11} 
- c_{\phi} s_{\phi} 
\left( c_{\phi}^2 e^{ - i h_{3} x } - s_{\phi}^2 e^{ - i h_{1} x } \right) 
\left( H_{11} - H_{33} \right) 
\nonumber \\
&+& 
c_{\phi}^2 s_{\phi}^2 
\left( e^{ - i h_{3} x } + e^{ - i h_{1} x } \right)
\left( H_{13} + H_{31} \right) 
\biggr] 
\nonumber \\
&+& \frac{ \Delta_{b} }{ h_{3} - h_{1} }
\left( e^{ - i h_{3} x } - e^{ - i h_{1} x } \right) 
\biggl[
c_{\phi} s_{\phi} \cos 2 \phi \left( H_{11} - H_{33} \right) + H_{31} 
- 2 c_{\phi}^2 s_{\phi}^2 \left( H_{13} + H_{31} \right)  
\biggr]. 
\label{Xi-31-result}
\end{eqnarray} 
\begin{eqnarray} 
&& \left( \tilde{S}_{ \text{EV} }^{(1)} \right)_{12} = 
\Delta_{b} \left[
\frac{ e^{ - i h_{2} x } - e^{ -i h_{1} x } }{ ( h_{2} - h_{1} ) } 
\left( c_{\phi}^2 H_{12} - c_{\phi} s_{\phi} H_{32} \right) 
+ 
\frac{ e^{ -i h_{3} x } - e^{ - i h_{2} x }  }{ ( h_{3} - h_{2} ) } 
\left( s_{\phi}^2 H_{12} + c_{\phi} s_{\phi} H_{32} \right) 
\right]. 
\nonumber \\
\label{Xi-12-result}
\end{eqnarray}
\begin{eqnarray} 
&& \left( \tilde{S}_{ \text{EV} }^{(1)} \right)_{22} = 
(-i) ( \Delta_{b} x ) e^{ - i h_{2} x } H_{22}.  
\label{Xi-22-result}
\end{eqnarray}
\begin{eqnarray} 
&& \left( \tilde{S}_{ \text{EV} }^{(1)} \right)_{32} = 
\Delta_{b} \left[ 
- \frac{ e^{ - i h_{2} x } - e^{ -i h_{1} x } }{ ( h_{2} - h_{1} ) } 
\left( c_{\phi} s_{\phi} H_{12} - s_{\phi}^2 H_{32} \right) 
+ 
\frac{ e^{ -i h_{3} x } - e^{ - i h_{2} x }  }{ ( h_{3} - h_{2} ) } 
\left( c_{\phi} s_{\phi} H_{12} + c_{\phi}^2 H_{32} \right) 
\right]. 
\nonumber \\
\label{Xi-32-result}
\end{eqnarray}
\begin{eqnarray} 
&& \left( \tilde{S}_{ \text{EV} }^{(1)} \right)_{13} = 
(-i) ( \Delta_{b} x ) 
\biggl[ 
c_{\phi} s_{\phi} 
\left( e^{ - i h_{3} x } - e^{ - i h_{1} x } \right) H_{33} 
+ c_{\phi} s_{\phi} 
\left( s_{\phi}^2 e^{ - i h_{3} x } - c_{\phi}^2 e^{ - i h_{1} x } \right) 
\left( H_{11} - H_{33} \right) 
\nonumber \\
&+& 
c_{\phi}^2 s_{\phi}^2 
\left( e^{ - i h_{3} x } + e^{ - i h_{1} x } \right)
\left( H_{13} + H_{31} \right) 
\biggr] 
\nonumber \\
&+& \frac{ \Delta_{b} }{ h_{3} - h_{1} }
\left( e^{ - i h_{3} x } - e^{ - i h_{1} x } \right) 
\biggl[
c_{\phi} s_{\phi} \cos 2 \phi \left( H_{11} - H_{33} \right) 
+ H_{13} - 2 c_{\phi}^2 s_{\phi}^2 \left( H_{13} + H_{31} \right) 
\biggr]. 
\label{Xi-13-result}
\end{eqnarray}
\begin{eqnarray} 
&& \left( \tilde{S}_{ \text{EV} }^{(1)} \right)_{23} = 
\Delta_{b} \left[ 
- \frac{ e^{ - i h_{2} x } - e^{ -i h_{1} x } }{ ( h_{2} - h_{1} ) } 
\left( c_{\phi} s_{\phi} H_{21} - s_{\phi}^2 H_{23} \right) 
+ 
\frac{ e^{ -i h_{3} x } - e^{ - i h_{2} x }  }{ ( h_{3} - h_{2} ) } 
\left( c_{\phi} s_{\phi} H_{21} + c_{\phi}^2 H_{23} \right) 
\right]. 
\nonumber \\
\label{Xi-23-result}
\end{eqnarray}
\begin{eqnarray} 
&& \left( \tilde{S}_{ \text{EV} }^{(1)} \right)_{33} = 
(-i) ( \Delta_{b} x )  
\biggl[ 
\left( c_{\phi}^2 e^{ - i h_{1} x } + s_{\phi}^2 e^{ - i h_{3} x } \right) H_{33} 
+ c_{\phi}^2 s_{\phi}^2 \left( e^{ - i h_{3} x } + e^{ - i h_{1} x } \right) \left( H_{11} - H_{33} \right) 
\nonumber \\
&+& 
c_{\phi} s_{\phi} 
\left( c_{\phi}^2 e^{ - i h_{3} x } - s_{\phi}^2 e^{ - i h_{1} x } \right) \left( H_{13} + H_{31} \right) 
\biggr] 
\nonumber \\
&-& \frac{ \Delta_{b} }{ h_{3} - h_{1} }
\left( e^{ - i h_{3} x } - e^{ - i h_{1} x } \right) 
\biggl[
2 c_{\phi}^2 s_{\phi}^2 \left( H_{11} - H_{33} \right) + c_{\phi} s_{\phi} \cos 2 \phi \left( H_{13} + H_{31} \right) 
\biggr]. 
\label{Xi-33-result}
\end{eqnarray}

\section{The oscillation probabilities in $\nu_{e} - \nu_{\mu}$ sector} 
\label{sec:e-mu-sector}

In this section we present $\nu$SM parts of $P(\nu_{e} \rightarrow \nu_{e})$ and $P(\nu_{\mu} \rightarrow \nu_{e})$, and then move to the first order expression of the former (for the latter, see eqs.~\eqref{P-mue-intrinsic-UV} and \eqref{P-mue-extrinsic-UV}). 

\subsection{$P(\nu_{e} \rightarrow \nu_{e})_{\text{ helio} }^{(0+1)}$ and $P(\nu_{\mu} \rightarrow \nu_{e})_{\text{ helio} }^{(0+1)}$: the ``simple and compact'' formulas}
\label{sec:MP-formulas-mue}

Since all the calculations for 
$P(\nu_{e} \rightarrow \nu_{e})_{\text{ helio} }^{(0+1)}$ and 
$P(\nu_{\mu} \rightarrow \nu_{e})_{\text{ helio} }^{(0+1)}$ are done in \cite{Minakata:2015gra} and described in detail in this reference we just present here the result:\footnote{
For more transparent expressions of the $\nu$SM part of the oscillation probability to make the symmetry discussed in section~\ref{sec:symmetry} more explicit, the readers are advised to visit appendix~B in ref.~\cite{Minakata:2015gra}. } 
\begin{eqnarray}
P(\nu_{e} \rightarrow \nu_{e})_{ \text{ helio} }^{(0+1)} &=&
1 -  \sin^2 2\phi  ~\sin^2 \frac{ ( h_{3} - h_{1} ) x }{2}, 
\label{Pee-helio-0th1st}
\end{eqnarray}
\begin{eqnarray} 
&& 
P(\nu_{\mu} \rightarrow \nu_{e})_{\text{ helio} }^{(0+1)}
\nonumber \\
&=& 
\left[ 
s^2_{23} \sin^2 2 \theta_{13}
+ 
4 \epsilon   
J_r \cos \delta 
\left\{ \frac{ ( h_{3} - h_{1} ) - ( \Delta_{ \text{ren} } - \Delta_{a} ) }{ ( h_{3} - h_{2} ) } \right\}
\right]
 \left(\frac{ \Delta_{ \text{ren} } }{ h_{3} - h_{1} } \right)^2 \sin^2 \frac{ ( h_{3} - h_{1} ) x }{ 2 } 
\nonumber \\[2mm]
&+& 8 \epsilon  
J_r 
\frac{ ( \Delta_{ \text{ren} } )^3 }{ ( h_{3} - h_{1} ) ( h_{3} - h_{2} ) ( h_{2} - h_{1} ) }
\sin \frac{ ( h_{3} - h_{1} ) x }{ 2 } 
\sin \frac{ ( h_{2} - h_{1} ) x }{ 2 } 
\cos \left( \delta + \frac{ ( h_{3} - h_{2} ) x }{ 2 } \right),
\nonumber \\
\label{P-mue-helio-0+1}
\end{eqnarray}
where $x$ is the baseline and $J_r$, the reduced Jarlskog factor \cite{Jarlskog:1985ht}, is defined as
\begin{eqnarray} 
J_r \equiv c_{12} s_{12} c_{23} s_{23} c^2_{13} s_{13}.
\label{Jarlskog-def}
\end{eqnarray}
For simplified notations such as $\Delta_{ \text{ren} } \equiv \frac{
  \Delta m^2_{ \text{ren} } }{ 2E }$, and $\Delta_{a} \equiv \frac{ a
}{ 2E }$, see
sections~\ref{sec:mass-basis}~and~\ref{sec:tilde-basis}. $h_{i}$
($i=1,2,3$) denote the eigenvalues of $\tilde{H}^{(0)}$ as defined in
eq.~\eqref{hat-hamiltonian1}.

\subsection{First order $\nu_{e} \rightarrow \nu_{e}$ disppearance probability $P(\nu_{e} \rightarrow \nu_{e})^{(1)}$} 
\label{sec:P-int-ext-UV-1st-ee-mue}

The $\alpha$ parameter dependent unitary and non-unitary contributions, $P(\nu_{e}  \rightarrow \nu_{e})_{ \text{EV } }^{(1)}$ and $P(\nu_{e}  \rightarrow \nu_{e})_{ \text{UV } }^{(1)}$, are given respectively by 
\begin{eqnarray} 
&& P(\nu_{e}  \rightarrow \nu_{e})_{ \text{EV } }^{(1)} 
\nonumber \\
&=& 
- \sin^2 2 \phi 
\biggl\{ 
\cos 2 \phi 
\left[ 
\left\{ \left( \frac{ \Delta_{a} }{ \Delta_{b} } - 1 \right) \alpha_{ee} + \alpha_{\mu \mu} \right\} 
- c_{23}^2 ( \alpha_{\mu \mu} - \alpha_{\tau \tau} ) 
+ c_{23} s_{23} \mbox{Re} \left( e^{i \delta } \alpha_{\tau \mu} \right)
\right] 
\nonumber \\
&+& 
\sin 2 \phi 
\left[
s_{23} \mbox{Re} \left( e^{- i \delta } \alpha_{\mu e} \right) 
+ c_{23} \mbox{Re} \left( \alpha_{\tau e} \right) 
\right] 
\biggr\} 
( \Delta_{b} x ) \sin ( h_{3} - h_{1} ) x 
\nonumber \\
&+& 
4 \cos 2\phi \sin 2 \phi 
\biggl\{
\sin 2 \phi 
\left[ 
\left\{ \left( \frac{ \Delta_{a} }{ \Delta_{b} } - 1 \right) \alpha_{ee} + \alpha_{\mu \mu} \right\} 
- c_{23}^2 ( \alpha_{\mu \mu} - \alpha_{\tau \tau} ) 
+ c_{23} s_{23} \mbox{Re} \left( e^{i \delta } \alpha_{\tau \mu} \right)
\right] 
\nonumber \\
&-&
\cos 2 \phi 
\left[
s_{23} \mbox{Re} \left( e^{- i \delta } \alpha_{\mu e} \right) 
+ c_{23} \mbox{Re} \left( \alpha_{\tau e} \right) 
\right]
\biggr\} 
\frac{ \Delta_{b} }{ h_{3} - h_{1} } 
\sin^2 \frac{ ( h_{3} - h_{1} ) x }{2}, 
\label{Pee-intrinsic-UV-1st}
\end{eqnarray}
\begin{eqnarray}
P(\nu_{e} \rightarrow \nu_{e})^{(1)}_{ \text{UV} } &=& 
- 4 \alpha_{ee} 
\left[
1 - \sin^2 2 \phi 
\sin^2 \frac{ ( h_{3} - h_{1} ) x }{2} 
\right]. 
\label{Pee-extrinsic-UV-1st}
\end{eqnarray}
How to obtain the oscillation probabilities in the channels $\nu_{e} \rightarrow \nu_{\mu}$ and $\nu_{e} \rightarrow \nu_{\tau}$ are discussed in section~\ref{sec:unitarity-nue}.

\section{The oscillation probabilities in $\nu_{\mu} - \nu_{\tau}$ sector} 
\label{sec:mu-tau-sector} 

In this section, starting from $\nu$SM parts of the oscillation probabilities $P(\nu_{\mu} \rightarrow \nu_{\mu})$ and $P(\nu_{\mu} \rightarrow \nu_{\tau})$ \cite{Minakata:2015gra} to be self-contained, we present the first order expression of $\nu_{\mu} \rightarrow \nu_{\tau}$, the EV and UV parts.\footnote{
Similarly, if necessary, one can compute $P(\nu_{\tau} \rightarrow
\nu_{\tau})_{ \text{ EV } }^{(1)}$ and $P(\nu_{\tau} \rightarrow
\nu_{\tau})_{ \text{ UV } }^{(1)}$ in the same way. The former can
be used to verify unitarity in $\nu_{\tau}$ row with the other
probabilities $P(\nu_{\tau} \rightarrow \nu_{e})_{ \text{ EV }
}^{(1)}$ and $P(\nu_{\tau} \rightarrow \nu_{\mu})_{ \text{ EV }
}^{(1)}$. 
The latter can be obtained by generalized T transformation from \eqref{P-mutau-intrinsic-UV}, and the former by the 2-3 rotation \eqref{23-rotation} from $P(\nu_{\mu} \rightarrow \nu_{e})_{ \text{ EV } }^{(1)}$ in eq.~\eqref{P-mue-intrinsic-UV}. }

\subsection{$P(\nu_{\mu} \rightarrow \nu_{\mu})_{\text{ helio} }^{(0+1)}$ and 
$P(\nu_{\mu} \rightarrow \nu_{\tau})_{\text{ helio} }^{(0+1)}$: the ``simple and compact'' formula } 
\label{sec:MP-formulas-mutau}

We discuss $\nu_{\mu} \rightarrow \nu_{\mu}$ and $\nu_{\mu} \rightarrow \nu_{\tau}$ channels in parallel, and present $P(\nu_{\mu} \rightarrow \nu_{\mu})_{\text{ helio} }^{(0+1)}$ and $P(\nu_{\mu} \rightarrow \nu_{\tau})_{\text{ helio} }^{(0+1)}$ in forms which may be convenient to verify unitarity:\footnote{
The formulas written here may be more reader friendly compared to the
ones in ref.~\cite{Minakata:2015gra} which are presented in a
condensed and abstract fashion.  }
\begin{eqnarray}
&& P(\nu_{\mu} \rightarrow \nu_{\mu})_{\text{ helio} }^{(0+1)}
\nonumber \\ 
&=& 
1 - 
\left[
s^4_{23} \sin^2 2\phi 
+ 8 \epsilon J_r \cos \delta ~s^2_{23} 
~\frac{ (\Delta_{ \text{ren} })^2 
\left\{ ( h_{3} - h_{1} ) - ( \Delta_{ \text{ren} } - \Delta_{a} ) 
\right\} }{ ( h_{3} - h_{1} )^2 ( h_{3} - h_{2} ) } 
\right]
\sin^2 \frac{ ( h_{3} - h_{1} ) x}{ 2 } 
\nonumber \\ 
&-& 
\left[
\sin^2 2\theta_{23} c^2_{\phi} 
- 4 \epsilon \left(J_r \cos \delta/c^2_{13} \right)  ~\cos 2\theta_{23}  
~\frac{ \Delta_{ \text{ren} } \left\{ ( h_{3} - h_{1} ) - ( \Delta_{ \text{ren} } + \Delta_{ a } ) \right\} }{ ( h_{3} - h_{1} ) ( h_{3} - h_{2} ) } 
\right]
\sin^2 \frac{ ( h_{3} - h_{2} ) x}{ 2 } 
\nonumber \\ 
&-& 
\left[
\sin^2 2\theta_{23} s^2_{\phi} 
- 4 \epsilon \left(J_r \cos \delta /c^2_{13} \right)  ~  \cos 2\theta_{23}  
~ \frac{ \Delta_{ \text{ren} } \left\{ ( h_{3} - h_{1} ) + ( \Delta_{ \text{ren} } + \Delta_{ a } ) \right\} }{ ( h_{3} - h_{1} ) ( h_{1} - h_{2} ) } 
\right]
\sin^2 \frac{ ( h_{1} - h_{2} ) x}{ 2 } 
\nonumber\\ 
&-& 
16 \epsilon J_r \cos \delta ~s^2_{23} 
~ \frac{ (\Delta_{ \text{ren} })^3 }{ ( h_{3} - h_{1} ) ( h_{3} - h_{2} ) ( h_{1} - h_{2} ) }
\sin \frac{ ( h_{3} - h_{1} ) x}{ 2 } 
\sin \frac{ ( h_{1} - h_{2} ) x}{ 2 } 
\cos \frac{ ( h_{3} - h_{2} ) x}{ 2 }. 
\nonumber \\
\label{P-mumu-helio}
\end{eqnarray}
\begin{eqnarray}
&& P(\nu_{\mu} \rightarrow \nu_{\tau})_{\text{ helio} }^{(0+1)}
\nonumber \\ 
&=& 
- \left[ 
c^2_{23} s^2_{23} \sin^2 2\phi  
+ 4 \epsilon J_r \cos \delta 
\cos 2\theta_{23}
\frac{ (\Delta_{ \text{ren} })^2   
\left\{ ( h_{3} - h_{1} ) - ( \Delta_{ \text{ren} } - \Delta_{ a } ) 
\right\} }{ ( h_{3} - h_{1} )^2 ( h_{3} - h_{2} ) } 
\right]
\sin^2 \frac{ ( h_{3} - h_{1} ) x}{ 2 } 
\nonumber \\ 
&+& 
\left[
\sin^2 2\theta_{23} c_\phi^2 
- 4 \epsilon \left(J_r \cos \delta/c^2_{13} \right) ~\cos 2\theta_{23} 
~\frac{ \Delta_{ \text{ren} } \left\{ ( h_{3} - h_{1} ) - ( \Delta_{ \text{ren} } + \Delta_{ a } ) \right\} }{ ( h_{3} - h_{1} ) ( h_{3} - h_{2} ) } 
\right]
\sin^2 \frac{ ( h_{3} - h_{2} ) x}{ 2 } 
\nonumber \\ 
&+& 
\left[
\sin^2 2\theta_{23} s_\phi^2 
- 4 \epsilon \left(J_r \cos \delta/c^2_{13} \right) ~\cos 2\theta_{23} 
~\frac{ \Delta_{ \text{ren} } \left\{ ( h_{3} - h_{1} ) + ( \Delta_{ \text{ren} } + \Delta_{ a } ) \right\} }{ ( h_{3} - h_{1} ) ( h_{1} - h_{2} ) } 
\right]
\sin^2 \frac{ ( h_{1} - h_{2} ) x}{ 2 } 
\nonumber\\ 
&-& 
8 \epsilon J_r  
\frac{ (\Delta_{ \text{ren} })^3 }{ ( h_{3} - h_{1} ) ( h_{3} - h_{2} ) ( h_{1} - h_{2} ) }
\sin \frac{ ( h_{3} - h_{1} ) x}{ 2 } 
\sin \frac{ ( h_{1} - h_{2} ) x}{ 2 } 
\nonumber \\ 
& & \quad \quad \quad \quad \quad \quad \quad \quad \quad \quad \quad \times
\left[
 \cos 2\theta_{23}  ~\cos \delta \cos \frac{ ( h_{3} - h_{2} ) x}{ 2 } 
- \sin \delta \sin \frac{ ( h_{3} - h_{2} ) x}{ 2 } 
\right]. 
\label{P-mutau-helio}
\end{eqnarray}

\subsection{The oscillation probability $P(\nu_{\mu} \rightarrow \nu_{\tau})_{ \text{ EV } }^{(1)}$ and $P(\nu_{\mu} \rightarrow \nu_{\tau})_{ \text{ UV } }^{(1)}$}
\label{sec:P-mu-tau-1st}

The unitary part of the first-order appearance oscillation probability $P(\nu_{\mu} \rightarrow \nu_{\tau})$ reads 
\begin{eqnarray} 
&& 
P(\nu_{\mu} \rightarrow \nu_{\tau})^{(1)}_{ \text{EV} } 
\nonumber \\
&=&
\sin^2 2\theta_{23}
\left[
\cos 2 \theta_{23} \left( \alpha_{\tau \tau} - \alpha_{\mu \mu} \right) 
+ \sin 2 \theta_{23} \mbox{Re} \left( e^{i \delta } \alpha_{\tau \mu} \right) 
\right]
( \Delta_{b} x )
\biggl\{
c_{\phi}^2 \sin ( h_{3} - h_{2} ) x 
- s_{\phi}^2 \sin ( h_{2} - h_{1} ) x 
\biggr\}
\nonumber \\
&-& 
\sin^2 2\theta_{23}
c_{\phi}^2 s_{\phi}^2 
\left[ 
\left\{ \left( \frac{ \Delta_{a} }{ \Delta_{b} } - 1 \right) \alpha_{ee} + \alpha_{\mu \mu} \right\} 
- c_{23}^2 ( \alpha_{\mu \mu} - \alpha_{\tau \tau} ) 
+ c_{23} s_{23} \mbox{Re} \left( e^{i \delta } \alpha_{\tau \mu} \right)
\right] 
\nonumber \\
&\times& 
( \Delta_{b} x )
\biggl\{
\sin ( h_{3} - h_{2} ) x 
- \sin ( h_{2} - h_{1} ) x 
+ \cos 2\phi \sin ( h_{3} - h_{1} ) x
\biggr\}
\nonumber \\
&+& 
\sin^2 2\theta_{23}
c_{\phi} s_{\phi} 
\left[
s_{23} \mbox{Re} \left( e^{- i \delta } \alpha_{\mu e} \right) 
+ c_{23} \mbox{Re} \left( \alpha_{\tau e} \right) 
\right]
\nonumber \\
&\times& 
( \Delta_{b} x )
\biggl\{ 
c_{\phi}^2 \sin ( h_{3} - h_{2} ) x 
+ s_{\phi}^2 \sin ( h_{2} - h_{1} ) x 
- 2 c_{\phi}^2 s_{\phi}^2 \sin ( h_{3} - h_{1} ) x 
\biggr\} 
\nonumber \\
&-& 
2 \sin 2\theta_{23} 
\cos 2\theta_{23} 
\nonumber \\
&\times&
\biggl\{
c_{\phi} s_{\phi} \left[
c_{23} \mbox{Re} \left( e^{- i \delta } \alpha_{\mu e} \right) - s_{23} \mbox{Re} \left( \alpha_{\tau e} \right) 
\right]
- \cos 2\theta_{23} s_{\phi}^2 \mbox{Re} \left( e^{ i \delta } \alpha_{\tau \mu} \right) 
- \sin 2\theta_{23} s_{\phi}^2 ( \alpha_{\mu \mu} - \alpha_{\tau \tau} ) 
\biggr\}
\nonumber \\
&\times& 
\frac{ \Delta_{b} }{ h_{2} -  h_{1} } 
\biggl\{
- c_{\phi}^2 \sin^2 \frac{ ( h_{3} - h_{2} ) x }{2} 
+ c_{\phi}^2 \sin^2 \frac{ ( h_{3} - h_{1} ) x }{2} 
- ( 1 + s_{\phi}^2 ) \sin^2 \frac{ ( h_{2} - h_{1} ) x }{2} 
\biggr\}
\nonumber \\
&+& 
2 \sin 2\theta_{23}  
\cos 2\theta_{23} 
\nonumber \\
&\times& 
\biggl\{ 
c_{\phi} s_{\phi} \left[
c_{23} \mbox{Re} \left( e^{- i \delta } \alpha_{\mu e} \right) - s_{23} \mbox{Re} \left( \alpha_{\tau e} \right) 
\right] 
+ \cos 2\theta_{23} c_{\phi}^2 \mbox{Re} \left( e^{ i \delta } \alpha_{\tau \mu} \right) 
+ \sin 2\theta_{23} c_{\phi}^2 ( \alpha_{\mu \mu} - \alpha_{\tau \tau} ) 
\biggr\}
\nonumber \\
&\times&
\frac{ \Delta_{b} }{ h_{3} - h_{2} } 
\biggl\{
( 1 + c_{\phi}^2 ) \sin^2 \frac{ ( h_{3} - h_{2} ) x }{2} 
- s_{\phi}^2 \sin^2 \frac{ ( h_{3} - h_{1} ) x }{2}
+ s_{\phi}^2 \sin^2 \frac{ ( h_{2} - h_{1} ) x }{2}
\biggr\}
\nonumber \\
&+& 
\sin^2 2\theta_{23} 
\sin 2 \phi 
\biggl\{
\sin 2 \phi 
\left[ 
\left\{ \left( \frac{ \Delta_{a} }{ \Delta_{b} } - 1 \right) \alpha_{ee} + \alpha_{\mu \mu} \right\} 
- c_{23}^2 ( \alpha_{\mu \mu} - \alpha_{\tau \tau} ) 
+ c_{23} s_{23} \mbox{Re} \left( e^{i \delta } \alpha_{\tau \mu} \right)
\right] 
\nonumber \\
&-&
\cos 2 \phi 
\left[
s_{23} \mbox{Re} \left( e^{- i \delta } \alpha_{\mu e} \right) 
+ c_{23} \mbox{Re} \left( \alpha_{\tau e} \right) 
\right]
\biggr\}
\nonumber \\
&\times&
\frac{ \Delta_{b} }{ h_{3} - h_{1} } 
\biggl\{
\sin^2 \frac{ ( h_{3} - h_{2} ) x }{2} 
+ ( c_{\phi}^2 - s_{\phi}^2 ) \sin^2 \frac{ ( h_{3} - h_{1} ) x }{2}
- \sin^2 \frac{ ( h_{2} - h_{1} ) x }{2}
\biggr\}
\nonumber \\
&-& 
4 \sin 2\theta_{23} 
c_{\phi}^2
\biggl\{ 
c_{\phi} s_{\phi} 
\left[ c_{23} \mbox{Im} \left( e^{- i \delta } \alpha_{\mu e} \right) - s_{23} \mbox{Im} \left( \alpha_{\tau e} \right) \right]
+ s_{\phi}^2 \mbox{Im} \left( e^{ i \delta } \alpha_{\tau \mu} \right) 
\biggr\}
\nonumber \\
&\times&
\frac{ \Delta_{b} }{ h_{2} -  h_{1} } 
\sin \frac{ ( h_{3} - h_{1} ) x }{2} 
\sin \frac{ ( h_{1} - h_{2} ) x }{2} 
\sin \frac{ ( h_{2} - h_{3} ) x }{2}
\nonumber \\
&+& 
4 \sin 2\theta_{23} 
s_{\phi}^2 
\biggl\{
c_{\phi} s_{\phi} \left[ 
c_{23} \mbox{Im} \left( e^{- i \delta } \alpha_{\mu e} \right) - s_{23} \mbox{Im} \left( \alpha_{\tau e} \right) 
\right] 
- c_{\phi}^2 \mbox{Im} \left( e^{ i \delta } \alpha_{\tau \mu} \right) 
\biggr\}
\nonumber \\
&\times&
\frac{ \Delta_{b} }{ h_{3} -  h_{2} } 
\sin \frac{ ( h_{3} - h_{1} ) x }{2} 
\sin \frac{ ( h_{1} - h_{2} ) x }{2} 
\sin \frac{ ( h_{2} - h_{3} ) x }{2}. 
\label{P-mutau-intrinsic-UV}
\end{eqnarray}
While, the first order non-unitary contribution to $P(\nu_{\mu} \rightarrow
\nu_{\tau})$ is given by
\begin{eqnarray} 
&& P(\nu_{\mu} \rightarrow \nu_{\tau})^{(1)}_{ \text{UV} } 
\nonumber \\
&=& 
- \sin 2\theta_{23} \sin 2\phi
\left[
c_{23} \mbox{Re} \left( e^{- i \delta } \alpha_{\mu e} \right) 
+ s_{23} \mbox{Re} \left( \alpha_{\tau e} \right) 
\right]
\nonumber \\
&\times& 
\left\{
\cos 2 \phi \sin^2 \frac{ ( h_{3} - h_{1} ) x }{2} 
+ \sin^2 \frac{ ( h_{3} - h_{2} ) x }{2}
- \sin^2 \frac{ ( h_{2} - h_{1} ) x }{2} 
\right\} 
\nonumber \\
&+& 
\sin 2\theta_{23} \sin^2 2\phi
\left[
s^2_{23} \mbox{Re} \left( e^{ i \delta } \alpha_{\tau \mu} \right) 
+ c_{23} s_{23} ( \alpha_{\mu \mu} + \alpha_{\tau \tau} ) 
\right]
\sin^2 \frac{ ( h_{3} - h_{1} ) x }{2} 
\nonumber \\
&+& 
2 \sin 2\theta_{23} 
\left[
\cos 2 \theta_{23} \mbox{Re} \left( e^{ i \delta } \alpha_{\tau \mu} \right) 
- \sin 2\theta_{23} ( \alpha_{\mu \mu} + \alpha_{\tau \tau} ) 
\right]
\biggl\{ 
c^2_{\phi} \sin^2 \frac{ ( h_{3} - h_{2} ) x }{2}
+ s^2_{\phi} \sin^2 \frac{ ( h_{2} - h_{1} ) x }{2} 
\biggr\}
\nonumber \\
&-& 
2 \sin 2\theta_{23} \sin 2\phi 
\left[
c_{23} \mbox{Im} \left( e^{- i \delta } \alpha_{\mu e} \right) 
- s_{23} \mbox{Im} \left( \alpha_{\tau e} \right) 
\right]
\sin \frac{ ( h_{3} - h_{1} ) x }{2} 
\sin \frac{ ( h_{1} - h_{2} ) x }{2} 
\sin \frac{ ( h_{2} - h_{3} ) x }{2} 
\nonumber \\
&+& 
\sin 2\theta_{23}
\mbox{Im} \left( e^{ i \delta } \alpha_{\tau \mu} \right) 
\biggl\{
c^2_{\phi} \sin ( h_{3} - h_{2} ) x 
- s^2_{\phi} \sin ( h_{2} - h_{1} ) x 
\biggr\}.
\label{P-mutau-extrinsic-UV}
\end{eqnarray}
Here in $P(\nu_{\mu} \rightarrow \nu_{\tau})_{ \text{ UV } }^{(1)}$ we encounter the diagonal $\alpha$ parameter correlation $\alpha_{\mu \mu} + \alpha_{\tau \tau}$. It looks in parallel with the case of $P(\nu_{\mu} \rightarrow \nu_{e})_{ \text{ UV } }^{(1)}$ in \eqref{P-mue-extrinsic-UV} which displays $\alpha_{ee} + \alpha_{\mu \mu}$ correlation. The origin of these peculiar correlations is not understood by the present authors.  

\subsection{Perturbative unitarity yes and no of the first order oscillation probability in $\nu_{\mu}$ row}
\label{sec:unitarity-nu-mu-row}

Given the expressions of the oscillation probabilities in eqs.~\eqref{P-mue-intrinsic-UV},~\eqref{P-mumu-intrinsic-UV},~and~\eqref{P-mutau-intrinsic-UV}, it is straightforward to prove perturbative unitarity for neutrino evolution to first order in $\alpha_{\beta \gamma}$ in $\nu_{\mu}$ row
\begin{eqnarray} 
&& 
P(\nu_{\mu} \rightarrow \nu_{e})_{ \text{ EV } }^{(1)} 
+ P(\nu_{\mu} \rightarrow \nu_{\mu})_{ \text{ EV } }^{(1)} 
+ P(\nu_{\mu} \rightarrow \nu_{\tau})_{ \text{ EV } }^{(1)} =0.
\label{unitarity-numu-row}
\end{eqnarray}
On the other hand, the non-unitary part in first order in $\alpha$ parameters in the oscillation probabilities in $\nu_{\mu}$ row, eqs.~\eqref{P-mue-extrinsic-UV},~\eqref{P-mumu-extrinsic-UV},~and~\eqref{P-mutau-extrinsic-UV}, gives no indication even for a partial cancellation. Clearly the extrinsic UV corrections do not respect unitarity, as in the case of $\nu_{e}$ row.

\section{$\alpha$ matrix multiplication to the lozenge matrix}
\label{sec:alpha-multiplication}

One can easily show that $\alpha Y$ and $Y \alpha^{\dagger}$ are the lozenge position $e^{ \pm i \delta }$ matrices by simple calculation, as shown below. Notice that inside the square parenthesis only the quantities with the canonical phase combination live.
\begin{eqnarray}
\hspace{-14mm} 
\alpha Y =  
\left[
\begin{array}{ccc}
\left[ \alpha_{ee} Y_{e e} \right] & 
\left[ \alpha_{ee} Y_{e \mu} \right] e^{- i \delta} & 
\left[ \alpha_{ee} Y_{e \tau} \right] \\
\left[ e^{ - i \delta} \alpha_{\mu e} Y_{e e} + \alpha_{\mu \mu} Y_{\mu e} \right] e^{ i \delta} & 
\left[ e^{- i \delta} \alpha_{\mu e} Y_{e \mu} + \alpha_{\mu \mu} Y_{\mu \mu} \right] & 
\left[ e^{ - i \delta} \alpha_{\mu e} Y_{e \tau} + \alpha_{\mu \mu} Y_{\mu \tau} \right] e^{ i \delta} \\
\left[ \alpha_{\tau e} Y_{e e} + e^{ i \delta} \alpha_{\tau \mu} Y_{\mu e} + \alpha_{\tau \tau} Y_{\tau e} \right] & 
\left[ \alpha_{\tau e} Y_{e \mu} + e^{ i \delta} \alpha_{\tau \mu} Y_{\mu \mu} + \alpha_{\tau \tau} Y_{\tau \mu} \right] e^{- i \delta} & 
\left[ \alpha_{\tau e} Y_{e \tau} + e^{ i \delta} \alpha_{\tau \mu} Y_{\mu \tau} + \alpha_{\tau \tau} Y_{\tau \tau} \right] \\
\end{array}
\right], 
\nonumber \\
\label{alpha-Y}
\end{eqnarray}
\begin{eqnarray}
&& Y \alpha^{\dagger} =
\left[
\begin{array}{ccc}
\left[ \alpha_{ee} Y_{e e} \right] & 
\left[ \left( e^{- i \delta} \alpha_{\mu e} \right)^* Y_{e e} + \alpha_{\mu \mu} Y_{e \mu} \right] e^{- i \delta} & 
\left[ \alpha_{\tau e}^* Y_{e e} + \left( e^{ i \delta} \alpha_{\tau \mu} \right)^* Y_{e \mu} + \alpha_{\tau \tau} Y_{e \tau} \right] \\
\left[ \alpha_{ee} Y_{\mu e} \right] e^{ i \delta} & 
\left[ \left( e^{- i \delta} \alpha_{\mu e} \right)^* Y_{\mu e} + \alpha_{\mu \mu} Y_{\mu \mu} \right] & 
\left[ \alpha_{\tau e}^* Y_{\mu e} + \left( e^{ i \delta} \alpha_{\tau \mu} \right)^* Y_{\mu \mu} + \alpha_{\tau \tau} Y_{\mu \tau} \right] e^{ i \delta} \\
\left[ \alpha_{ee} Y_{\tau e} \right] & 
\left[ \left( e^{- i \delta} \alpha_{\mu e} \right)^* Y_{\tau e} + \alpha_{\mu \mu} Y_{\tau \mu} \right] e^{- i \delta} & 
\left[ \alpha_{\tau e}^* Y_{\tau e} + \left( e^{ i \delta} \alpha_{\tau \mu} \right)^* Y_{\tau \mu} + \alpha_{\tau \tau} Y_{\tau \tau} \right] \\
\end{array}
\right]. 
\nonumber \\
\label{Y-alpha}
\end{eqnarray}

\section{Identifying the relevant variables}
\label{sec:relevant-variables}

When non-unitarity is introduced the number of parameters increases from six ($\nu$SM) to fifteen (adding nine $\alpha$ parameters), a growth by a factor of 2.5. Here, we look for a possibility of reducing the number of parameters by finding an extra small parameter by which the oscillation probability can be expanded. By the estimate $a / \Delta m^2_{ \text{ren} } \lsim 0.1$ (assuming $Y_{e} = 0.5$) for $E=1$ GeV and $\rho = 3~\text{g/cm}^3$, $\sin \phi$ can be approximated as $s_{13}$. Since the measured value of $\theta_{13}$ is small, $s_{13} = 0.148$, which is the one from the largest statistics measurement \cite{Adey:2018zwh},\footnote{
See some recent global fits \cite{Capozzi:2018ubv,Esteban:2018azc,deSalas:2018bym} for the similar values of $s_{13}$.  }
it can be used as another expansion parameter. Then, we can expand the
probability formulas in terms of $s_{\phi} \equiv \sin \phi \simeq
s_{13}$ to first order, assuming $\rho E \ll 10~\text{GeV
  g/cm}^3$.\footnote{
Though we take a short cut here, one can formulate a systematic expansion by $s_{13}$, called ``$\sqrt{\epsilon}$ perturbation theory'' \cite{Asano:2011nj,Minakata:2009sr}.
}

Given the oscillation probability formulas tabulated in table~\ref{tab:frac-uncert},  it is easy to expand $P(\nu_{\beta} \rightarrow \nu_{\alpha})$ to first order in $s_{\phi}$. Then, we count the $\alpha$ parameters that remain in the zeroth- and the first-order formulas. The results of this exercise are presented in table~\ref{tab:frac-uncert}.

\begin{table}[h!]
\vglue -0.2cm
\begin{center}
\caption{ The UV $\alpha$ parameters which are present in the UV related part of the first order probability $P(\nu_{\beta} \rightarrow \nu_{\alpha})^{(1)}$ to
  zeroth (second column) and to the first order (third column) in
  $\sin \phi$. The results for anti-neutrino channels are the same as
  the corresponding neutrino channels. 
}
\label{tab:frac-uncert} 
\vglue 0.2cm
\begin{tabular}{c|c|c}
\hline 
channel & 
parameters in $P(\nu_{\beta} \rightarrow \nu_{\alpha})^{(1)}_{ \text{UV} }$ & parameters in $P(\nu_{\beta} \rightarrow \nu_{\alpha})^{(1)}_{ \text{UV} }$ \\
             & in zeroth order in $s_{\phi}$ & to first order in $s_{\phi}$ \\
\hline 
$\nu_{e} \rightarrow \nu_{e}$ & $\alpha_{ee}$ & left col. plus $\mbox{Re} \left( e^{- i \delta } \alpha_{\mu e} \right)$, $\mbox {Re} \left( \alpha_{\tau e} \right) $ \\
\hline
$\nu_{e} \rightarrow \nu_{\mu}$, $\nu_{\mu} \rightarrow \nu_{e}$ & 
does not apply & 
$\mbox {Re} \left( e^{ - i \delta } \alpha_{\mu e} \right)$, 
$\mbox {Im} \left( e^{ - i \delta } \alpha_{\mu e} \right)$, \\
$\nu_{e} \rightarrow \nu_{\tau}$, $\nu_{\tau} \rightarrow \nu_{e}$  &  & $\mbox {Re} \left( \alpha_{\tau e} \right) $, 
$\mbox {Im} \left( \alpha_{\tau e} \right) $ \\
\hline 
$\nu_{\mu} \rightarrow \nu_{\mu}$ & 
$\alpha_{\mu \mu}$, $\alpha_{\tau \tau}$, $\mbox{Re} \left( e^{ i \delta } \alpha_{\tau \mu} \right)$ & 
left col. plus 
$\mbox{Re} \left( e^{- i \delta } \alpha_{\mu e} \right)$, 
$\mbox{Re} \left( \alpha_{\tau e} \right)$ \\
\hline 
$\nu_{\mu} \rightarrow \nu_{\tau}$, $\nu_{\tau} \rightarrow \nu_{\mu}$ & 
$\alpha_{\mu \mu}$, $\alpha_{\tau \tau}$, 
$\mbox{Re} \left( e^{ i \delta } \alpha_{\tau \mu} \right)$, 
$\mbox{Im} \left( e^{ i \delta } \alpha_{\tau \mu} \right)$ & 
left col. plus 
$\mbox{Re} \left( e^{- i \delta } \alpha_{\mu e} \right)$, \\
   &     & 
$\mbox{Im} \left( e^{- i \delta } \alpha_{\mu e} \right)$, 
$\mbox{Re} \left( \alpha_{\tau e} \right)$, 
$\mbox{Im} \left( \alpha_{\tau e} \right)$ \\
\hline 
\end{tabular}
\end{center}
\vglue -0.4cm
\end{table}

A few remarks are in order: First of all, we should note that in the appearance channels, $\nu_{\mu} \rightarrow \nu_{e}$ and $\nu_{\mu} \rightarrow \nu_{\tau}$, all the nine UV parameters come in in propagation in matter if we do not expand in terms of $\sin \phi$. When expended by $\sin \phi$ to first order, reduction of number of parameters is effective for $\nu_{\mu} \rightarrow \nu_{e}$ and $\nu_{e} \rightarrow \nu_{\tau}$ channels, only four parameters out of nine. 
On the other hand, reduction of number of parameters to first order in
$\sin \phi$ is not so effective for $\nu_{\mu} \rightarrow \nu_{\mu}$
and $\nu_{\mu} \rightarrow \nu_{\tau}$ channels, missing only a single
parameter $\alpha_{e e}$.


\begin{thebibliography}{99}

\bibitem{Maki:1962mu}
  Z.~Maki, M.~Nakagawa and S.~Sakata,
``Remarks on the unified model of elementary particles,''
  Prog.\ Theor.\ Phys.\  {\bf 28} (1962) 870.
  doi:10.1143/PTP.28.870

\bibitem{Kajita:2016cak}
  T.~Kajita,
``Nobel Lecture: Discovery of atmospheric neutrino oscillations,''
  Rev.\ Mod.\ Phys.\  {\bf 88} (2016) no.3,  030501.
  doi:10.1103/RevModPhys.88.030501

\bibitem{McDonald:2016ixn}
  A.~B.~McDonald,
``Nobel Lecture: The Sudbury Neutrino Observatory: Observation of flavor change for solar neutrinos,''
  Rev.\ Mod.\ Phys.\  {\bf 88} (2016) no.3,  030502.
  doi:10.1103/RevModPhys.88.030502

\bibitem{Kobayashi:1973fv}
  M.~Kobayashi and T.~Maskawa,
``CP Violation in the Renormalizable Theory of Weak Interaction,''
  Prog.\ Theor.\ Phys.\  {\bf 49} (1973) 652.
  doi:10.1143/PTP.49.652

\bibitem{Abe:2018wpn}
  K.~Abe {\it et al.} [T2K Collaboration],
 ``Search for CP Violation in Neutrino and Antineutrino Oscillations by the T2K Experiment with $2.2\times10^{21}$ Protons on Target,''
  Phys.\ Rev.\ Lett.\  {\bf 121} (2018) no.17,  171802
  doi:10.1103/PhysRevLett.121.171802
  [arXiv:1807.07891 [hep-ex]].

\bibitem{Abe:2019vii}
  K.~Abe {\it et al.} [T2K Collaboration],
 ``Constraint on the Matter-Antimatter Symmetry-Violating Phase in Neutrino Oscillations,''
  arXiv:1910.03887 [hep-ex].

\bibitem{Acero:2019ksn}
  M.~A.~Acero {\it et al.} [NOvA Collaboration],
 ``First Measurement of Neutrino Oscillation Parameters using Neutrinos and Antineutrinos by NOvA,''
  Phys.\ Rev.\ Lett.\  {\bf 123} (2019) no.15,  151803
  doi:10.1103/PhysRevLett.123.151803
  [arXiv:1906.04907 [hep-ex]].

\bibitem{Minakata:2001qm}
  H.~Minakata and H.~Nunokawa,
``Exploring neutrino mixing with low-energy superbeams,''
  JHEP {\bf 0110} (2001) 001
  doi:10.1088/1126-6708/2001/10/001
  [hep-ph/0108085].

\bibitem{Abe:2017aap}
  K.~Abe {\it et al.} [Super-Kamiokande Collaboration],
 ``Atmospheric neutrino oscillation analysis with external constraints in Super-Kamiokande I-IV,''
  Phys.\ Rev.\ D {\bf 97} (2018) no.7,  072001
  doi:10.1103/PhysRevD.97.072001
  [arXiv:1710.09126 [hep-ex]].

\bibitem{Capozzi:2018ubv}
  F.~Capozzi, E.~Lisi, A.~Marrone and A.~Palazzo,
``Current unknowns in the three neutrino framework,''
  Prog.\ Part.\ Nucl.\ Phys.\  {\bf 102} (2018) 48
  doi:10.1016/j.ppnp.2018.05.005
  [arXiv:1804.09678 [hep-ph]].

\bibitem{Esteban:2018azc}
  I.~Esteban, M.~C.~Gonzalez-Garcia, A.~Hernandez-Cabezudo, M.~Maltoni and T.~Schwetz,
 ``Global analysis of three-flavour neutrino oscillations: synergies and tensions in the determination of $\theta_{23}$, $\delta_{CP}$, and the mass ordering,''
  JHEP {\bf 1901} (2019) 106
  doi:10.1007/JHEP01(2019)106
  [arXiv:1811.05487 [hep-ph]].

\bibitem{deSalas:2018bym}
  P.~F.~De Salas, S.~Gariazzo, O.~Mena, C.~A.~Ternes and M.~Tórtola,
``Neutrino Mass Ordering from Oscillations and Beyond: 2018 Status and Future Prospects,''
  Front.\ Astron.\ Space Sci.\  {\bf 5} (2018) 36
  doi:10.3389/fspas.2018.00036
  [arXiv:1806.11051 [hep-ph]].

\bibitem{Abe:2015zbg}
  K.~Abe {\it et al.} [Hyper-Kamiokande Proto- Collaboration],
``Physics potential of a long-baseline neutrino oscillation experiment using a J-PARC neutrino beam and Hyper-Kamiokande,''
  PTEP {\bf 2015} (2015) 053C02
  doi:10.1093/ptep/ptv061
  [arXiv:1502.05199 [hep-ex]].

\bibitem{Abi:2020evt}
  B.~Abi {\it et al.} [DUNE Collaboration],
``Deep Underground Neutrino Experiment (DUNE), Far Detector Technical Design Report, Volume II DUNE Physics,''
  arXiv:2002.03005 [hep-ex].

\bibitem{Baussan:2013zcy}
  E.~Baussan {\it et al.} [ESSnuSB Collaboration],
``A very intense neutrino super beam experiment for leptonic CP violation discovery based on the European spallation source linac,''
  Nucl.\ Phys.\ B {\bf 885} (2014) 127
  doi:10.1016/j.nuclphysb.2014.05.016
  [arXiv:1309.7022 [hep-ex]].

\bibitem{An:2015jdp}
  F.~An {\it et al.} [JUNO Collaboration],
``Neutrino Physics with JUNO,''
  J.\ Phys.\ G {\bf 43} (2016) no.3,  030401
  doi:10.1088/0954-3899/43/3/030401
  [arXiv:1507.05613 [physics.ins-det]].

\bibitem{Abe:2016ero}
  K.~Abe {\it et al.} [Hyper-Kamiokande proto- Collaboration],
``Physics potentials with the second Hyper-Kamiokande detector in Korea,''
  PTEP {\bf 2018} (2018) no.6,  063C01
  doi:10.1093/ptep/pty044
  [arXiv:1611.06118 [hep-ex]].

\bibitem{Kajita:2006bt}
  T.~Kajita, H.~Minakata, S.~Nakayama and H.~Nunokawa,
``Resolving eight-fold neutrino parameter degeneracy by two identical detectors with different baselines,''
  Phys.\ Rev.\ D {\bf 75} (2007) 013006
  doi:10.1103/PhysRevD.75.013006
  [hep-ph/0609286].

\bibitem{Kumar:2017sdq}
  S.~Ahmed {\it et al.} [ICAL Collaboration],
``Physics Potential of the ICAL detector at the India-based Neutrino Observatory (INO),''
  Pramana {\bf 88} (2017) no.5,  79
  doi:10.1007/s12043-017-1373-4
  [arXiv:1505.07380 [physics.ins-det]].

\bibitem{TheIceCube-Gen2:2016cap}
  M.~G.~Aartsen {\it et al.} [IceCube Collaboration],
 ``PINGU: A Vision for Neutrino and Particle Physics at the South Pole,''
  J.\ Phys.\ G {\bf 44} (2017) no.5,  054006
  doi:10.1088/1361-6471/44/5/054006
  [arXiv:1607.02671 [hep-ex]].

\bibitem{Adrian-Martinez:2016zzs}
  S.~Adrián-Martínez {\it et al.},
``Intrinsic limits on resolutions in muon- and electron-neutrino charged-current events in the KM3NeT/ORCA detector,''
  JHEP {\bf 1705} (2017) 008
  doi:10.1007/JHEP05(2017)008
  [arXiv:1612.05621 [physics.ins-det]].





\bibitem{Antusch:2006vwa}
  S.~Antusch, C.~Biggio, E.~Fernandez-Martinez, M.~B.~Gavela and J.~Lopez-Pavon,
 ``Unitarity of the Leptonic Mixing Matrix,''
  JHEP {\bf 0610} (2006) 084
  doi:10.1088/1126-6708/2006/10/084
  [hep-ph/0607020].

\bibitem{Schechter:1980gr}
  J.~Schechter and J.~W.~F.~Valle,
``Neutrino Masses in SU(2) x U(1) Theories,''
  Phys.\ Rev.\ D {\bf 22} (1980) 2227.
  doi:10.1103/PhysRevD.22.2227

\bibitem{Barger:1980tfa}
  V.~D.~Barger, P.~Langacker, J.~P.~Leveille and S.~Pakvasa,
 ``Consequences of Majorana and Dirac Mass Mixing for Neutrino Oscillations,''
  Phys.\ Rev.\ Lett.\  {\bf 45} (1980) 692.
  doi:10.1103/PhysRevLett.45.692

\bibitem{Fong:2016yyh}
  C.~S.~Fong, H.~Minakata and H.~Nunokawa,
 ``A framework for testing leptonic unitarity by neutrino oscillation experiments,''
  JHEP {\bf 1702} (2017) 114
  doi:10.1007/JHEP02(2017)114
  [arXiv:1609.08623 [hep-ph]].

\bibitem{Fong:2017gke}
  C.~S.~Fong, H.~Minakata and H.~Nunokawa,
``Non-unitary evolution of neutrinos in matter and the leptonic unitarity test,''
  JHEP {\bf 1902} (2019) 015
  doi:10.1007/JHEP02(2019)015
  [arXiv:1712.02798 [hep-ph]].

\bibitem{Minakata:2015gra}
  H.~Minakata and S.~J.~Parke,
 ``Simple and Compact Expressions for Neutrino Oscillation Probabilities in Matter,''
  JHEP {\bf 1601} (2016) 180
  doi:10.1007/JHEP01(2016)180
  [arXiv:1505.01826 [hep-ph]].

\bibitem{Arafune:1997hd}
  J.~Arafune, M.~Koike and J.~Sato,
``CP violation and matter effect in long baseline neutrino oscillation experiments,''
  Phys.\ Rev.\ D {\bf 56} (1997) 3093
   Erratum: [Phys.\ Rev.\ D {\bf 60} (1999) 119905]
  doi:10.1103/PhysRevD.60.119905, 10.1103/PhysRevD.56.3093
  [hep-ph/9703351].

\bibitem{Cervera:2000kp}
  A.~Cervera, A.~Donini, M.~B.~Gavela, J.~J.~Gomez Cadenas, P.~Hernandez, O.~Mena and S.~Rigolin,
 ``Golden measurements at a neutrino factory,''
  Nucl.\ Phys.\ B {\bf 579} (2000) 17
   Erratum: [Nucl.\ Phys.\ B {\bf 593} (2001) 731]
  doi:10.1016/S0550-3213(00)00606-4, 10.1016/S0550-3213(00)00221-2
  [hep-ph/0002108].

\bibitem{Escrihuela:2015wra}
  F.~J.~Escrihuela, D.~V.~Forero, O.~G.~Miranda, M.~Tórtola and J.~W.~F.~Valle,
``On the description of non-unitary neutrino mixing,''
  Phys.\ Rev.\ D {\bf 92} (2015) no.5,  053009
  doi:10.1103/PhysRevD.92.053009
  [arXiv:1503.08879 [hep-ph]].

\bibitem{Tanabashi:2018oca}
  M.~Tanabashi {\it et al.} [Particle Data Group],
``Review of Particle Physics,''
  Phys.\ Rev.\ D {\bf 98} (2018) no.3,  030001.
  doi:10.1103/PhysRevD.98.030001

\bibitem{Miranda:2016wdr}
  O.~G.~Miranda, M.~Tortola and J.~W.~F.~Valle,
 ``New ambiguity in probing CP violation in neutrino oscillations,''
  Phys.\ Rev.\ Lett.\  {\bf 117} (2016) no.6,  061804
  doi:10.1103/PhysRevLett.117.061804
  [arXiv:1604.05690 [hep-ph]].

\bibitem{Abe:2017jit}
  Y.~Abe, Y.~Asano, N.~Haba and T.~Yamada,
``Heavy neutrino mixing in the T2HK, the T2HKK and an extension of the T2HK with a detector at Oki Islands,''
  Eur.\ Phys.\ J.\ C {\bf 77} (2017) no.12,  851
  doi:10.1140/epjc/s10052-017-5294-7
  [arXiv:1705.03818 [hep-ph]].

\bibitem{Blennow:2016jkn}
  M.~Blennow, P.~Coloma, E.~Fernandez-Martinez, J.~Hernandez-Garcia and J.~Lopez-Pavon,
``Non-Unitarity, sterile neutrinos, and Non-Standard neutrino Interactions,''
  JHEP {\bf 1704} (2017) 153
  doi:10.1007/JHEP04(2017)153
  [arXiv:1609.08637 [hep-ph]].

\bibitem{FernandezMartinez:2007ms}
  E.~Fernandez-Martinez, M.~B.~Gavela, J.~Lopez-Pavon and O.~Yasuda,
``CP-violation from non-unitary leptonic mixing,''
  Phys.\ Lett.\ B {\bf 649} (2007) 427
  doi:10.1016/j.physletb.2007.03.069
  [hep-ph/0703098].

\bibitem{Goswami:2008mi}
  S.~Goswami and T.~Ota,
 ``Testing non-unitarity of neutrino mixing matrices at neutrino factories,''
  Phys.\ Rev.\ D {\bf 78} (2008) 033012
  doi:10.1103/PhysRevD.78.033012
  [arXiv:0802.1434 [hep-ph]].

\bibitem{Antusch:2009pm}
  S.~Antusch, M.~Blennow, E.~Fernandez-Martinez and J.~Lopez-Pavon,
 ``Probing non-unitary mixing and CP-violation at a Neutrino Factory,''
  Phys.\ Rev.\ D {\bf 80} (2009) 033002
  doi:10.1103/PhysRevD.80.033002
  [arXiv:0903.3986 [hep-ph]].

\bibitem{Antusch:2009gn}
  S.~Antusch, S.~Blanchet, M.~Blennow and E.~Fernandez-Martinez,
``Non-unitary Leptonic Mixing and Leptogenesis,''
  JHEP {\bf 1001} (2010) 017
  doi:10.1007/JHEP01(2010)017
  [arXiv:0910.5957 [hep-ph]].

\bibitem{Antusch:2014woa}
  S.~Antusch and O.~Fischer,
``Non-unitarity of the leptonic mixing matrix: Present bounds and future sensitivities,''
  JHEP {\bf 1410} (2014) 094
  doi:10.1007/JHEP10(2014)094
  [arXiv:1407.6607 [hep-ph]].

\bibitem{Fernandez-Martinez:2016lgt}
  E.~Fernandez-Martinez, J.~Hernandez-Garcia and J.~Lopez-Pavon,
``Global constraints on heavy neutrino mixing,''
  JHEP {\bf 1608} (2016) 033
  doi:10.1007/JHEP08(2016)033
  [arXiv:1605.08774 [hep-ph]].

\bibitem{Ge:2016xya}
  S.~F.~Ge, P.~Pasquini, M.~Tortola and J.~W.~F.~Valle,
 ``Measuring the leptonic CP phase in neutrino oscillations with nonunitary mixing,''
  Phys.\ Rev.\ D {\bf 95} (2017) no.3,  033005
  doi:10.1103/PhysRevD.95.033005
  [arXiv:1605.01670 [hep-ph]].

\bibitem{Dutta:2016vcc}
  D.~Dutta and P.~Ghoshal,
 ``Probing CP violation with T2K, NO$\nu$A and DUNE in the presence of non-unitarity,''
  JHEP {\bf 1609} (2016) 110
  doi:10.1007/JHEP09(2016)110
  [arXiv:1607.02500 [hep-ph]].

\bibitem{Dutta:2016czj}
  D.~Dutta, P.~Ghoshal and S.~Roy,
 ``Effect of Non Unitarity on Neutrino Mass Hierarchy determination at DUNE, NO$\nu$A and T2K,''
  Nucl.\ Phys.\ B {\bf 920} (2017) 385
  doi:10.1016/j.nuclphysb.2017.04.018
  [arXiv:1609.07094 [hep-ph]].

\bibitem{Pas:2016qbg}
  H.~Päs and P.~Sicking,
``Discriminating sterile neutrinos and unitarity violation with CP invariants,''
  Phys.\ Rev.\ D {\bf 95} (2017) no.7,  075004
  doi:10.1103/PhysRevD.95.075004
  [arXiv:1611.08450 [hep-ph]].

\bibitem{Escrihuela:2016ube}
  F.~J.~Escrihuela, D.~V.~Forero, O.~G.~Miranda, M.~Tórtola and J.~W.~F.~Valle,
``Probing CP violation with non-unitary mixing in long-baseline neutrino oscillation experiments: DUNE as a case study,''
  New J.\ Phys.\  {\bf 19} (2017) no.9,  093005
  doi:10.1088/1367-2630/aa79ec
  [arXiv:1612.07377 [hep-ph]].

\bibitem{Rout:2017udo}
  J.~Rout, M.~Masud and P.~Mehta,
``Can we probe intrinsic CP and T violations and nonunitarity at long baseline accelerator experiments?,''
  Phys.\ Rev.\ D {\bf 95} (2017) no.7,  075035
  doi:10.1103/PhysRevD.95.075035
  [arXiv:1702.02163 [hep-ph]].

\bibitem{Li:2018jgd}
  Y.~F.~Li, Z.~z.~Xing and J.~y.~Zhu,
``Indirect unitarity violation entangled with matter effects in reactor antineutrino oscillations,''
  Phys.\ Lett.\ B {\bf 782} (2018) 578
  doi:10.1016/j.physletb.2018.05.079
  [arXiv:1802.04964 [hep-ph]].

\bibitem{Parke:2015goa}
  S.~Parke and M.~Ross-Lonergan,
 ``Unitarity and the three flavor neutrino mixing matrix,''
  Phys.\ Rev.\ D {\bf 93} (2016) no.11,  113009
  doi:10.1103/PhysRevD.93.113009
  [arXiv:1508.05095 [hep-ph]].


\bibitem{Wolfenstein:1977ue}
  L.~Wolfenstein,
``Neutrino Oscillations in Matter,''
  Phys.\ Rev.\ D {\bf 17} (1978) 2369.
  doi:10.1103/PhysRevD.17.2369

\bibitem{Valle:1987gv}
  J.~W.~F.~Valle,
 ``Resonant Oscillations of Massless Neutrinos in Matter,''
  Phys.\ Lett.\ B {\bf 199} (1987) 432.
  doi:10.1016/0370-2693(87)90947-6

\bibitem{Broncano:2002rw}
  A.~Broncano, M.~B.~Gavela and E.~E.~Jenkins,
``The Effective Lagrangian for the seesaw model of neutrino mass and leptogenesis,''
  Phys.\ Lett.\ B {\bf 552} (2003) 177
   Erratum: [Phys.\ Lett.\ B {\bf 636} (2006) 332]
  doi:10.1016/j.physletb.2006.04.003, 10.1016/S0370-2693(02)03130-1
  [hep-ph/0210271].

\bibitem{Xing:2007zj}
  Z.~z.~Xing,
 ``Correlation between the Charged Current Interactions of Light and Heavy Majorana Neutrinos,''
  Phys.\ Lett.\ B {\bf 660} (2008) 515
  doi:10.1016/j.physletb.2008.01.038
  [arXiv:0709.2220 [hep-ph]].

\bibitem{Xing:2011ur}
  Z.~z.~Xing,
``A full parametrization of the 6 X 6 flavor mixing matrix in the presence of three light or heavy sterile neutrinos,''
  Phys.\ Rev.\ D {\bf 85} (2012) 013008
  doi:10.1103/PhysRevD.85.013008
  [arXiv:1110.0083 [hep-ph]].

\bibitem{Li:2015oal}
  Y.~F.~Li and S.~Luo,
``Neutrino Oscillation Probabilities in Matter with Direct and Indirect Unitarity Violation in the Lepton Mixing Matrix,''
  Phys.\ Rev.\ D {\bf 93} (2016) no.3,  033008
  doi:10.1103/PhysRevD.93.033008
  [arXiv:1508.00052 [hep-ph]].

\bibitem{Kikuchi:2008vq}
  T.~Kikuchi, H.~Minakata and S.~Uchinami,
 ``Perturbation Theory of Neutrino Oscillation with Nonstandard Neutrino Interactions,''
  JHEP {\bf 0903} (2009) 114
  doi:10.1088/1126-6708/2009/03/114
  [arXiv:0809.3312 [hep-ph]].

\bibitem{Martinez-Soler:2019nhb}
  I.~Martinez-Soler and H.~Minakata,
``Perturbing neutrino oscillations around the solar resonance,''
 PTEP {\bf 2019} (2019) no.7, 073B07 (28 pages)
 [arXiv:1904.07853 [hep-ph]].

\bibitem{Denton:2016wmg}
  P.~B.~Denton, H.~Minakata and S.~J.~Parke,
``Compact Perturbative Expressions For Neutrino Oscillations in Matter,''
  JHEP {\bf 1606} (2016) 051
  doi:10.1007/JHEP06(2016)051
  [arXiv:1604.08167 [hep-ph]].


\bibitem{Akhmedov:2004ny}
  E.~K.~Akhmedov, R.~Johansson, M.~Lindner, T.~Ohlsson and T.~Schwetz,
 ``Series expansions for three flavor neutrino oscillation probabilities in matter,''
  JHEP {\bf 0404} (2004) 078
  doi:10.1088/1126-6708/2004/04/078
  [hep-ph/0402175].

\bibitem{Kimura:2002wd}
  K.~Kimura, A.~Takamura and H.~Yokomakura,
``Exact formulas and simple CP dependence of neutrino oscillation probabilities in matter with constant density,''
  Phys.\ Rev.\ D {\bf 66} (2002) 073005
  doi:10.1103/PhysRevD.66.073005
  [hep-ph/0205295].

\bibitem{Agarwalla:2013tza}
  S.~K.~Agarwalla, Y.~Kao and T.~Takeuchi,
``Analytical approximation of the neutrino oscillation matter effects at large $\theta_{13}$,''
  JHEP {\bf 1404} (2014) 047
  doi:10.1007/JHEP04(2014)047
  [arXiv:1302.6773 [hep-ph]].
  
\bibitem{Mikheev:1986gs}
  S.~P.~Mikheev and A.~Y.~Smirnov,
``Resonance Amplification of Oscillations in Matter and Spectroscopy of Solar Neutrinos,''
  Sov.\ J.\ Nucl.\ Phys.\  {\bf 42} (1985) 913
   [Yad.\ Fiz.\  {\bf 42} (1985) 1441].

\bibitem{Martinez-Soler:2019noy}
  I.~Martinez-Soler and H.~Minakata,
``On the Nature of Correlation between Neutrino-SM CP Phase and Unitarity Violating New Physics Parameters,''
  arXiv:1908.04855 [hep-ph].

\bibitem{Ohlsson:2012kf}
  T.~Ohlsson,
 ``Status of non-standard neutrino interactions,''
  Rept.\ Prog.\ Phys.\  {\bf 76} (2013) 044201
  doi:10.1088/0034-4885/76/4/044201
  [arXiv:1209.2710 [hep-ph]].

\bibitem{Miranda:2015dra}
  O.~G.~Miranda and H.~Nunokawa,
``Non standard neutrino interactions: current status and future prospects,''
  New J.\ Phys.\  {\bf 17} (2015) no.9,  095002
  doi:10.1088/1367-2630/17/9/095002
  [arXiv:1505.06254 [hep-ph]].

\bibitem{Farzan:2017xzy}
  Y.~Farzan and M.~Tortola,
``Neutrino oscillations and Non-Standard Interactions,''
  Front.\ in Phys.\  {\bf 6} (2018) 10
  doi:10.3389/fphy.2018.00010
  [arXiv:1710.09360 [hep-ph]].

\bibitem{GonzalezGarcia:2001mp}
  M.~C.~Gonzalez-Garcia, Y.~Grossman, A.~Gusso and Y.~Nir,
``New CP violation in neutrino oscillations,''
  Phys.\ Rev.\ D {\bf 64} (2001) 096006
  doi:10.1103/PhysRevD.64.096006
  [hep-ph/0105159].

\bibitem{Huber:2002bi}
  P.~Huber, T.~Schwetz and J.~W.~F.~Valle,
 ``Confusing nonstandard neutrino interactions with oscillations at a neutrino factory,''
  Phys.\ Rev.\ D {\bf 66} (2002) 013006
  doi:10.1103/PhysRevD.66.013006
  [hep-ph/0202048].

\bibitem{Jarlskog:1985ht}
  C.~Jarlskog,
``Commutator of the Quark Mass Matrices in the Standard Electroweak Model and a Measure of Maximal CP Violation,''
  Phys.\ Rev.\ Lett.\  {\bf 55} (1985) 1039.
  doi:10.1103/PhysRevLett.55.1039

\bibitem{Adey:2018zwh}
  D.~Adey {\it et al.} [Daya Bay Collaboration],
``Measurement of the Electron Antineutrino Oscillation with 1958 Days of Operation at Daya Bay,''
  Phys.\ Rev.\ Lett.\  {\bf 121} (2018) no.24,  241805
  doi:10.1103/PhysRevLett.121.241805
  [arXiv:1809.02261 [hep-ex]].

\bibitem{Minakata:2009sr}
  H.~Minakata,
``Large-Theta(13) Perturbation Theory of Neutrino Oscillation,''
  Acta Phys.\ Polon.\ B {\bf 40} (2009) 3023
  [arXiv:0910.5545 [hep-ph]].

\bibitem{Asano:2011nj}
  K.~Asano and H.~Minakata, 
``Large-Theta(13) Perturbation Theory of Neutrino Oscillation for Long-Baseline Experiments,''
  JHEP {\bf 1106} (2011) 022
  doi:10.1007/JHEP06(2011)022
  [arXiv:1103.4387 [hep-ph]].


\end{thebibliography}
\end{document}